\documentclass[12pt]{iopart}

\usepackage{iopams}
\usepackage{graphicx}
\usepackage{mathrsfs}
\usepackage{color}
\usepackage[table]{xcolor}
\usepackage{hyperref}
\hypersetup{colorlinks, citecolor=black, filecolor=black, linkcolor=black, urlcolor=black}

\expandafter\let\csname equation*\endcsname\relax
\expandafter\let\csname endequation*\endcsname\relax
\usepackage{amsmath}
\allowdisplaybreaks[1]

\definecolor{grayish}{RGB}{230,230,230}

\newcommand{\refEq}[1] {(\ref{#1})}

\newcommand{\Sin}[1]{\ensuremath{\sin \left( #1 \right)}}
\newcommand{\Cos}[1]{\ensuremath{\cos \left( #1 \right)}}
\newcommand{\Sec}[1]{\ensuremath{\sec \left( #1 \right)}}

\newcommand{\ArcTan}[1]{\ensuremath{\text{arctan} \left( #1 \right)}}
\newcommand{\Exp}[1]{\ensuremath{\exp \left( #1 \right)}}

\newcommand{\Nabla}{\ensuremath{\vec{\nabla}}}
\newcommand{\dlpdtheta}{\ensuremath{\frac{\partial l_{p}}{\partial \theta}}}
\newcommand{\dlpdthetaPrime}{\ensuremath{\frac{\partial l_{p}}{\partial \theta'}}}

\begin{document}

\title[Scaling of momentum flux with poloidal shaping mode number]{Scaling of up-down asymmetric turbulent momentum flux with poloidal shaping mode number in tokamaks}

\author{Justin Ball and Felix I. Parra}

\address{Rudolf Peierls Centre for Theoretical Physics, University of Oxford, Oxford OX1 3NP, United Kingdom}
\address{Culham Centre for Fusion Energy, Culham Science Centre, Abingdon OX14 3DB, United Kingdom}
\ead{Justin.Ball@physics.ox.ac.uk}

\begin{abstract}

Breaking the up-down symmetry of tokamaks removes a constraint limiting intrinsic momentum transport, and hence toroidal rotation, to be small. Using gyrokinetic theory, we study the effect of different up-down asymmetric flux surface shapes on the turbulent transport of momentum. This is done by perturbatively expanding the gyrokinetic equation in large flux surface shaping mode number. It is found that the momentum flux generated by shaping that lacks mirror symmetry (which is necessarily up-down asymmetric) has a power law scaling with the shaping mode number. However, the momentum flux generated by mirror symmetric flux surface shaping (even if it is up-down asymmetric) decays exponentially with large shaping mode number. These scalings are consistent with nonlinear local gyrokinetic simulations and indicate that low mode number shaping effects (e.g. elongation, triangularity) are optimal for creating rotation. Additionally it suggests that breaking the mirror symmetry of flux surfaces may generate significantly more toroidal rotation.

\end{abstract}



\section{Introduction}
\label{sec:introduction}

Bulk toroidal rotation has been shown to be beneficial for plasma performance in tokamaks. It can stabilize the resistive wall mode, which allows violation of the Troyon limit \cite{StraitExpRWMstabilizationD3D1995, SabbaghExpRWMstabilizationNSTX2002, deVriesRotMHDStabilization1996, ReimerdesRWMmachineComp2006}, a fundamental constraint on how much plasma pressure can be confined with a given magnetic field \cite{TroyonMHDLimit1984}. Exceeding the Troyon limit directly improves the economic viability of a tokamak power plant \cite{KrakowskiPowerPlantPhysFromEcon1988, NajmabadiARIES_AT2006}. Furthermore, a strong gradient in toroidal rotation can reduce energy transport by shearing turbulent eddies \cite{RitzRotShearTurbSuppression1990, BurrellShearTurbStabilization1997, BarnesPrandtlNum2011, HighcockRotationBifurcation2010, ParraMomentumTransitions2011}.

There are several mechanisms that currently generate rotation in tokamaks. Beams of neutral particles \cite{SuckewerRotationNBI1979} and radio frequency waves \cite{InceLowHybridRotation2009} are commonly used to heat the plasma, but can also drive toroidal momentum. This externally injected momentum is significant in current experiments, but is expected to diminish in larger devices \cite{LiuITERrwmStabilization2004}.

Self-generated momentum transport, driven by plasma turbulence, is observed, even in the absence of external injection. It is called ``intrinsic'' momentum transport, but it is generally weak, creating rotation less than a tenth of plasma sound speed \cite{RiceExpIntrinsicRotMeas2007, ParraEmpiricalRotScaling2012}. Recently however, the strength of intrinsic rotation was explained through a symmetry of the gyrokinetic model \cite{CattoLinearizedGyrokinetics1978, FriemanNonlinearGyrokinetics1982}, a set of equations that are believed to govern turbulence in the core of tokamaks \cite{McKeeTurbulenceScale2001}. This symmetry constrains the turbulent transport of momentum to be on the order of $\rho_{\ast} \equiv \rho_{i} / a \ll 1$, the ratio of the ion gyroradius to the tokamak minor radius \cite{ParraUpDownSym2011, SugamaUpDownSym2011, CamenenPRLSim2009}. Reference \cite{ParraUpDownSym2011} shows that, in the absence of preexisting rotation, this constraint holds as long as the tokamak flux surfaces are up-down symmetric (i.e. have mirror symmetry about the midplane). Further investigation suggests that breaking the up-down symmetry of the magnetic geometry is a practical means to generate significant plasma rotation \cite{CamenenPRLExp2010, BallMomUpDownAsym2014, BallMastersThesis2013, BallShapingPenetration2015}. Hence it appears that up-down asymmetry is the most promising method to generate significant intrinsic momentum transport in a reactor-scale, initially stationary plasma \cite{CamenenRadialVariation2011}.

Subsequent work has demonstrated a new symmetry of the gyrokinetic model \cite{BallMirrorSymArg2016}. This symmetry means that poloidally translating all high order flux surface shaping effects by a single tilt angle has little effect on the transport properties of the equilibrium. This has important consequences for intrinsic rotation generated by up-down asymmetry because it creates a distinction between mirror symmetric tokamaks and non-mirror symmetric tokamaks, which we will explore in depth. ``Mirror symmetric'' refers to tokamaks that have flux surfaces with reflectional symmetry about some line in the poloidal plane. When the line of symmetry is the midplane the mirror symmetric tokamak can also be said to be ``up-down symmetric.'' ``Non-mirror symmetric'' tokamaks have flux surfaces that do not have reflectional symmetry about any line in the poloidal plane.

In this work we compare the intrinsic momentum transport in magnetic geometries with different up-down asymmetric shaping effects. In section \ref{sec:gyrokineticAnalysis}, we present the electrostatic gyrokinetic model and give a generalized version of the local Miller equilibrium, appropriate for specifying unusual up-down asymmetric configurations. Then, we expand the gyrokinetic equation order-by-order in large shaping mode number to compare the momentum flux generated by different types of flux surface shaping. In doing so we will present two distinct arguments concerning the momentum flux generated by the local equilibrium. First, in section \ref{subsec:practicalNonMirrorSymShaping}, we calculate how the momentum flux scales with the shaping effect mode number given a specific set of simplified, non-mirror symmetric geometries. This is designed to give a concrete illustration of the more abstract and general scaling argument for non-mirror symmetric geometries presented in section \ref{subsec:genNonMirrorSymShaping}. Second, in section \ref{subsec:mirrorSymShaping}, we apply the symmetry presented in reference \cite{BallMirrorSymArg2016} to establish the scaling of momentum flux with shaping mode number in mirror symmetric (but still up-down asymmetric) configurations. Then in section \ref{sec:numResults} we compare the analytic results of section \ref{sec:gyrokineticAnalysis} to nonlinear local gyrokinetic simulations. Lastly, section \ref{sec:conclusions} gives a summary, a broad interpretation of the analytic scalings, and some concluding remarks.

\section{Analytic gyrokinetic analysis}
\label{sec:gyrokineticAnalysis}

Gyrokinetics \cite{LeeGenFreqGyro1983, LeeParticleSimGyro1983, DubinHamiltonianGyro1983, HahmGyrokinetics1988, SugamaGyroTransport1996, SugamaHighFlowGyro1998, BrizardGyroFoundations2007, ParraGyrokineticLimitations2008, ParraLagrangianGyro2011, AbelGyrokineticsDeriv2012} is a theoretical framework used to study plasma behavior with perpendicular wavenumbers comparable to the ion gyroradius ($k_{\perp} \rho_{i} \sim 1$) and timescales much slower than the particle cyclotron frequencies ($\omega \ll \Omega_{i} \ll \Omega_{e}$). Fundamentally, gyrokinetics relies on an expansion in $\rho_{\ast} \equiv \rho_{i} / a \ll 1$, where $\rho_{i}$ is the ion gyroradius and $a$ is the tokamak minor radius. These particular scales have been experimentally shown to be appropriate for modeling turbulence \cite{McKeeTurbulenceScale2001}. In deriving the gyrokinetic equations, we expand the distribution function, $f_{s} = f_{s 0} + f_{s 1} + \ldots$, and assume the perturbation is small compared to the background ($f_{s 1} \sim \rho_{\ast} f_{s 0}$) \cite{HowesAstroGyro2006}. Additionally, for tokamak plasmas, axisymmetry implies radially confined orbits and the transport timescale usually exceeds the collisional timescale. As a result, the lowest order distribution function is assumed to be Maxwellian ($f_{s 0} = F_{Ms}$). Here
\begin{align}
   F_{M s} \equiv n_{s} \left( \frac{m_{s}}{2 \pi T_{s}} \right)^{3/2} \text{exp} \left( - \frac{m_{s} w^{2}}{2 T_{s}} \right) \label{eq:maxwellianDef}
\end{align}
is the Maxwellian distribution function for species $s$, $n_{s}$ is the particle density, $m_{s}$ is the particle mass, $T_{s}$ is the temperature, and $\vec{w}$ is the velocity in the frame rotating with the plasma. In this work we will choose to neglect both electromagnetic effects (for simplicity) and pre-existing rotation (because we are interested in generating rotation in a stationary plasma).

Given these assumptions, we can change the coordinates of the Fokker-Plank and quasineutrality equations from real-space coordinates to the guiding center position, i.e. the average position of the particle as it spirals around a magnetic field line. Then we can average over the gyrophase angle $\varphi$, i.e. the angle that determines the particle location on its circular motion perpendicular to the magnetic field. This gives the two governing equations of electrostatic gyrokinetics: the gyrokinetic equation and a modified version of the quasineutrality equation.

The electrostatic gyrokinetic equation, in the absence of rotation and collisions, can be Fourier-analyzed in the directions perpendicular to the magnetic field and written as \cite{ParraUpDownSym2011}
\begin{align}
   \frac{\partial h_{s}}{\partial t} &+ w_{||} \hat{b} \cdot \Nabla \theta \left. \frac{\partial h_{s}}{\partial \theta} \right|_{w_{||}, \mu} + i \left[ \left( w^{2}_{||} + \frac{B}{m_{s}} \mu \right) \left( k_{\psi} v_{d s \psi} + k_{\alpha} v_{d s \alpha} \right) - w^{2}_{||} \left( k_{\alpha} \frac{\mu_{0}}{\Omega_{s} B} \frac{d p}{d \psi} \right) \right] h_{s}  \nonumber \\
   &+ a_{s ||} \left. \frac{\partial h_{s}}{\partial w_{||}} \right|_{\theta, \mu}+ \left\{ \langle \phi \rangle_{\varphi}, h_{s} \right\} - \frac{Z_{s} e F_{M s}}{T_{s}} \frac{\partial \langle \phi \rangle_{\varphi}}{\partial t} \label{eq:gyrokineticEq} \\
   &+ i k_{\alpha} \langle \phi \rangle_{\varphi} F_{M s} \left[ \frac{1}{n_{s}} \frac{d n_{s}}{d \psi} + \left( \frac{m_{s} w^{2}}{2 T_{s}} - \frac{3}{2} \right) \frac{1}{T_{s}} \frac{d T_{s}}{d \psi} \right] = 0 \nonumber
\end{align}
in the $\left( k_{\psi}, k_{\alpha}, \theta, w_{||}, \mu, \varphi, t \right)$ coordinate system. Here $h_{s}$ is the Fourier-analyzed nonadiabatic portion of the distribution function, $t$ is the time, $w_{||}$ is the component of the velocity parallel to $\hat{b} \equiv \vec{B} / B$, the magnetic field unit vector, $\theta$ is the usual cylindrical poloidal angle shown in figure \ref{fig:coordinateSystem}, $\mu \equiv m_{s} w_{\perp}^{2} / 2 B$ is the magnetic moment, $k_{\psi}$ is the radial wavenumber, $k_{\alpha}$ is the wavenumber within the flux surface and perpendicular to the magnetic field, $\mu_{0}$ is the permeability of free space, $\Omega_{s} \equiv Z_{s} e B / m_{s}$ is the gyrofrequency, $p$ is the plasma pressure, $\psi$ is the poloidal magnetic flux, $\phi$ is the Fourier-analyzed electrostatic potential, $Z_{s}$ is the particle charge number, and $e$ is the electric charge of the proton. The magnetic drift coefficients are given by
\begin{align}
   v_{d s \psi} \equiv& - \frac{I \left( \psi \right)}{\Omega_{s} B} \hat{b} \cdot \Nabla \theta \left. \frac{\partial B}{\partial \theta} \right|_{\psi} \label{eq:driftVelPsi}  \\
   v_{d s \alpha} \equiv& - \frac{1}{\Omega_{s}} \left( \left. \frac{\partial B}{\partial \psi} \right|_{\theta} - \left. \frac{\partial B}{\partial \theta} \right|_{\psi} \frac{\hat{b} \cdot \left( \Nabla \theta \times \Nabla \alpha \right)}{B} \right) , \label{eq:driftVelAlpha}
\end{align}
where $I \left( \psi \right) = R B_{\zeta}$ is the toroidal field flux function,
\begin{align}
   \alpha \equiv \zeta - \int_{\theta_{\alpha} \left( \psi \right)}^{\theta} d \theta' A_{\alpha} \left( \psi, \theta' \right) \label{eq:alphaDef}
\end{align}
is a coordinate that selects a particular field line from a given flux surface,
\begin{align}
   A_{\alpha} \left( \psi, \theta \right) \equiv \frac{I \left( \psi \right)}{R^{2} \vec{B} \cdot \Nabla \theta} \label{eq:IalphaDef}
\end{align}
is the integrand in the definition of $\alpha$, and $\theta_{\alpha} \left( \psi \right)$ is a free function that determines the field line selected by $\alpha = 0$ on each flux surface. The parallel acceleration is given by
\begin{align}
   a_{s ||} \equiv - \frac{\mu}{m_{s}} \hat{b} \cdot \Nabla \theta \left. \frac{\partial B}{\partial \theta} \right|_{\psi} . \label{eq:parAccelDef}
\end{align}
The nonlinear term is
\begin{align}
  \left\{ \langle \phi \rangle_{\varphi}, h_{s} \right\} = \sum_{k_{\psi}', k_{\alpha}'} \left( k_{\psi}' k_{\alpha} - k_{\psi} k_{\alpha}' \right) \left\langle \phi \right\rangle_{\varphi} \left( k_{\psi}', k_{\alpha}' \right) h_{s} \left( k_{\psi} - k_{\psi}', k_{\alpha} - k_{\alpha}' \right) \label{eq:nonlinearTerm}
\end{align}
and the gyroaverage is given by
\begin{align}
   \left\langle \ldots \right\rangle_{\varphi} = J_{0} \left( \frac{k_{\perp} \sqrt{2 \mu B}}{\Omega_{s} \sqrt{m_{s}}} \right) \left( \ldots \right) ,
\end{align}
where
\begin{align}
   k_{\perp} = \sqrt{k_{\psi}^{2} \left| \Nabla \psi \right|^{2} + 2 k_{\psi} k_{\alpha} \Nabla \psi \cdot \Nabla \alpha + k_{\alpha}^{2} \left| \Nabla \alpha \right|^{2}} \label{eq:kperpDef}
\end{align}
is the perpendicular wavenumber and $J_{n} \left( \ldots \right)$ is the $n$th order Bessel function of the first kind.

The quasineutrality equation can be Fourier-analyzed and written as \cite{ParraUpDownSym2011}
\begin{align}
  \phi = \left( \sum_{s} \frac{Z_{s}^{2} e^{2} n_{s}}{T_{s}} \right)^{-1} \sum_{s} \frac{2 \pi Z_{s} e B}{m_{s}} \int dw_{||} d \mu \langle h_{s} \rangle_{\varphi} . \label{eq:quasineut}
\end{align}
Solving the gyrokinetic and quasineutrality equations for $h_{s}$ and $\phi$ allows us to calculate the electrostatic, turbulent flux of toroidal angular momentum according to \cite{ParraUpDownSym2011}
\begin{align}
  \Pi_{s} \equiv& - \left\langle R \left\langle \left\langle \int d^{3} w \breve{h}_{s} m_{s} R \left( \vec{w} \cdot \hat{e}_{\zeta} \right) \left( \delta \vec{E} \cdot \hat{e}_{\zeta} \right) \right\rangle_{\psi} \right\rangle_{\Delta \psi} \right\rangle_{\Delta t} \\
   =& \frac{4 \pi^{2} i}{V'} \left\langle \sum_{k_{\psi}, k_{\alpha}} k_{\alpha} \oint d \theta J B \phi \left( k_{\psi}, k_{\alpha} \right) \int dw_{||} d \mu ~ h_{s} \left( - k_{\psi}, - k_{\alpha} \right) \right. \label{eq:momFlux} \\
   &\times \left. \left( \frac{I}{B} w_{||} J_{0} \left( k_{\perp} \rho_{s} \right) + \frac{i}{\Omega_{s}} \frac{k^{\psi}}{B} \frac{\mu B}{m_{s}} \frac{2 J_{1} \left( k_{\perp} \rho_{s} \right)}{k_{\perp} \rho_{s}} \right) \right\rangle_{\Delta t} ,\nonumber
\end{align}
where $\breve{h}_{s}$ is the non-adiabatic perturbed distribution function (before Fourier analysis), $\delta \vec{E}$ is the turbulent electric field, $\left\langle \ldots \right\rangle_{\psi} \equiv \left( 2 \pi / V' \right) \oint_{0}^{2 \pi} d \theta J \left( \ldots \right)$ is the flux surface average, $V' \equiv 2 \pi \oint_{0}^{2 \pi} d\theta J$,
\begin{align}
   J \equiv \left| \Nabla \psi \cdot \left( \Nabla \theta \times \Nabla \zeta \right) \right|^{-1} = \left( \vec{B} \cdot \Nabla \theta \right)^{-1} \label{eq:jacobian}
\end{align}
is the Jacobian, $\left\langle \ldots \right\rangle_{\Delta \psi} \equiv \Delta \psi^{-1} \int_{\Delta \psi} \left( \ldots \right)$ is the coarse-grain average over a radial distance $\Delta \psi$ (which is larger than the scale of the turbulence, but smaller than the scale of the device), $\left\langle \ldots \right\rangle_{\Delta t} \equiv \Delta t^{-1} \int_{\Delta t} \left( \ldots \right)$ is the coarse-grain average over a time $\Delta t$ (which is longer than the turbulent decorrelation time), and $k^{\psi} \equiv \vec{k}_{\perp} \cdot \Nabla \psi = k_{\psi} \left| \Nabla \psi \right|^{2} + k_{\alpha} \Nabla \psi \cdot \Nabla \alpha$.

We note that the following eight coefficients contain all the information about the flux surface geometry: $\hat{b} \cdot \Nabla \theta$, $B$, $v_{d s \psi}$, $v_{d s \alpha}$, $a_{s ||}$, $\left| \Nabla \psi \right|^{2}$, $\Nabla \psi \cdot \Nabla \alpha$, and $\left| \Nabla \alpha \right|^{2}$. In an up-down symmetric tokamak, the coefficients $v_{d s \psi}$, $a_{s ||}$, and $\Nabla \psi \cdot \Nabla \alpha$ are necessarily odd in $\theta$, while $\hat{b} \cdot \Nabla \theta$, $B$, $v_{d s \alpha}$, $\left| \Nabla \psi \right|^{2}$, and $\left| \Nabla \alpha \right|^{2}$ are even. As shown in reference \cite{ParraUpDownSym2011}, the parity of the geometric coefficients in an up-down symmetric tokamak has important consquences for overall symmetry properties of the gyrokinetic equations. The equations become invariant to the $\left( k_{\psi}, k_{\alpha}, \theta, w_{||}, \mu, t \right) \rightarrow \left( - k_{\psi}, k_{\alpha}, - \theta, - w_{||}, \mu, t \right)$ coordinate system transformation, which is not true in up-down asymmetric devices. This symmetry means that, given any solution $h_{s} \left( k_{\psi}, k_{\alpha}, \theta, w_{||}, \mu, t \right)$, we can construct a second solution $- h_{s} \left( - k_{\psi}, k_{\alpha}, - \theta, - w_{||}, \mu, t \right)$ that will also satisfy the gyrokinetic equations. From equation \refEq{eq:momFlux} we see that this second solution will have a momentum flux that cancels that of the original. These two solutions are each valid for different initial conditions, but since the tokamak is presumed to be chaotic, both solutions will arise within a turbulent decorrelation time (statistically speaking). This demonstrates that, in the gyrokinetic limit, the time-averaged momentum flux must be zero in an up-down symmetric tokamak.

In subsection \ref{subsec:MillerEquil} we present a local MHD equilibrium specification that is appropriate for flux surfaces with arbitrary shaping. Then in subsection \ref{subsec:asymptoticExpansion} we briefly preface the asymptotic expansion of the gyrokinetic model in large shaping mode number. In subsection \ref{subsec:practicalNonMirrorSymShaping}, we first calculate the geometric coefficients in the large aspect ratio limit from the MHD equilibrium for a realistic, but simple example geometry. Using this geometry we expand the gyrokinetic equations to determine how the momentum flux scales with the mode number of the symmetry-breaking effect. This concrete example serves to illustrate the derivation for a general geometry without expanding in aspect ratio, which is detailed in subsection \ref{subsec:genNonMirrorSymShaping}. Lastly, in subsection \ref{subsec:mirrorSymShaping} we explain why mirror symmetric geometries are a special case and should be expected to have weak momentum transport.

\subsection{Up-down asymmetric local Miller equilibrium}
\label{subsec:MillerEquil}

We will calculate the local value of the geometric coefficients that appear in the gyrokinetic equation by using the local Miller geometry model \cite{MillerGeometry1998}. The Miller equilibrium model is a way of specifying the local tokamak equilibrium in the vicinity of a single flux surface of interest. The local equilibrium is completely described by the shape of the flux surface of interest (labeled by $r_{\psi 0}$), how this shape changes with $r_{\psi}$ (the minor radial coordinate), and four scalar quantities. Traditionally $B_{0}$ (the on-axis toroidal magnetic field), $q$ (the safety factor), $\hat{s} \equiv \left( r_{\psi 0} / q \right) dq/dr_{\psi}$ (the magnetic shear), and $dp/dr_{\psi}$ (the pressure gradient) are used. Typically, a combination of vertical elongation and positive triangularity are used to specify the flux surface shape, but in this work we will use a completely general flux surface shape specification (similar to that presented in reference \cite{BallMirrorSymArg2016}).

\begin{figure}
 \begin{center}
  \includegraphics[width=0.4\textwidth]{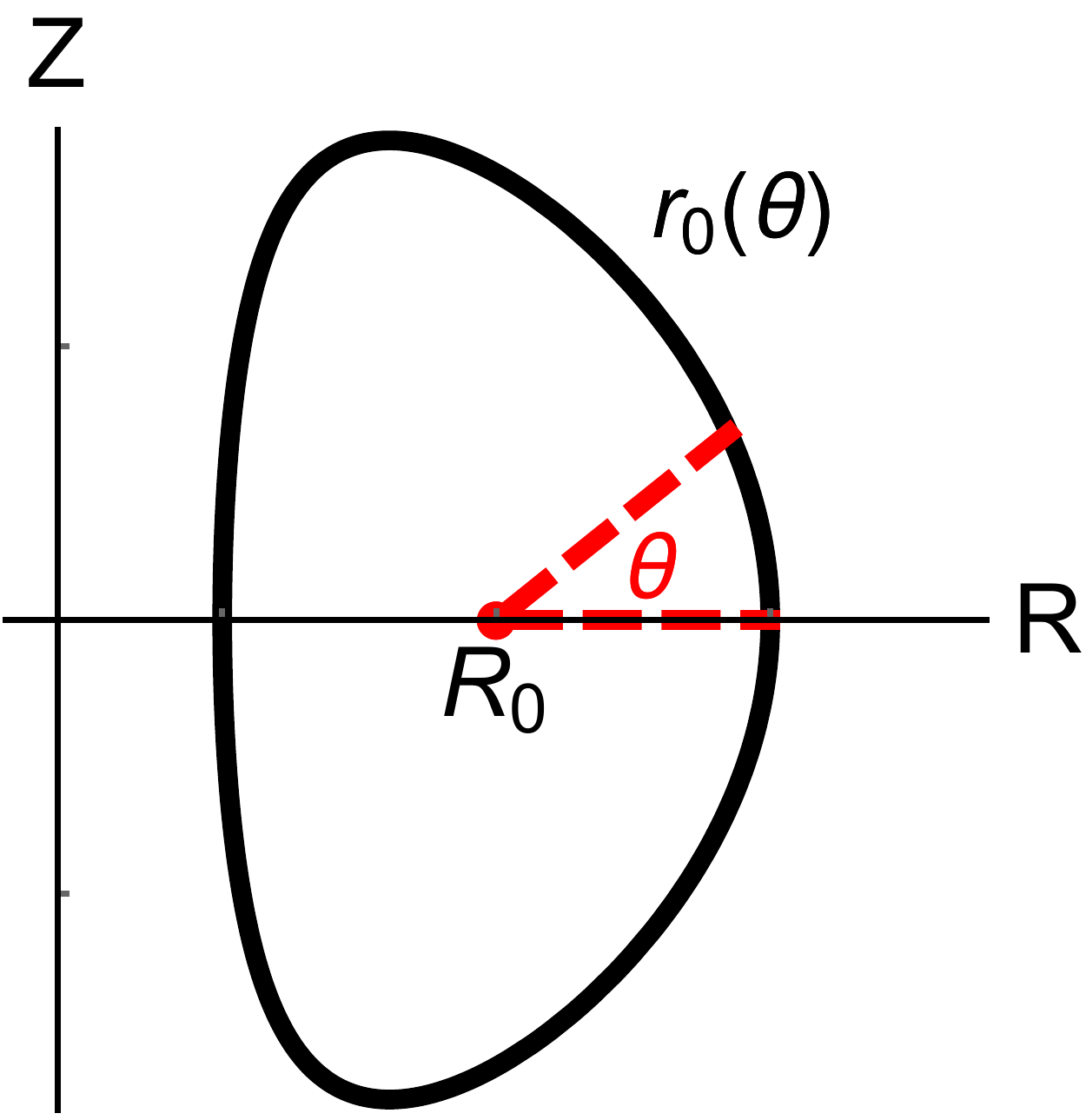}
 \end{center}
 \caption{An example flux surface of interest, $r_{0} \left( \theta \right)$, needed by equation \refEq{eq:gradShafranovLocalEq} for the Miller local equilibrium model. The $\left( R, Z, \zeta \right)$ coordinate system is defined such that the toroidal angle $\zeta$ and the plasma current are coming out of the page.}
 \label{fig:coordinateSystem}
\end{figure}

Since we know that flux surfaces must be periodic in poloidal angle, we are free to Fourier analyze and express them as an infinite series of shaping modes. We will choose to specify the shape of the flux surface of interest in polar form (see figure \ref{fig:coordinateSystem}) as 
\begin{align}
   r_{0} \left( \theta \right) =& r \left( r_{\psi 0}, \theta \right) = r_{\psi 0} \left( 1 - \sum_{m} \frac{\Delta_{m} - 1}{\Delta_{m} + 1} \Cos{m \left( \theta + \theta_{t m} \right)} \right) , \label{eq:fluxSurfaceSpec}
\end{align}
where $m$ is the shaping mode number. Note that this is a completely general Fourier decomposition. The strength of each shaping effect is set by the parameter $\Delta_{m}$. If only one shaping effect is present then $\Delta_{m} = b / a$, where $b$ and $a$ are the maximum and minimum distance of the flux surface from the magnetic axis respectively. When $m=2$, this definition reduces to the usual elongation (typically denoted by $\kappa$). The tilt angles, $\theta_{t m}$, control the relative strength of the sine and cosine terms for every $m$. Lastly, the flux surface label $r_{\psi}$ determines the constant Fourier term. Note the distinction between $a$ (the minimum distance of a flux surface from the magnetic axis) and $r_{\psi}$ (a flux surface label that, as we will see, is defined through equation \refEq{eq:gradShafranovLocalEq}).

Differentiating equation \refEq{eq:fluxSurfaceSpec} radially, we find
\begin{align}
   \left. \frac{\partial r}{\partial r_{\psi}} \right|_{\psi_{0}} =& 1 - \sum_{m} \delta \Delta_{m} \Cos{m \left( \theta + \delta \theta_{t m} \right)} , \label{eq:fluxSurfaceChangeSpec}
\end{align}
where
\begin{align}
   \delta \Delta_{m} \equiv& \sqrt{\left( \frac{\Delta_{m} - 1}{\Delta_{m} + 1} + \frac{2 r_{\psi 0}}{\left( \Delta_{m} + 1 \right)^{2}} \frac{d \Delta_{m}}{d r_{\psi}} \right)^{2} + \left( m r_{\psi 0} \frac{\Delta_{m} - 1}{\Delta_{m} + 1} \frac{d \theta_{t m}}{d r_{\psi}} \right)^{2}} \\
   \delta \theta_{t m} \equiv& \theta_{t m} + \frac{1}{m} \ArcTan{ \left. m r_{\psi 0} \left( \Delta_{m} - 1 \right) \frac{d \theta_{t m}}{d r_{\psi}} \middle/ \left( \left(\Delta_{m} - 1 \right) + \frac{2 r_{\psi 0}}{\Delta_{m} + 1} \frac{d \Delta_{m}}{d r_{\psi}} \right) \right.}
\end{align}
for each $m$. Note that for the local equilibrium all radially varying quantities are evaluated at $r_{\psi} = r_{\psi 0}$ (or equivalently $\psi = \psi_{0}$), the flux surface of interest. The change in the strength, $d \Delta_{m} / d r_{\psi}$, and tilt, $d \theta_{t m} / d r_{\psi}$, of each mode would be determined by the global MHD equilibrium. This is governed by the Grad-Shafranov equation \cite{GradGradShafranovEq1958} and requires the entire radial current profile. In \ref{app:MHDequil} we derive these quantities using a constant current profile in the limits of large aspect ratio and weak shaping. However, in the local Miller equilibrium model the radial variation of the flux surface shape is an input used to construct the poloidal magnetic field. After calculating the poloidal field, the Grad-Shafranov equation is used to calculate all higher order radial derivatives and approximate the global equilibrium.

In summary, the flux surface geometry for the Miller local equilibrium model is completely specified by equations \refEq{eq:fluxSurfaceSpec}, \refEq{eq:fluxSurfaceChangeSpec}, and
\begin{align}
   r \left( r_{\psi}, \theta \right) =& r_{0} \left( \theta \right) + \left. \frac{\partial r}{\partial r_{\psi}} \right|_{\psi_{0}} \left( r_{\psi} - r_{\psi 0} \right) \label{eq:gradShafranovLocalEq} \\
   R \left( r_{\psi}, \theta \right) =& R_{0} + r \left( r_{\psi}, \theta \right) \Cos{\theta} \label{eq:geoMajorRadial} \\
   Z \left( r_{\psi}, \theta \right) =& r \left( r_{\psi}, \theta \right) \Sin{\theta} , \label{eq:geoAxial}
\end{align}
where $r_{\psi 0}$, $\Delta_{m}$, $\theta_{t m}$, $\delta \Delta_{m}$, $\delta \theta_{t m}$, $R_{0}$, $q$, $\hat{s}$, $d p / d r_{\psi}$, and $B_{0}$ (the tokamak major radius) are inputs. We note that if $r_{0} \left( \theta \right) = r_{0} \left( - \theta + \theta_{0} \right)$ and $\left. \partial r / \partial r_{\psi} \right|_{\psi_{0}, \theta} = \left. \partial r / \partial r_{\psi} \right|_{\psi_{0}, -\theta + \theta_{0}}$ for any $\theta_{0}$, then the tokamak is mirror symmetric, otherwise it is non-mirror symmetric. Similarly, if $r_{0} \left( \theta \right) = r_{0} \left( - \theta \right)$ and $\left. \partial r / \partial r_{\psi} \right|_{\psi_{0}, \theta} = \left. \partial r / \partial r_{\psi} \right|_{\psi_{0}, -\theta}$, then the tokamak is up-down symmetric (as well as mirror symmetric), otherwise it is up-down asymmetric.

The full calculation of all eight geometric coefficients is shown in \ref{app:genGeoCoeff}, but for brevity here we will only calculate them to lowest order in $\epsilon \equiv a / R_{0} \ll 1$ (i.e. the inverse aspect ratio). To lowest order in aspect ratio $B \rightarrow B_{0}$ and $a_{|| s} \rightarrow 0$, so we can focus on the other six ($\hat{b} \cdot \Nabla \theta$, $v_{d s \psi}$, $v_{d s \alpha}$, $\left| \Nabla \psi \right|^{2}$, $\Nabla \psi \cdot \Nabla \alpha$, and $\left| \Nabla \alpha \right|^{2}$). In this limit the momentum flux, given by equation \refEq{eq:momFlux}, becomes
\begin{align}
  \Pi_{s} =& \frac{2 \pi i R_{0} B_{0}}{\oint d\theta \left( \hat{b} \cdot \Nabla \theta \right)^{-1}} \sum_{k_{\psi}, k_{\alpha}} k_{\alpha} \oint d\theta \left( \hat{b} \cdot \Nabla \theta \right)^{-1} \label{eq:momFluxSimple} \\
  &\times \int dw_{||} d\mu ~ w_{||} J_{0} \left( k_{\perp} \rho_{s} \right) \phi \left( k_{\psi}, k_{\alpha} \right) h_{s} \left( - k_{\psi}, - k_{\alpha} \right) \nonumber
\end{align}
to lowest order in $\epsilon \ll 1$. For ease of notation we will not use $q$, $\hat{s}$, or $d p / d r_{\psi}$ as inputs to the Miller local equilibrium model. Instead, we will choose to replace $q$ by $d \psi/ d r_{\psi}$ (see equation \refEq{eq:dpsidrpsi}). Also, when we expand to lowest order in aspect ratio, we will find that we can replace both $dp/dr_{\psi}$ and $\hat{s}$ (derived from $d I / d \psi$) with
\begin{align}
  \hat{s}' \equiv& 2 + r_{\psi 0} \left( \frac{d \psi}{d r_{\psi}} \right)^{-1} \left( \mu_{0} R_{0}^{2} \left( \frac{d \psi}{d r_{\psi}} \right)^{-1} \frac{d p}{d r_{\psi}} + R_{0} B_{0} \frac{d I}{d \psi} \right) \label{eq:shiftedShatDef} \\
  =& 2 - r_{\psi 0} \left( \frac{d \psi}{d r_{\psi}} \right)^{-1} \mu_{0} j_{\zeta} R_{0} , \nonumber
\end{align}
where $I \left( \psi \right) \equiv R B_{\zeta}$ is the toroidal field flux function, $j_{\zeta}$ is the current density in the toroidal direction, and $R$ is the major radial coordinate. We can make this replacement because the toroidal current, which appears on the right side of the Grad-Shafranov equation (see equation \refEq{eq:gradShafranov}), is a flux function to lowest order in aspect ratio. We note that if the flux surfaces are exactly circular and $d p / d r_{\psi} = 0$, then $\hat{s}' = \hat{s}$.

Using our geometry specification given by equations \refEq{eq:fluxSurfaceSpec} through \refEq{eq:geoAxial} and employing
\begin{align}
   \Nabla u_{1} =& \frac{\partial \vec{r} / \partial u_{2} \times \partial \vec{r} / \partial u_{3}}{\partial \vec{r} / \partial u_{1} \cdot \left( \partial \vec{r} / \partial u_{2} \times \partial \vec{r} / \partial u_{3} \right)} , \label{eq:gradIdentity}
\end{align}
where $\left( u_{1}, u_{2}, u_{3} \right)$ is a cyclic permutation of $\left( r_{\psi}, \theta, \zeta \right)$, we can directly calculate the poloidal field,
\begin{align}
   \vec{B}_{p} =& \frac{d \psi}{d r_{\psi}} \Nabla \zeta \times \Nabla r_{\psi} . \label{eq:poloidalFieldDef}
\end{align}
This allows us to calculate $\hat{b} \cdot \Nabla \theta$,
\begin{align}
   v_{d s \psi} =& \frac{m_{s}}{Z_{s} e} \hat{b} \cdot \Nabla \theta \frac{\partial R}{\partial \theta}  , \label{eq:driftVelPsiAspect}
\end{align}
and $\left| \Nabla \psi \right|^{2} = R^{2} B_{p}^{2}$ to lowest order in aspect ratio. However, $\Nabla \alpha$ contains second-order radial derivatives, which are not specified. The Miller model determines them by ensuring that the Grad-Shafranov equation \cite{GradGradShafranovEq1958},
\begin{align}
  R^{2} \Nabla \cdot \left( \frac{\Nabla \psi}{R^{2}} \right) = \mu_{0} j_{\zeta} R = -\mu_{0} R^{2} \frac{dp}{d \psi} - I \frac{d I}{d \psi} , \label{eq:gradShafranov}
\end{align}
is satisfied. With considerable work (shown in \ref{app:genGeoCoeff}), we can use the Grad-Shafranov equation to calculate that
\begin{align}
   \Nabla \alpha =& \frac{\partial \alpha}{\partial \psi} \Nabla \psi + \frac{\partial \alpha}{\partial \theta} \Nabla \theta , \label{eq:gradAlphaAspect}
\end{align}
where
\begin{align}
   \frac{\partial \alpha}{\partial \psi} =& - \left. \int_{\theta_{\alpha}}^{\theta} \right|_{\psi} d \theta' \frac{\partial A_{\alpha}}{\partial \psi} + A_{\alpha} \left( \psi, \theta_{\alpha} \right) \frac{d \theta_{\alpha}}{d \psi} \label{eq:gradAlphaPsiAspectUnsimplified} \\
   =& - \left. \int_{\theta_{\alpha}}^{\theta} \right|_{\psi} d \theta' \left( \frac{\partial A_{\alpha}}{\partial \psi} \right)_{\text{orthog}} + \left[ \frac{B_{0}}{R_{0}^{3} B_{p}^{3}} \dlpdthetaPrime \Nabla \psi \cdot \Nabla \theta' \right]_{\theta' = \theta_{\alpha}}^{\theta' = \theta} \label{eq:gradAlphaPsiAspect} \\
   &+ \left( \frac{B_{0}}{R_{0} B_{p}} \dlpdtheta \right)_{\theta = \theta_{\alpha}} \frac{d \theta_{\alpha}}{d \psi} \nonumber
\end{align}
and
\begin{align}
   \frac{\partial \alpha}{\partial \theta} =& - A_{\alpha} \left( \psi, \theta \right) = - \frac{B_{0}}{R_{0} B_{p}} \dlpdtheta
\end{align}
to lowest order in aspect ratio. Here
\begin{align}
   \left( \frac{\partial A_{\alpha}}{\partial \psi} \right)_{\text{orthog}} = \frac{B_{0}}{R_{0}^{2} B_{p}^{2}} \dlpdthetaPrime \left( \frac{d \psi}{d r_{\psi}} \frac{\hat{s}' - 2}{r_{\psi 0} R_{0} B_{p}} + 2 \kappa_{p} \right) \label{eq:gradAlphaAspectIntegrand}
\end{align}
is the part of $\partial A_{\alpha} / \partial \psi$ that remains if the $\left( r_{\psi}, \theta, \zeta \right)$ coordinate system is orthogonal,
\begin{align}
   \kappa_{p} \equiv& - \left( \hat{b}_{p} \cdot \Nabla \hat{b}_{p} \right) \cdot \frac{\Nabla \psi}{\left| \Nabla \psi \right|} = \left( \dlpdtheta \right)^{-3} \left( \frac{\partial R}{\partial \theta} \frac{\partial^{2} Z}{\partial \theta^{2}} - \frac{\partial^{2} R}{\partial \theta^{2}} \frac{\partial Z}{\partial \theta}  \right) \label{eq:poloidalCurv}
\end{align}
is the poloidal magnetic field curvature (defined such that the inwards normal direction is positive), $\hat{b}_{p} \equiv \vec{B}_{p} / B_{p}$ is the poloidal field unit vector, and $l_{p}$ is the poloidal arc length, defined such that
\begin{align}
   \dlpdtheta =& \sqrt{\frac{\partial \vec{r}}{\partial \theta} \cdot \frac{\partial \vec{r}}{\partial \theta}} = \sqrt{\left( \frac{\partial R}{\partial \theta} \right)^{2} + \left( \frac{\partial Z}{\partial \theta} \right)^{2}} . \label{eq:dLpdtheta}
\end{align}
The form of equation \refEq{eq:gradAlphaAspectIntegrand} is useful because it does not contain any radial derivatives (except $d \psi/ d r_{\psi}$ which is an input to the calculation) and distinguishes the important term: the poloidal curvature. This allows us to find $\Nabla \psi \cdot \Nabla \alpha$, $\left| \Nabla \alpha \right|^{2}$, and
\begin{align}
   v_{d s \alpha} =& \frac{1}{\Omega_{s}} \left( \frac{B_{0}}{R_{0}} \frac{\partial R}{\partial \psi} + \frac{\partial R}{\partial \theta} \frac{\partial \alpha}{\partial \psi} \Nabla \psi \cdot \left( \Nabla \theta \times \Nabla \zeta \right) \right) \label{eq:driftVelAlphaAspect}
\end{align}
to lowest order in aspect ratio.

\subsection{Asymptotic expansion ordering}
\label{subsec:asymptoticExpansion}

We know from reference \cite{ParraUpDownSym2011} that, unless the up-down symmetry of the geometric coefficients is broken, the time-averaged momentum flux will always be zero to lowest order in $\rho_{\ast} \equiv \rho_{i} / a \ll 1$. We will investigate the consequences of breaking the up-down symmetry using different shaping effects. To do this we will expand equations \refEq{eq:gyrokineticEq}, \refEq{eq:quasineut}, and \refEq{eq:momFluxSimple} in $m \gg 1$ using
\begin{align}
  h_{s} =& h_{s 0} + h_{s 1} + h_{s 2} + h_{s 3} + \ldots \\
  \phi =& \phi_{0} + \phi_{1} + \phi_{2} + \phi_{3} + \ldots ,
\end{align}
where the subscript indicates the order in $m^{-1} \ll 1$. This expansion separates the long spatial scale coordinate $\theta$, from the short spatial scale coordinate
\begin{align}
  z \equiv m \theta .
\end{align}
Distinguishing the variation on each scale, e.g. $h_{s} \left( \theta, z \right)$ and $\phi \left( \theta, z \right)$, means that
\begin{align}
  \left. \frac{\partial}{\partial \theta} \right|_{w_{||}, \mu} = \left. \frac{\partial}{\partial \theta} \right|_{z, w_{||}, \mu} + m \left. \frac{\partial}{\partial z} \right|_{\theta, w_{||}, \mu} .
\end{align}
Ultimately we will only be interested in large scale phenomena, so we will need to average quantities in $z$ using
\begin{align}
  \overline{\left( \ldots \right)} \equiv \frac{1}{2 \pi} \left. \oint_{-\pi}^{\pi} \right|_{\theta} dz \left( \ldots \right) , \label{eq:zAvg}
\end{align}
but we must still manipulate the $z$-dependent portion, given by
\begin{align}
  \widetilde{\left( \ldots \right)} \equiv \left( \ldots \right) - \overline{\left( \ldots \right)} . \label{eq:zDep}
\end{align}

\subsection{Practical non-mirror symmetric shaping in the gyrokinetic model}
\label{subsec:practicalNonMirrorSymShaping}

In this section we will expand the large aspect ratio gyrokinetic, quasineutrality, and momentum flux equations order-by-order to determine the scaling of the momentum flux with $m \gg 1$. Hence formally we require that $\epsilon \ll 1$ for the aspect ratio expansion and also that $\epsilon \ll m^{-1} \ll 1$ for the subsidiary expansion in shaping mode number. We perform the calculation for $\epsilon \sim 1$ in subsection \ref{subsec:genNonMirrorSymShaping}. The shape of the flux surface of interest (and how it changes with radius) is completely specified by equations \refEq{eq:fluxSurfaceSpec} through \refEq{eq:geoAxial}. We will choose the ordering
\begin{align}
  \Delta_{m} - 1 \sim m^{-2} \label{eq:weakShapingOrdering}
\end{align}
because it is physical (see \ref{app:maxShaping}), straightforward to treat analytically, and arises naturally from experimental flux surface shapes. For regular polygons, $\Delta_{m} - 1 = \Sec{\pi / m} - 1 \sim m^{-2}$, so we see that exceeding this scaling necessarily leads to flux surfaces with convex regions. With the exception of ``bean-shaped'' tokamaks \cite{GrimmBeanShape1985}, practically all configurations have purely concave flux surfaces, so we know they respect this scaling. We can determine how to order the radial derivative by balancing it against the poloidal derivative in the Grad-Shafranov equation (see equations \refEq{eq:gradShafranovLowestOrderNormalized} and \refEq{eq:shapingRadialDeriv}) to find
\begin{align}
  \frac{d \Delta_{m}}{d r_{\psi}} \sim \frac{m \left( \Delta_{m} - 1 \right)}{r_{\psi 0}} . \label{eq:changeShapeOrdering}
\end{align}
Lastly, we take
\begin{align}
   \frac{d \theta_{t m}}{d r_{\psi}} = 0
\end{align}
to lowest order in aspect ratio (as seen in \ref{app:MHDequil} and reference \cite{BallMastersThesis2013}), so $\delta \theta_{t m} = \theta_{t m}$.

In this calculation, we will use flux surfaces with simple shaping that is not mirror symmetric. To create these flux surfaces, we include only two fast shaping effects, $m$ and $n$, in equations \refEq{eq:fluxSurfaceSpec} and \refEq{eq:fluxSurfaceChangeSpec}. They are free to have different strengths, $\Delta_{m}$ and $\Delta_{n}$, and tilt angles, $\theta_{t m}$ and $\theta_{t n}$. However, we order $n - m \sim 1$ (implying that $n \sim m$), $\Delta_{n} - 1 \sim \Delta_{m} - 1 \sim m^{-2}$, and $d \Delta_{n} / d r_{\psi} \sim d \Delta_{m} / d r_{\psi}$. Given these orderings equations \refEq{eq:fluxSurfaceSpec} and \refEq{eq:fluxSurfaceChangeSpec} become 
\begin{align}
   r_{0} \left( \theta \right) =& r_{\psi 0} \left( 1 - \frac{\Delta_{m} - 1}{2} \Cos{z_{m s}} - \frac{\Delta_{n} - 1}{2} \Cos{z_{n s}} \right) + O \left( m^{-4} r_{\psi 0} \right) \label{eq:fluxSurfaceSpecSimplified} \\
   \left. \frac{\partial r}{\partial r_{\psi}} \right|_{\psi_{0}} =& 1 - \frac{r_{\psi 0}}{2} \frac{d \Delta_{m}}{d r_{\psi}} \Cos{z_{m s}} - \frac{r_{\psi 0}}{2} \frac{d \Delta_{n}}{d r_{\psi}} \Cos{z_{n s}} + O \left( m^{-2} \right) , \label{eq:fluxSurfaceChangeSpecSimplified}
\end{align}
where
\begin{align}
  z_{m s} \equiv& m \left( \theta + \theta_{t m} \right) \label{eq:zmsDef} \\
  z_{n s} \equiv& n \left( \theta + \theta_{t n} \right) . \label{eq:znsDef}
\end{align}

\subsubsection{Geometric coefficients.}
\label{subsubsec:geoCoeffs}

To lowest order, $O \left( 1 \right)$, the geometric coefficients are those of a circular tokamak and are entirely independent of the short spatial scale coordinate, $z$. To next order the coefficients depend on $z$, but are algebraically intensive to find. The full expressions for all six coefficients (and several intermediate quantities that are useful in the derivation) are given in \ref{app:nonMirrorGeoCoeffs}, but here we will only derive $v_{d s \alpha}$ to serve as an illustrative example. This coefficient signifies the magnetic drifts in the direction perpendicular to the magnetic field, but still within the flux surface. We will start with equations \refEq{eq:gradShafranovLocalEq}, \refEq{eq:geoMajorRadial}, \refEq{eq:geoAxial}, \refEq{eq:fluxSurfaceSpecSimplified}, and \refEq{eq:fluxSurfaceChangeSpecSimplified} and use them to construct all of the quantities appearing in equation \refEq{eq:driftVelAlphaAspect}, the expression for $v_{d s \alpha}$ to lowest order in aspect ratio.

It will be sufficient to calculate all quantities to $O \left( m^{-1} \right)$ with the exception of $\partial R / \partial \theta$ and $\partial Z / \partial \theta$ because they appear in the poloidal curvature with an extra poloidal derivative (see equation \refEq{eq:poloidalCurv}). This extra derivative creates an additional factor of $m$, which boosts $O \left( m^{-2} \right)$ effects to $O \left( m^{-1} \right)$. Directly differentiating equations \refEq{eq:geoMajorRadial} and \refEq{eq:geoAxial} we find
\begin{align}
   \frac{\partial R}{\partial \theta} =& \frac{d r_{0}}{d \theta} \Cos{\theta} - r_{0} \Sin{\theta} \label{eq:partialRpartialTheta} \\
   \frac{\partial Z}{\partial \theta} =& \frac{d r_{0}}{d \theta} \Sin{\theta} + r_{0} \Cos{\theta} \label{eq:partialZpartialTheta} \\
      \frac{\partial^{2} R}{\partial \theta^{2}} =& \frac{d^{2} r_{0}}{d \theta^{2}} \Cos{\theta} - 2 \frac{d r_{0}}{d \theta} \Sin{\theta} - r_{0} \Cos{\theta} \label{eq:partialSqRpartialThetaSq} \\
   \frac{\partial^{2} Z}{\partial \theta^{2}} =& \frac{d^{2} r_{0}}{d \theta^{2}} \Sin{\theta} + 2 \frac{d r_{0}}{d \theta} \Cos{\theta} - r_{0} \Sin{\theta} , \label{eq:partialSqZpartialThetaSq}
\end{align}
where
\begin{align}
   \frac{d r_{0}}{d \theta} =& \frac{r_{\psi 0}}{2} \Big( m \left( \Delta_{m} - 1 \right) \Sin{z_{m s}} + n \left( \Delta_{n} - 1 \right) \Sin{z_{n s}} \Big) + O \left( m^{-3} r_{\psi 0} \right) \\
   \frac{d^{2} r_{0}}{d \theta^{2}} =& \frac{r_{\psi 0}}{2} \Big( m^{2} \left( \Delta_{m} - 1 \right) \Cos{z_{m s}} + n^{2} \left( \Delta_{n} - 1 \right) \Cos{z_{n s}} \Big) + O \left( m^{-2} r_{\psi 0} \right) .
\end{align}
From this point forward we will only need quantities to $O \left( m^{-1} \right)$ to accurately capture the up-down symmetry breaking. Substituting equations \refEq{eq:partialRpartialTheta} and \refEq{eq:partialZpartialTheta} into equation \refEq{eq:dLpdtheta} gives
\begin{align}
   \dlpdtheta =& r_{\psi 0} + O \left( m^{-2} r_{\psi 0} \right) . \label{eq:nonMirrordLpdTheta}
\end{align}
We can now substitute equations \refEq{eq:partialRpartialTheta} through \refEq{eq:nonMirrordLpdTheta} into equation \refEq{eq:poloidalCurv} to find
\begin{align}
   \kappa_{p} =& \frac{1}{r_{\psi 0}} \left( 1 - \frac{1}{r_{\psi 0}} \frac{d^{2} r_{0}}{d \theta^{2}} \right) + O \left( \frac{m^{-2}}{r_{\psi 0}} \right) \label{eq:nonMirrorPolCurv} \\
   =& \frac{1}{r_{\psi 0}} \left( 1 - \frac{1}{2} \left[ m^{2} \left( \Delta_{m} - 1 \right) \Cos{z_{m s}} + n^{2} \left( \Delta_{n} - 1 \right) \Cos{z_{n s}} \right] \right) + O \left( \frac{m^{-2}}{r_{\psi 0}} \right) .
\end{align}

Next we will calculate
\begin{align}
   \frac{\partial R}{\partial r_{\psi}} =& \left. \frac{\partial r}{\partial r_{\psi}} \right|_{\psi_{0}} \Cos{\theta} \label{eq:partialRpartialrPsi} \\
   \frac{\partial Z}{\partial r_{\psi}} =& \left. \frac{\partial r}{\partial r_{\psi}} \right|_{\psi_{0}} \Sin{\theta} \label{eq:partialZpartialrPsi}
\end{align}
straightforwardly from equations \refEq{eq:geoMajorRadial} and \refEq{eq:geoAxial}. We can determine $\Nabla r_{\psi}$ through equation \refEq{eq:gradIdentity} and
\begin{align}
   \Nabla \theta =& \frac{\hat{e}_{\theta}}{r_{0} \left( \theta \right)} \label{eq:gradTheta} \\
   \Nabla \zeta =& \frac{\hat{e}_{\zeta}}{R} = \frac{\hat{e}_{\zeta}}{R_{0}} + O \left( \frac{\epsilon}{R_{0}} \right) \label{eq:gradZeta}
\end{align}
directly, where $\hat{e}_{\theta}$ and $\hat{e}_{\zeta}$ are the poloidal and toroidal angle unit vectors respectively. With this we can find the coordinate scalar triple product to be
\begin{align}
   \Nabla \psi \cdot \left( \Nabla \theta \times \Nabla \zeta \right) = \frac{1}{J} = \frac{1}{r_{\psi 0} R_{0}} \frac{d \psi}{d r_{\psi}} \left( \left. \frac{\partial r}{\partial r_{\psi}} \right|_{\psi_{0}} \right)^{-1} + O \left( m^{-2} \frac{B_{0}}{R_{0}} \right) , \label{eq:tripleProd}
\end{align}
which is needed to calculate the second term of equation \refEq{eq:driftVelAlphaAspect}. Since we are using $d \psi / d r_{\psi}$ as an input instead of $q$, it is simple to find $\partial R / \partial \psi$ from equation \refEq{eq:partialRpartialrPsi} in order to calculate the first term of equation \refEq{eq:driftVelAlphaAspect}.

At this point we see that we have calculated all of the quantities appearing in equation \refEq{eq:driftVelAlphaAspect}, except for $\partial \alpha / \partial \psi$. This is specified by equation \refEq{eq:gradAlphaPsiAspect} and is made up of three terms. All of the terms require that we know
\begin{align}
   B_{p} = \frac{1}{J} \frac{\partial l_{p}}{\partial \theta} = \frac{1}{R_{0}} \frac{d \psi}{d r_{\psi}} \left( \left. \frac{\partial r}{\partial r_{\psi}} \right|_{\psi_{0}} \right)^{-1} + O \left( m^{-2} B_{p} \right) , \label{eq:BpAspect}
\end{align}
which is found using equations \refEq{eq:poloidalFieldDef}, \refEq{eq:dLpdtheta}, \refEq{eq:gradZeta}, \refEq{eq:gradIdentity}, and \refEq{eq:tripleProd}. Using equations \refEq{eq:nonMirrordLpdTheta}, \refEq{eq:nonMirrorPolCurv}, and \refEq{eq:BpAspect}, we can calculate the integrand (see equation \refEq{eq:gradAlphaAspectIntegrand}) that appears in the first term to be
\begin{align}
   \left( \frac{\partial A_{\alpha}}{\partial \psi} \right)_{\text{orthog}} =& B_{0} \left( \frac{d \psi}{d r_{\psi}} \right)^{-2} \left( \left. \frac{\partial r}{\partial r_{\psi}} \right|_{\psi_{0}} \right)^{2} \label{eq:nonMirrorAlphaIntegrand} \\
   &\times \left[ \left( \hat{s}' - 2 \right) \left. \frac{\partial r}{\partial r_{\psi}} \right|_{\psi_{0}} + 2 \left( 1 - \frac{1}{r_{\psi 0}} \frac{d^{2} r_{0}}{d \theta'^{2}} \right) \right] + O \left( \frac{m^{-2}}{r_{\psi 0}^{2} B_{0}} \right) \nonumber
\end{align}
to lowest order in aspect ratio. Finding the indefinite integral of equation \refEq{eq:nonMirrorAlphaIntegrand} is straightforward and is explicitly given in \ref{app:nonMirrorGeoCoeffs}. The second term of equation \refEq{eq:gradAlphaPsiAspect} is found to be
\begin{align}
   \frac{B_{0}}{R_{0}^{3} B_{p}^{3}} \dlpdthetaPrime \Nabla \psi \cdot \Nabla \theta' = - \frac{B_{0}}{r_{\psi 0}} \left( \frac{d \psi}{d r_{\psi}} \right)^{-2} \left( \left. \frac{\partial r}{\partial r_{\psi}} \right|_{\psi_{0}} \right)^{2} \frac{d r_{0}}{d \theta'} + O \left( \frac{m^{-2}}{r_{\psi 0}^{2} B_{0}} \right)
\end{align}
by substituting equations \refEq{eq:nonMirrordLpdTheta}, \refEq{eq:gradTheta}, \refEq{eq:gradIdentity}, and \refEq{eq:BpAspect}. At this point, by specifying the free parameter
\begin{align}
   \frac{d \theta_{\alpha}}{d \psi} =& \left( \frac{B_{0}}{R_{0} B_{p}} \dlpdtheta \right)_{\theta = \theta_{\alpha}}^{-1} \label{eq:nonMirrorThetaAlphaDerivCond} \\
   &\times \left[ - \left. \int_{\theta_{0}}^{\theta_{\alpha}} \right|_{\psi} d \theta' \left( \frac{\partial A_{\alpha}}{\partial \psi} \right)_{\text{orthog}} + \left( \frac{B_{0}}{R_{0}^{3} B_{p}^{3}} \dlpdthetaPrime \Nabla \psi \cdot \Nabla \theta' \right)_{\theta' = \theta_{\alpha}} \right] , \nonumber
\end{align}
we can use the third term of equation \refEq{eq:gradAlphaPsiAspect} to eliminate all of the terms in $\partial \alpha / \partial \psi$ that do not depend on $\theta$. Here $\theta_{0}$ is defined such that the resulting integral does not have a term that is constant in poloidal angle. Additionally, we choose $\theta_{\alpha} \left( \psi_{0} \right) = 0$ for simplicity. Given this choice, equation \refEq{eq:gradAlphaPsiAspect} becomes
\begin{align}
   \frac{\partial \alpha}{\partial \psi} =& - \left. \int_{\theta_{0}}^{\theta} \right|_{\psi} d \theta' \left( \frac{\partial A_{\alpha}}{\partial \psi} \right)_{\text{orthog}} - \frac{B_{0}}{r_{\psi 0}} \left( \frac{d \psi}{d r_{\psi}} \right)^{-2} \left( \left. \frac{\partial r}{\partial r_{\psi}} \right|_{\psi_{0}} \right)^{2} \frac{d r_{0}}{d \theta} + O \left( \frac{m^{-2}}{r_{\psi 0}^{2} B_{0}} \right) . \label{eq:nonMirrorAlphaIntegral}
\end{align}
Substituting equations \refEq{eq:partialRpartialTheta}, \refEq{eq:partialRpartialrPsi}, \refEq{eq:tripleProd}, and \refEq{eq:nonMirrorAlphaIntegral} into equation \refEq{eq:driftVelAlphaAspect} gives
\begin{align}
   v_{d s \alpha} =& \frac{B_{0}}{R_{0} \Omega_{s}} \left( \frac{d \psi}{d r_{\psi}} \right)^{-1} \Bigg[ \frac{d r_{0}}{d r_{\psi}} \Cos{\theta} + \frac{1}{r_{\psi 0}} \frac{d r_{0}}{d r_{\psi}} \frac{d r_{0}}{d \theta} \Sin{\theta} \\
   &+ \frac{1}{B_{0}} \left( \frac{d \psi}{d r_{\psi}} \right)^{2} \left( \Sin{\theta} - \frac{1}{r_{\psi 0}} \frac{\partial r_{0}}{\partial \theta} \Cos{\theta} \right) \left( \frac{d r_{0}}{d r_{\psi}} \right)^{-1} \left. \int_{\theta_{0}}^{\theta} \right|_{\psi} d \theta' \left( \frac{\partial A_{\alpha}}{\partial \psi} \right)_{\text{orthog}} \Bigg] \nonumber \\
   &+ O \left( \frac{m^{-2}}{r_{\psi 0} R_{0} \Omega_{s}} \right) . \nonumber
\end{align}
To lowest order, this is the usual result for circular flux surfaces,
\begin{align}
   v_{d s \alpha 0} =& \frac{B_{0}}{R_{0} \Omega_{s}} \left( \frac{d \psi}{d r_{\psi}} \right)^{-1} \left( \Cos{\theta} + \hat{s}' \theta \Sin{\theta} \right) . \nonumber
\end{align}
To next order this is a complicated expression with the form of
\begin{align}
   v_{d s \alpha 1} =& D_{1} \theta \Sin{\theta} + \left( D_{2} \Sin{\theta} + D_{3} \theta \Cos{\theta} \right) \left( D_{4} \Sin{z_{m s}} + D_{5} \Sin{z_{n s}} \right) \nonumber \\
   &+ \left( D_{6} \Cos{\theta} + D_{7} \theta \Sin{\theta} \right) \left( D_{8} \Cos{z_{m s}} + D_{9} \Cos{z_{n s}} \right) + D_{10} \Sin{\theta} \label{eq:nonMirrorAlphaDrift} \\
   &\times \left[ \Sin{\left( n - m \right) \theta} \Cos{m \left( \theta_{t m} - \theta_{t n} \right)} - \Cos{\left( n - m \right) \theta} \Sin{m \left( \theta_{t m} - \theta_{t n} \right)} \right] . \nonumber
\end{align}
where $D_{i}$ are constants (the full expression is given in \ref{app:nonMirrorGeoCoeffs}). Even after averaging over $z$ the last term remains, which has a coefficient of
\begin{align}
   D_{10} = \frac{r_{\psi 0}}{\left( n - m \right)} \left( m^{2} \left( \Delta_{m} - 1 \right) \frac{d \Delta_{n}}{d r_{\psi}} + n^{2} \left( \Delta_{n} - 1 \right) \frac{d \Delta_{m}}{d r_{\psi}} \right) . \label{eq:symBreakCoeff}
\end{align}
As we will show shortly, this term, which does not disappear after averaging over $z$, breaks the up-down symmetry of the gyrokinetic equations to $O \left( m^{-1} \right)$.

\ref{app:nonMirrorGeoCoeffs} gives the full expressions for all six geometric coefficients to lowest order in aspect ratio. We find those that do not depend on $\Nabla \alpha$ (i.e. $v_{d s \psi}$ and $\left| \Nabla \psi \right|^{2}$) are up-down symmetric in $\theta$ to $O \left( m^{-1} \right)$. However, the other three coefficients (i.e. $v_{d s \alpha}$, $\Nabla \psi \cdot \Nabla \alpha$, and $\left| \Nabla \alpha \right|^{2}$) lose their symmetry at $O \left( m^{-1} \right)$. The symmetry breaking terms arise from the interaction between $\kappa_{p}$ and $B_{p}^{-2}$ in equation \refEq{eq:gradAlphaAspectIntegrand}. Since $m \gg 1$ the second order derivatives in $\kappa_{p}$ (see equation \refEq{eq:poloidalCurv}) brings the effect of shaping from $O \left( m^{-2} \right)$ to $O \left( 1 \right)$. This shaping can then beat with the $O \left( m^{-1} \right)$ shaping in $B_{p}^{-2}$ and break the symmetry of the geometric coefficients to $O \left( m^{-1} \right)$. We note that $\kappa_{p}$ is ``normal'' curvature (i.e. perpendicular to the flux surface), as opposed to ``geodesic'' curvature (i.e. within the flux surface) \cite{RafiqNormalGeodesicCurv2005}. The importance of $\kappa_{p}$ is surprising because it arises from the {\it poloidal} field, not the {\it toroidal} field. Usually the focus is on the ``normal'' curvature of the {\it toroidal} field because it generates the largest contribution to the total field line curvature that appears in the magnetic drifts.

Ultimately, this beating between $\kappa_{p}$ and $B_{p}^{-2}$ is the dominate mechanism that breaks the up-down symmetry of the geometric coefficients to lowest order in aspect ratio. It is a subtle effect because it enters through the integral in $\partial \alpha / \partial \psi$ (see equations \refEq{eq:gradAlphaPsiAspect} and \refEq{eq:gradAlphaAspectIntegrand}), which is contained in $\Nabla \alpha$ (see equation \refEq{eq:gradAlphaAspect}). However, it does not enter into the magnetic drifts. From studying these equations we can see that this mechanism acts through altering the local magnetic shear (but without modifying the total magnetic shear). Therefore, in the perfect $m \gg 1$ limit, adding a small amount of non-mirror symmetric shaping modifies local field line pitch from one flux surface to the next (without changing the field line spacing). This perturbs the local cross-sectional shape (i.e. the shape in the plane perpendicular to the field line) of the turbulent eddies as they wrap around the torus. Specifically, it tilts the eddy cross-sectional shape a small amount one way or the other, depending on the location along the field line. This non-mirror symmetric perturbation to the eddy is then acted on by the original mirror symmetric magnetic drifts.

The interaction of $\kappa_{p}$ and $B_{p}^{-2}$ certainly breaks the up-down symmetry of the geometric coefficients and generates momentum flux, but it is still unclear at what order. By expanding the gyrokinetic and quasineutrality equations order-by-order in $m^{-1} \ll 1$ we will connect the symmetry-breaking of the geometric coefficients to symmetry-breaking of the distribution function and non-zero momentum flux.

\subsubsection{$O \left( m \right)$ gyrokinetic equation.}

Expanding equation \refEq{eq:gyrokineticEq} to lowest order in $m \gg 1$ gives
\begin{align}
  w_{||} \left( \hat{b} \cdot \Nabla \theta \right)_{0} m \left. \frac{\partial \widetilde{h}_{s 0}}{\partial z} \right|_{\theta, w_{||}, \mu} = 0 . \label{eq:gyrokinEqOminus1}
\end{align}
We see from equation \refEq{eq:gradparO0} that $\left( \hat{b} \cdot \Nabla \theta \right)_{0}$ is a constant, so integrating over $z$ gives
\begin{align}
  \overline{h}_{s 0} =& h_{s 0} \label{eq:hs0avg} \\
  \widetilde{h}_{s 0} =& 0 . \label{eq:hs0tilde}
\end{align}

\subsubsection{$O \left( 1 \right)$ quasineutrality equation.}

Expanding equation \refEq{eq:quasineut} to lowest order in $m \gg 1$ gives
\begin{align}
   \phi_{0} =& \left( \sum_{s} \frac{Z_{s}^{2} e^{2} n_{s}}{T_{s}} \right)^{-1} \sum_{s} \frac{2 \pi Z_{s} e B_{0}}{m_{s}} \int dw_{||} d \mu \left( J_{0} \left( k_{\perp} \rho_{s} \right) \right)_{0} h_{s 0} . \label{eq:phi0}
\end{align}
Using equations \refEq{eq:hs0avg} and \refEq{eq:FLRO0} we see that
\begin{align}
  \overline{\phi}_{0} =& \phi_{0} = \left( \sum_{s} \frac{Z_{s}^{2} e^{2} n_{s}}{T_{s}} \right)^{-1} \sum_{s} \frac{2 \pi Z_{s} e B_{0}}{m_{s}} \int dw_{||} d \mu \overline{\left( J_{0} \left( k_{\perp} \rho_{s} \right) \right)}_{0} \overline{h}_{s 0} \label{eq:phi0avg} \\
  \widetilde{\phi}_{0} =& 0 . \label{eq:phi0tilde}
\end{align}

\subsubsection{$O \left( 1 \right)$ gyrokinetic equation.}

Expanding equation \refEq{eq:gyrokineticEq} to $O \left( 1 \right)$ gives
\begin{align}
  \frac{\partial h_{s 0}}{\partial t} &+ w_{||} \left( \hat{b} \cdot \Nabla \theta \right)_{0} \left( \left. \frac{\partial h_{s 0}}{\partial \theta} \right|_{z, w_{||}, \mu} + m \left. \frac{\partial \widetilde{h}_{s 1}}{\partial z} \right|_{\theta, w_{||}, \mu} \right) + i \left( w^{2}_{||} + \frac{B_{0}}{m_{s}} \mu \right) \left( k_{\psi} v_{d s \psi 0} + k_{\alpha} v_{d s \alpha 0} \right) h_{s 0} \nonumber \\
 &+ \left\{ \left( J_{0} \left( k_{\perp} \rho_{s} \right) \right)_{0} \phi_{0}, h_{s 0} \right\} - \frac{Z_{s} e F_{M s}}{T_{s}} \frac{\partial}{\partial t} \Big( \left( J_{0} \left( k_{\perp} \rho_{s} \right) \right)_{0} \phi_{0} \Big) \label{eq:gyrokinEqO0} \\
 &+ i k_{\alpha} \left( J_{0} \left( k_{\perp} \rho_{s} \right) \right)_{0} \phi_{0}  F_{M s} \left[ \frac{1}{n_{s}} \frac{d n_{s}}{d \psi} + \left( \frac{m_{s} w^{2}}{2 T_{s}} - \frac{3}{2} \right) \frac{1}{T_{s}} \frac{d T_{s}}{d \psi} \right] = 0 . \nonumber
\end{align}
Averaging over $z$ after using equations \refEq{eq:hs0avg}, \refEq{eq:phi0avg}, and \refEq{eq:gradparO0} through \refEq{eq:FLRO0} gives
\begin{align}
  \frac{\partial \overline{h}_{s 0}}{\partial t} &+ w_{||} \overline{\left( \hat{b} \cdot \Nabla \theta \right)}_{0} \left. \frac{\partial \overline{h}_{s 0}}{\partial \theta} \right|_{z, w_{||}, \mu} + i \left( w^{2}_{||} + \frac{B_{0}}{m_{s}} \mu \right) \left( k_{\psi} \overline{v}_{d s \psi 0} + k_{\alpha} \overline{v}_{d s \alpha 0} \right) \overline{h}_{s 0} \nonumber \\
 &+ \left\{ \overline{\left( J_{0} \left( k_{\perp} \rho_{s} \right) \right)}_{0} \overline{\phi}_{0}, \overline{h}_{s 0} \right\} - \frac{Z_{s} e F_{M s}}{T_{s}} \frac{\partial}{\partial t} \left( \overline{\left( J_{0} \left( k_{\perp} \rho_{s} \right) \right)}_{0} \overline{\phi}_{0} \right) \label{eq:gyrokinEqO0avg} \\
 &+ i k_{\alpha} \overline{\left( J_{0} \left( k_{\perp} \rho_{s} \right) \right)}_{0} \overline{\phi}_{0}  F_{M s} \left[ \frac{1}{n_{s}} \frac{d n_{s}}{d \psi} + \left( \frac{m_{s} w^{2}}{2 T_{s}} - \frac{3}{2} \right) \frac{1}{T_{s}} \frac{d T_{s}}{d \psi} \right] = 0 , \nonumber
\end{align}
which does not depend on $z$. From equations \refEq{eq:gradparO0} through \refEq{eq:FLRO0} we see that equations \refEq{eq:phi0avg} and \refEq{eq:gyrokinEqO0avg} are unchanged by the $\left( k_{\psi}, k_{\alpha}, \theta, w_{||}, \mu, t \right) \rightarrow \left( - k_{\psi}, k_{\alpha}, - \theta, - w_{||}, \mu, t \right)$ coordinate system transformation when $\overline{h}_{s 0} \rightarrow - \overline{h}_{s 0}$ and $\overline{\phi}_{0} \rightarrow - \overline{\phi}_{0}$. This symmetry of the $O \left( 1 \right)$ gyrokinetic equations is important because, in exactly up-down symmetric tokamaks, it can be used to show that the momentum flux is zero (see the discussion immediately preceding equation \refEq{eq:momFluxSimple}).

Subtracting equation \refEq{eq:gyrokinEqO0avg} from equation \refEq{eq:gyrokinEqO0} we find
\begin{align}
  m w_{||} \overline{\left( \hat{b} \cdot \Nabla \theta \right)}_{0} \left. \frac{\partial \widetilde{h}_{s 1}}{\partial z} \right|_{\theta, w_{||}, \mu} = 0 .
\end{align}
Therefore, we know that
\begin{align}
  \overline{h}_{s 1} =& h_{s 1} \label{eq:hs1avg} \\
  \widetilde{h}_{s 1} =& 0 . \label{eq:hs1tilde}
\end{align}

\subsubsection{$O \left( 1 \right)$ momentum transport.}

Expanding equation \refEq{eq:momFluxSimple} to lowest order gives
\begin{align}
  \Pi_{s 0} =& \frac{i R_{0} B_{0}}{\oint d\theta \left( \hat{b} \cdot \Nabla \theta \right)_{0}^{-1}} \sum_{k_{\psi}, k_{\alpha}} k_{\alpha} \oint d\theta \left( \hat{b} \cdot \Nabla \theta \right)_{0}^{-1} \label{eq:pi0} \\
  &\times \int dw_{||} d\mu w_{||} \left( J_{0} \left( k_{\perp} \rho_{s} \right) \right)_{0} \phi_{0} \left( k_{\psi}, k_{\alpha} \right) h_{s 0} \left( - k_{\psi}, - k_{\alpha} \right) . \nonumber
\end{align}
Using equations \refEq{eq:hs0avg}, \refEq{eq:phi0avg}, \refEq{eq:gradparO0}, and \refEq{eq:FLRO0} we find that
\begin{align}
    \Pi_{s 0} =& \frac{i R_{0} B_{0}}{2 \pi} \sum_{k_{\psi}, k_{\alpha}} k_{\alpha} \oint d\theta \int dw_{||} d\mu w_{||} \overline{\left( J_{0} \left( k_{\perp} \rho_{s} \right) \right)}_{0} \overline{\phi}_{0} \overline{h}_{s 0} . \label{eq:pi0avg}
\end{align}
Therefore by the $\left( k_{\psi}, k_{\alpha}, \theta, w_{||}, \mu, t \right) \rightarrow \left( - k_{\psi}, k_{\alpha}, - \theta, - w_{||}, \mu, t \right)$ symmetry outlined in reference \cite{ParraUpDownSym2011} we know that $\Pi_{s 0} = 0$ when averaged over a turbulent decorrelation time.

\subsubsection{$O \left( m^{-1} \right)$ quasineutrality equation.}

Equation \refEq{eq:quasineut}, expanded to $O \left( m^{-1} \right)$, is
\begin{align}
   \phi_{1} =& \left( \sum_{s} \frac{Z_{s}^{2} e^{2} n_{s}}{T_{s}} \right)^{-1} \sum_{s} \frac{2 \pi Z_{s} e B_{0}}{m_{s}} \int dw_{||} d \mu \Big( \left( J_{0} \left( k_{\perp} \rho_{s} \right) \right)_{1} h_{s 0} + \left( J_{0} \left( k_{\perp} \rho_{s} \right) \right)_{0} h_{s 1} \Big) . \label{eq:phi1}
\end{align}
Using equations \refEq{eq:hs0avg}, \refEq{eq:hs1avg}, and  \refEq{eq:FLRO0}, then averaging over $z$ gives
\begin{align}
   \overline{\phi}_{1} =& \left( \sum_{s} \frac{Z_{s}^{2} e^{2} n_{s}}{T_{s}} \right)^{-1} \sum_{s} \frac{2 \pi Z_{s} e B_{0}}{m_{s}} \int dw_{||} d \mu \left( \overline{\left( J_{0} \left( k_{\perp} \rho_{s} \right) \right)}_{1} \overline{h}_{s 0} + \overline{\left( J_{0} \left( k_{\perp} \rho_{s} \right) \right)}_{0} \overline{h}_{s 1} \right) . \label{eq:phi1avg}
\end{align}
Note that $\widetilde{\phi}_{1} \neq 0$.

\subsubsection{$O \left( m^{-1} \right)$ gyrokinetic equation.}

Expanding equation \refEq{eq:gyrokineticEq} to $O \left( m^{-1} \right)$, using equations \refEq{eq:hs0avg}, \refEq{eq:phi0avg}, \refEq{eq:hs1avg}, \refEq{eq:hs1tilde}, and \refEq{eq:gradparO0} through \refEq{eq:FLRO0}, gives
\begin{align}
  \frac{\partial \overline{h}_{s 1}}{\partial t} &+ w_{||} \left( \hat{b} \cdot \Nabla \theta \right)_{0} \left( \left. \frac{\partial \overline{h}_{s 1}}{\partial \theta} \right|_{z, w_{||}, \mu} + m \left. \frac{\partial \widetilde{h}_{s 2}}{\partial z} \right|_{\theta, w_{||}, \mu} \right) + i \left( w^{2}_{||} + \frac{B_{0}}{m_{s}} \mu \right) \left( k_{\psi} \overline{v}_{d s \psi 0} + k_{\alpha} \overline{v}_{d s \alpha 0} \right) \overline{h}_{s 1} \nonumber \\
 &+ \left\{ \overline{\left( J_{0} \left( k_{\perp} \rho_{s} \right) \right)}_{0} \overline{\phi}_{0}, \overline{h}_{s 1} \right\} + \left\{ \overline{\left( J_{0} \left( k_{\perp} \rho_{s} \right) \right)}_{0} \phi_{1}, \overline{h}_{s 0} \right\} - \frac{Z_{s} e F_{M s}}{T_{s}} \frac{\partial}{\partial t} \left( \overline{\left( J_{0} \left( k_{\perp} \rho_{s} \right) \right)}_{0} \phi_{1} \right) \nonumber \\
 &+ i k_{\alpha} \overline{\left( J_{0} \left( k_{\perp} \rho_{s} \right) \right)}_{0} \phi_{1}  F_{M s} \left[ \frac{1}{n_{s}} \frac{d n_{s}}{d \psi} + \left( \frac{m_{s} w^{2}}{2 T_{s}} - \frac{3}{2} \right) \frac{1}{T_{s}} \frac{d T_{s}}{d \psi} \right] \nonumber \\
 &= w_{||} \left( \hat{b} \cdot \Nabla \theta \right)_{1} \left. \frac{\partial \overline{h}_{s 0}}{\partial \theta} \right|_{z, w_{||}, \mu} - i \left( w^{2}_{||} + \frac{B_{0}}{m_{s}} \mu \right) \left( k_{\psi} v_{d s \psi 1} + k_{\alpha} v_{d s \alpha 1} \right) \overline{h}_{s 0} \label{eq:gyrokinEqO1simple} \\
 &- \left\{ \left( J_{0} \left( k_{\perp} \rho_{s} \right) \right)_{1} \overline{\phi}_{0}, \overline{h}_{s 0} \right\} + \frac{Z_{s} e F_{M s}}{T_{s}} \frac{\partial}{\partial t} \Big( \left( J_{0} \left( k_{\perp} \rho_{s} \right) \right)_{1} \overline{\phi}_{0} \Big) \nonumber \\
 &- i k_{\alpha} \left( J_{0} \left( k_{\perp} \rho_{s} \right) \right)_{1} \overline{\phi}_{0}  F_{M s} \left[ \frac{1}{n_{s}} \frac{d n_{s}}{d \psi} + \left( \frac{m_{s} w^{2}}{2 T_{s}} - \frac{3}{2} \right) \frac{1}{T_{s}} \frac{d T_{s}}{d \psi} \right] . \nonumber
\end{align}
Averaging over $z$ we find that
\begin{align}
  \frac{\partial \overline{h}_{s 1}}{\partial t} &+ w_{||} \hat{b} \cdot \Nabla \theta \left. \frac{\partial \overline{h}_{s 1}}{\partial \theta} \right|_{z, w_{||}, \mu} + i \left( w^{2}_{||} + \frac{B_{0}}{m_{s}} \mu \right) \left( k_{\psi} \overline{v}_{d s \psi 0} + k_{\alpha} \overline{v}_{d s \alpha 0} \right) \overline{h}_{s 1} \nonumber \\
 &+ \left\{ \overline{\left( J_{0} \left( k_{\perp} \rho_{s} \right) \right)}_{0} \overline{\phi}_{0}, \overline{h}_{s 1} \right\} + \left\{ \overline{\left( J_{0} \left( k_{\perp} \rho_{s} \right) \right)}_{0} \overline{\phi}_{1}, \overline{h}_{s 0} \right\} - \frac{Z_{s} e F_{M s}}{T_{s}} \frac{\partial}{\partial t} \left( \overline{\left( J_{0} \left( k_{\perp} \rho_{s} \right) \right)}_{0} \overline{\phi}_{1} \right) \nonumber \\
 &+ i k_{\alpha} \overline{\left( J_{0} \left( k_{\perp} \rho_{s} \right) \right)}_{0} \overline{\phi}_{1}  F_{M s} \left[ \frac{1}{n_{s}} \frac{d n_{s}}{d \psi} + \left( \frac{m_{s} w^{2}}{2 T_{s}} - \frac{3}{2} \right) \frac{1}{T_{s}} \frac{d T_{s}}{d \psi} \right] \nonumber \\
 &= w_{||} \overline{\left( \hat{b} \cdot \Nabla \theta \right)}_{1} \left. \frac{\partial \overline{h}_{s 0}}{\partial \theta} \right|_{z, w_{||}, \mu} - i \left( w^{2}_{||} + \frac{B_{0}}{m_{s}} \mu \right) \left( k_{\psi} \overline{v}_{d s \psi 1} + k_{\alpha} \overline{v}_{d s \alpha 1} \right) \overline{h}_{s 0} \label{eq:gyrokinEqO1avg} \\
 &- \left\{ \overline{\left( J_{0} \left( k_{\perp} \rho_{s} \right) \right)}_{1} \overline{\phi}_{0}, \overline{h}_{s 0} \right\} + \frac{Z_{s} e F_{M s}}{T_{s}} \frac{\partial}{\partial t} \left( \overline{\left( J_{0} \left( k_{\perp} \rho_{s} \right) \right)}_{1} \overline{\phi}_{0} \right) \nonumber \\
 &- i k_{\alpha} \overline{\left( J_{0} \left( k_{\perp} \rho_{s} \right) \right)}_{1} \overline{\phi}_{0}  F_{M s} \left[ \frac{1}{n_{s}} \frac{d n_{s}}{d \psi} + \left( \frac{m_{s} w^{2}}{2 T_{s}} - \frac{3}{2} \right) \frac{1}{T_{s}} \frac{d T_{s}}{d \psi} \right] . \nonumber
\end{align}
From equations \refEq{eq:alphaDriftO1} and \refEq{eq:gradPsiDotGradAlphaO1} through \refEq{eq:FLRO1def} we see that equations \refEq{eq:phi1avg} and \refEq{eq:gyrokinEqO1avg} are \textit{not} symmetric in $\left( k_{\psi}, k_{\alpha}, \theta, w_{||}, \mu, t \right) \rightarrow \left( - k_{\psi}, k_{\alpha}, - \theta, - w_{||}, \mu, t \right)$ when $\overline{h}_{s 0} \rightarrow - \overline{h}_{s 0}$, $\overline{\phi}_{0} \rightarrow - \overline{\phi}_{0}$, $\overline{h}_{s 1} \rightarrow - \overline{h}_{s 1}$, and $\overline{\phi}_{1} \rightarrow - \overline{\phi}_{1}$. This is due to both the drift term $\overline{v}_{d s \alpha 1}$ as well as $\left( \Nabla \psi \cdot \Nabla \alpha \right)_{1}$ and $\left| \Nabla \alpha \right|^{2}_{1}$ in $\overline{\left( J_{0} \left( k_{\perp} \rho_{s} \right) \right)}_{1}$ (which accounts for finite gyroradius effects).

\subsubsection{$O \left( m^{-1} \right)$ momentum transport.}

Expanding equation \refEq{eq:momFluxSimple} to $O \left( m^{-1} \right)$ and using equations \refEq{eq:gradparO0} and \refEq{eq:gradparO1}  gives
\begin{align}
  \Pi_{s 1} =& \frac{i R_{0} B_{0}}{2 \pi} \sum_{k_{\psi}, k_{\alpha}} k_{\alpha} \oint d\theta \int dw_{||} d\mu w_{||} \Bigg[ - \overline{\left( \hat{b} \cdot \Nabla \theta \right)}_{0}^{-1} \left( \hat{b} \cdot \Nabla \theta \right)_{1} \left( J_{0} \left( k_{\perp} \rho_{s} \right) \right)_{0} \phi_{0} h_{s 0} \nonumber \\
  &+ \left( J_{0} \left( k_{\perp} \rho_{s} \right) \right)_{1} \phi_{0} h_{s 0} + \left( J_{0} \left( k_{\perp} \rho_{s} \right) \right)_{0} \phi_{1} h_{s 0} + \left( J_{0} \left( k_{\perp} \rho_{s} \right) \right)_{0} \phi_{0} h_{s 1}  \Bigg] . \label{eq:pi1}
\end{align}
After applying equations \refEq{eq:FLRO0}, \refEq{eq:gradparO1}, \refEq{eq:hs0avg}, \refEq{eq:phi0avg}, and \refEq{eq:hs1avg} we find
\begin{align}
    \Pi_{s 1} =& i R_{0} B_{0} \sum_{k_{\psi}, k_{\alpha}} k_{\alpha} \oint d\theta \int dw_{||} d\mu w_{||} \left[ \overline{\left( J_{0} \left( k_{\perp} \rho_{s} \right) \right)}_{1} \overline{\phi}_{0} \overline{h}_{s 0} \right. \label{eq:pi1avg} \\
    &+ \left. \overline{\left( J_{0} \left( k_{\perp} \rho_{s} \right) \right)}_{0} \overline{\phi}_{1} \overline{h}_{s 0} + \overline{\left( J_{0} \left( k_{\perp} \rho_{s} \right) \right)}_{0} \overline{\phi}_{0} \overline{h}_{s 1} \right] . \nonumber
\end{align}
Since neither $\overline{\left( J_{0} \left( k_{\perp} \rho_{s} \right) \right)}_{1}$, $\overline{\phi}_{1}$, nor $\overline{h}_{s 1}$ have a definite parity in $\left( k_{\psi}, k_{\alpha}, \theta, w_{||}, \mu, t \right) \rightarrow \left( - k_{\psi}, k_{\alpha}, - \theta, - w_{||}, \mu, t \right)$, we cannot constrain $\Pi_{s 1}$ to be zero. This means that we expect the momentum flux to scale as $\Pi_{s} \sim m^{-1} \rho_{\ast}^{2} n_{i} R_{0} m_{i} v_{th i}^{2}$, where $v_{th i}$ is the ion thermal speed. Since the energy flux $Q_{s}$ is non-zero to lowest order in $m$ (i.e. circular flux surfaces still have a non-zero energy flux), we can also say that $\Pi_{s} / Q_{s} \sim m^{-1} R_{0} / v_{th i}$.

\subsection{General shaping in the gyrokinetic model}
\label{subsec:genNonMirrorSymShaping}

Section \ref{subsec:practicalNonMirrorSymShaping} showed that the momentum flux scales as $O \left( m^{-1} \right)$, given a specific non-mirror symmetric geometry (circular with two high-order cylindrical harmonic shaping effects) and a specific shaping ordering ($\Delta_{m} - 1 \sim m^{-2}$). However, this is a concrete, analytically tractable example of a more general argument. Here we will bound the symmetry breaking of the geometric coefficients by systematically ordering all of the quantities that compose them. We will make no presumptions about the slow spatial scale shaping (other than to assume up-down symmetry) nor will we order the size of the fast shaping (other than to assume $\Delta_{m} - 1 \ll 1$). We note that the analysis of this section does not use an expansion in aspect ratio.

\begin{table}
  \centering
  \caption{Scalings of the strength of fast plasma shaping effects for various geometric quantities, where $Q_{slow}$ is the geometric quantity in the absence of any fast shaping (i.e. $\Delta_{m} = 1$) and all quantities are evaluated at $r_{\psi} = r_{\psi 0}$.}
  \begin{tabular}{ l c c c }
    $Q$ & Reference & $\widetilde{Q} / Q_{slow}$ & $ \left( \overline{Q} - Q_{slow} \right) / Q_{slow}$ \\
    \hline
    $r$ & Eq. \refEq{eq:fluxSurfaceSpec} & $\left( \Delta_{m} - 1 \right)$ & $0$ \\
    $\partial r / \partial r_{\psi}$ & Eq. \refEq{eq:fluxSurfaceChangeSpec}, \refEq{eq:changeShapeOrdering} & $m \left( \Delta_{m} - 1 \right)$ & $0$ \\
    $R$ & Eq. \refEq{eq:geoMajorRadial} & $\left( \Delta_{m} - 1 \right)$ & $0$ \\
    $Z$ & Eq. \refEq{eq:geoAxial} & $\left( \Delta_{m} - 1 \right)$ & $0$ \\
    $\partial R / \partial r_{\psi}$ & Eq. \refEq{eq:geoMajorRadial} & $m \left( \Delta_{m} - 1 \right)$ & $0$ \\
    $\partial Z / \partial r_{\psi}$ & Eq. \refEq{eq:geoAxial} & $m \left( \Delta_{m} - 1 \right)$ & $0$ \\
    $\partial R / \partial \theta$ & Eq. \refEq{eq:geoMajorRadial} & $m \left( \Delta_{m} - 1 \right)$ & $0$ \\
    $\partial Z / \partial \theta$ & Eq. \refEq{eq:geoAxial} & $m \left( \Delta_{m} - 1 \right)$ & $0$ \\
    $\partial^{2} R / \partial \theta^{2}$ & Eq. \refEq{eq:geoMajorRadial} & $m^{2} \left( \Delta_{m} - 1 \right)$ & $0$ \\
    $\partial^{2} Z / \partial \theta^{2}$ & Eq. \refEq{eq:geoAxial} & $m^{2} \left( \Delta_{m} - 1 \right)$ & $0$ \\
    $\Nabla r_{\psi}$ & Eq. \refEq{eq:gradIdentity} & $m \left( \Delta_{m} - 1 \right)$ & $m^{2} \left( \Delta_{m} - 1 \right)^{2}$ \\
    $\Nabla \theta$ & Eq. \refEq{eq:fluxSurfaceSpec} & $\left( \Delta_{m} - 1 \right)$ & $\left( \Delta_{m} - 1 \right)^{2}$ \\
    $\Nabla \zeta$ & Eq. \refEq{eq:geoMajorRadial} & $\left( \Delta_{m} - 1 \right)$ & $\left( \Delta_{m} - 1 \right)^{2}$ \\
    $B_{\zeta}$ & Eq. \refEq{eq:millerTorField} & $\left( \Delta_{m} - 1 \right)$ & $\left( \Delta_{m} - 1 \right)^{2}$ \\
    $\Nabla \psi$ & Eq. \refEq{eq:gradIdentity} & $m \left( \Delta_{m} - 1 \right)$ & $m^{2} \left( \Delta_{m} - 1 \right)^{2}$ \\ \cline{4-4}
    $\left| \Nabla \psi \right|^{2}$ & Eq. \refEq{eq:gradIdentity} & $m \left( \Delta_{m} - 1 \right)$ & \multicolumn{1}{|c|}{$m^{2} \left( \Delta_{m} - 1 \right)^{2}$} \\ \cline{4-4}
    $B_{p}$ & Eq. \refEq{eq:poloidalFieldDef} & $m \left( \Delta_{m} - 1 \right)$ & $m^{2} \left( \Delta_{m} - 1 \right)^{2}$ \\ \cline{4-4}
    $B$ & Eqs. \refEq{eq:poloidalFieldDef}, \refEq{eq:millerTorField} & $m \left( \Delta_{m} - 1 \right)$ & \multicolumn{1}{|c|}{$m^{2} \left( \Delta_{m} - 1 \right)^{2}$} \\ \cline{4-4}
    $\hat{b} \cdot \Nabla \theta$ & Eqs. \refEq{eq:fluxSurfaceSpec}, \refEq{eq:poloidalFieldDef} & $m \left( \Delta_{m} - 1 \right)$ & $m^{2} \left( \Delta_{m} - 1 \right)^{2}$ \\
    $\partial B / \partial \theta$ & Eqs. \refEq{eq:poloidalFieldDef}, \refEq{eq:millerTorField} & $m^{2} \left( \Delta_{m} - 1 \right)$ & $m^{2} \left( \Delta_{m} - 1 \right)^{2}$ \\ \cline{4-4}
    $v_{d s \psi}$ & Eq. \refEq{eq:driftVelPsi} & $m^{2} \left( \Delta_{m} - 1 \right)$ & \multicolumn{1}{|c|}{$m^{3} \left( \Delta_{m} - 1 \right)^{2}$} \\ \cline{4-4}
    $a_{s ||}$ & Eq. \refEq{eq:parAccelDef} & $m^{2} \left( \Delta_{m} - 1 \right)$ & \multicolumn{1}{|c|}{$m^{3} \left( \Delta_{m} - 1 \right)^{2}$} \\ \cline{4-4}
    $d l_{p} / d \theta$ & Eq. \refEq{eq:dLpdtheta} & $m \left( \Delta_{m} - 1 \right)$ & $m^{2} \left( \Delta_{m} - 1 \right)^{2}$ \\
    $\kappa_{p}$ & Eq. \refEq{eq:poloidalCurv} & $m^{2} \left( \Delta_{m} - 1 \right)$ & $m^{2} \left( \Delta_{m} - 1 \right)^{2}$ \\
    $A_{\alpha}$ & Eq. \refEq{eq:IalphaDef} & $m \left( \Delta_{m} - 1 \right)$ & $m^{2} \left( \Delta_{m} - 1 \right)^{2}$ \\
    $\partial A_{\alpha} / \partial \psi$ & Eq. \refEq{eq:dIntegranddpsiRearrange} & $m^{2} \left( \Delta_{m} - 1 \right)$ & $m^{3} \left( \Delta_{m} - 1 \right)^{2}$ \\
    $\int d \theta ~ \partial A_{\alpha} / \partial \psi$ & Eq. \refEq{eq:gradAlphaInitial} & $m \left( \Delta_{m} - 1 \right)$ & $m^{3} \left( \Delta_{m} - 1 \right)^{2}$ \\
    $\Nabla \alpha$ & Eq. \refEq{eq:gradAlphaInitial} & $m \left( \Delta_{m} - 1 \right)$ & $m^{3} \left( \Delta_{m} - 1 \right)^{2}$ \\
    $\partial B / \partial r_{\psi}$ & Eqs. \refEq{eq:dBpdpsi}, \refEq{eq:dBtordpsi} & $m^{2} \left( \Delta_{m} - 1 \right)$ & $m^{3} \left( \Delta_{m} - 1 \right)^{2}$ \\ \cline{4-4}
    $v_{d s \alpha}$ & Eq. \refEq{eq:driftVelAlpha} & $m^{2} \left( \Delta_{m} - 1 \right)$ & \multicolumn{1}{|c|}{$m^{3} \left( \Delta_{m} - 1 \right)^{2}$} \\ \cline{4-4}
    $\Nabla \psi \cdot \Nabla \alpha$ & Eqs. \refEq{eq:gradIdentity}, \refEq{eq:gradAlphaInitial} & $m \left( \Delta_{m} - 1 \right)$ & \multicolumn{1}{|c|}{$m^{3} \left( \Delta_{m} - 1 \right)^{2}$} \\ \cline{4-4}
    $\left| \Nabla \alpha \right|^{2}$ & Eq. \refEq{eq:gradAlphaInitial} & $m \left( \Delta_{m} - 1 \right)$ & \multicolumn{1}{|c|}{$m^{3} \left( \Delta_{m} - 1 \right)^{2}$} \\ \cline{4-4}
  \end{tabular}
  \label{tab:shapingStrength}
\end{table}

Table \ref{tab:shapingStrength} gives a step-by-step summary of the results of the calculation. To begin, we must make some choices concerning the nature of the flux surface shape. The first two rows define the assumptions concerning the fast flux surface shaping. We require that the fast shaping must be periodic, that $\widetilde{r} \left( \theta, z \right) \sim O \left( \left( \Delta_{m} - 1 \right) r_{\psi 0} \right)$ on the flux surface of interest, and that $\widetilde{ \partial r / \partial r_{\psi}} \sim O \left( m \left( \Delta_{m} - 1 \right) \right)$ (which we discussed previously in arriving at equation \refEq{eq:changeShapeOrdering}). This is all consistent with equation \refEq{eq:gradShafranovLocalEq} used in the calculation of section \ref{subsec:practicalNonMirrorSymShaping}.

Now, we can derive the orderings for increasingly complex quantities and eventually find the geometric coefficients. For example, we can use equations \refEq{eq:geoMajorRadial} and \refEq{eq:geoAxial} to derive the order that shaping enters into $R$ and $Z$. We also know that when we take a poloidal derivative of $\overline{Q}$, a $z$-independent quantity, it remains of the same order. However, when we take a poloidal derivative of $\widetilde{Q}$, the $z$-dependent part of a quantity, it gains an additional factor of $m$. Therefore, $z$-dependent part of $\partial R / \partial \theta$ and $\partial Z / \partial \theta$ are larger than $\widetilde{R}$ and $\widetilde{Z}$ by a factor of $m$. Also, when we calculate quantities such as $\Nabla r_{\psi}$ (see equation \refEq{eq:gradIdentity}) we get beating between the different fast shaping effects. Therefore, when we Taylor expand in $m \gg 1$ and $\Delta_{m} - 1 \ll 1$, the shaping in the numerator and denominators of $\Nabla r_{\psi}$ can interact to produce terms that vary on the slow scale. This means that, when we use equation \refEq{eq:zAvg} to average over $z$, these slow terms remain and can break the up-down symmetry. On the other hand, $\Nabla \theta$ and $\Nabla \zeta$ are just $\hat{e}_{\theta} / r$ and $\hat{e}_{\zeta} / R$ respectively, so their scalings can be found by directly Taylor expanding equations \refEq{eq:fluxSurfaceSpec} and \refEq{eq:geoMajorRadial}.

As discussed at the end of section \ref{subsubsec:geoCoeffs}, the poloidal curvature, $\kappa_{p}$, turns out to produce the most important symmetry-breaking term. In equation \refEq{eq:poloidalCurv} we see the two poloidal derivatives that bring the effect of shaping up to $O \left( m^{2} \left( \Delta_{m} - 1 \right) \right)$. However because of the relationship between $R \left( r_{0} \left( \theta \right), \theta \right)$ and $Z \left( r_{0} \left( \theta \right), \theta \right)$, the beating between $\partial^{2} R / \partial \theta^{2}$ and $\partial Z / \partial \theta$ as well as $\partial^{2} Z / \partial \theta^{2}$ and $\partial R / \partial \theta$ cancels to $O\left( m^{3} \left( \Delta_{m} - 1 \right)^{2} \right)$ (which would be expected). Nevertheless, the poloidal curvature can still beat against the $O \left( m \left( \Delta_{m} - 1 \right) \right)$ shaping of $B_{p}^{-2}$ in equation \refEq{eq:gradAlphaAspectIntegrand}. This means that $\partial A_{\alpha} / \partial \psi$ (i.e. the integrand in $\Nabla \alpha$) contains $O \left( m^{3} \left( \Delta_{m} - 1 \right)^{2} \right)$ terms from the fast shaping that are independent of $z$ and break the up-down symmetry. When we take the integral to calculate $\Nabla \alpha$ the $z$-dependent terms lose a factor of $m$, but the $O \left( m^{3} \left( \Delta_{m} - 1 \right)^{2} \right)$ $z$-independent pieces are not altered. This means that the symmetry of the three geometric coefficients that contain $\Nabla \alpha$ is broken to $O \left( m^{3} \left( \Delta_{m} - 1 \right)^{2} \right)$.

We note that table \ref{tab:shapingStrength} only establishes an upper bound on the scaling of geometric quantities. It is always possible, especially when given a specific geometry, for the terms to vanish or become small, giving zero to the expected order. For example, unless the flux surfaces have low order shaping, the $z$-dependent portion of $\partial l_{p} / \partial \theta$ will scale as $\left( \Delta_{m} - 1 \right)$, rather than $m \left( \Delta_{m} - 1 \right)$. Similarly if the tokamak has a large aspect ratio or if the flux surfaces lack low order shaping the symmetry-breaking in $v_{d s \psi}$ and $a_{s ||}$ turns out to be $O \left( m^{2} \left( \Delta_{m} - 1 \right) \right)$, not $O \left( m^{3} \left( \Delta_{m} - 1 \right) \right)$. Lastly, in section \ref{subsec:mirrorSymShaping} we will see that, if the fast flux surface shaping has mirror symmetry, the geometric coefficients will maintain their symmetry to all orders.

We have just shown that, in general, the up-down symmetric breaking in the geometric coefficients can be no larger than $O \left( m^{3} \left( \Delta_{m} - 1 \right)^{2} \right)$. If we give $\Delta_{m} - 1$ a definite ordering in $m$, then we can expand the gyrokinetic equations (see equations \refEq{eq:gyrokineticEq}, \refEq{eq:quasineut}, and \refEq{eq:momFluxSimple}) as we did in the previous section. Keeping all terms of $O \left( m^{4} \left( \Delta_{m} - 1 \right)^{2} \right)$ or larger leaves us with a completely up-down symmetric system of equations. From the expansion in section \ref{subsec:practicalNonMirrorSymShaping} we know that these up-down symmetric equations determine the momentum flux to $O \left( m^{4} \left( \Delta_{m} - 1 \right)^{2} \right)$. Hence, we know that $\Pi_{s}$ can scale no stronger than $m^{3} \left( \Delta_{m} - 1 \right)^{2}$.

However, there is one case that requires special treatment. Thus far we have only assumed that $\Delta_{m} - 1 \ll 1$, which means we are free to use the ordering $\Delta_{m} - 1 \sim m^{-1}$. This ordering requires convex regions in the flux surface shape (see section \ref{subsec:practicalNonMirrorSymShaping}), but it does not necessarily introduce x-points into the plasma (see \ref{app:maxShaping}). When we adopt this ordering we see that the symmetry of the geometric coefficients is broken to $O \left( m \right)$, which causes problems when we try to repeat the order-by-order expansion performed in section \ref{subsec:practicalNonMirrorSymShaping}. Naively, as $\Nabla \psi \cdot \Nabla \alpha$ and $\left| \Nabla \alpha \right|^{2}$ become very large, we would expect the nonlinear and drive terms of the gyrokinetic equation to vanish (because $J_{0} \left( k_{\perp} \rho_{s} \right) \rightarrow 0$), meaning unstable solutions appear impossible. A more careful, sophisticated treatment of the Bessel functions (and the gyrokinetic equation as a whole) is beyond the scope of this paper. Regardless, we have established that the momentum flux must scale as $O \left( 1 \right)$ at the very least, because we know the that the symmetry of the $O \left( 1 \right)$ gyrokinetic equation is broken. The same argument applies for $\Delta_{m} - 1 \gtrsim m^{-3/2}$.

In summary we expect that non-mirror symmetric fast flux surface shaping will generate intrinsic momentum flux that scales as
\begin{align}
   \frac{\Pi_{s}}{Q_{s}} \sim m^{3} \left( \Delta_{m} - 1 \right)^{2} \frac{R_{0}}{v_{th i}} \label{eq:genMomFluxScaling}
\end{align}
when $\Delta_{m} - 1 \lesssim m^{-3/2}$. We note that normalizing the momentum flux by $Q_{s}$ (the energy flux) does not change the scalings because the $O \left( 1 \right)$ energy flux, that of circular flux surfaces, is non-zero. Equation \refEq{eq:genMomFluxScaling} is consistent with section \ref{subsec:practicalNonMirrorSymShaping}, where we used a $\Delta_{m} - 1 \sim m^{-2}$ ordering with a particular geometry specification to derive that $\Pi_{s} / Q_{s} \sim m^{-1} R_{0} / v_{th i}$.

\subsection{Mirror symmetric shaping}
\label{subsec:mirrorSymShaping}

In this section we will use the symmetry of the gyrokinetic model given in reference \cite{BallMirrorSymArg2016} to establish a scaling with $m \gg 1$ for the momentum flux generated by flux surfaces with mirror symmetry. A result of this symmetry is that a poloidal translation of all the high order shaping effects (those of order $m$) by a single tilt angle only has an exponentially small effect on the turbulent transport. Since up-down symmetric configurations generate no momentum flux and all mirror symmetric geometries can be created by tilting an up-down symmetric configuration, we conclude that the momentum flux from mirror symmetric flux surfaces cannot scale more strongly than $\Pi_{s} \sim \Exp{- \beta m^{\gamma}}$, where $\beta$ and $\gamma$ are both positive and do not depend on $m$.

This exponential scaling is true for all flux surfaces that have mirror symmetry about any line in the poloidal plane, not just those with mirror symmetry about the midplane (i.e. up-down symmetry). This argument only relies on the conditions needed for the symmetry, namely $m \gg 1$. It does \textit{not} presume that the flux surface shaping is weak.

This argument is consistent with the results from sections \ref{subsec:practicalNonMirrorSymShaping} and \ref{subsec:genNonMirrorSymShaping} because we must set $\Delta_{n} = 1$, $\theta_{t n} = \theta_{t m}$, or $n = m$ to create a mirror symmetric configuration. When we do so all of the symmetry-breaking terms cancel (see equations \refEq{eq:nonMirrorAlphaDrift} and \refEq{eq:symBreakCoeff}).

\section{Numerical results}
\label{sec:numResults}

\begin{figure}
 \centering

 \includegraphics[width=0.18\textwidth]{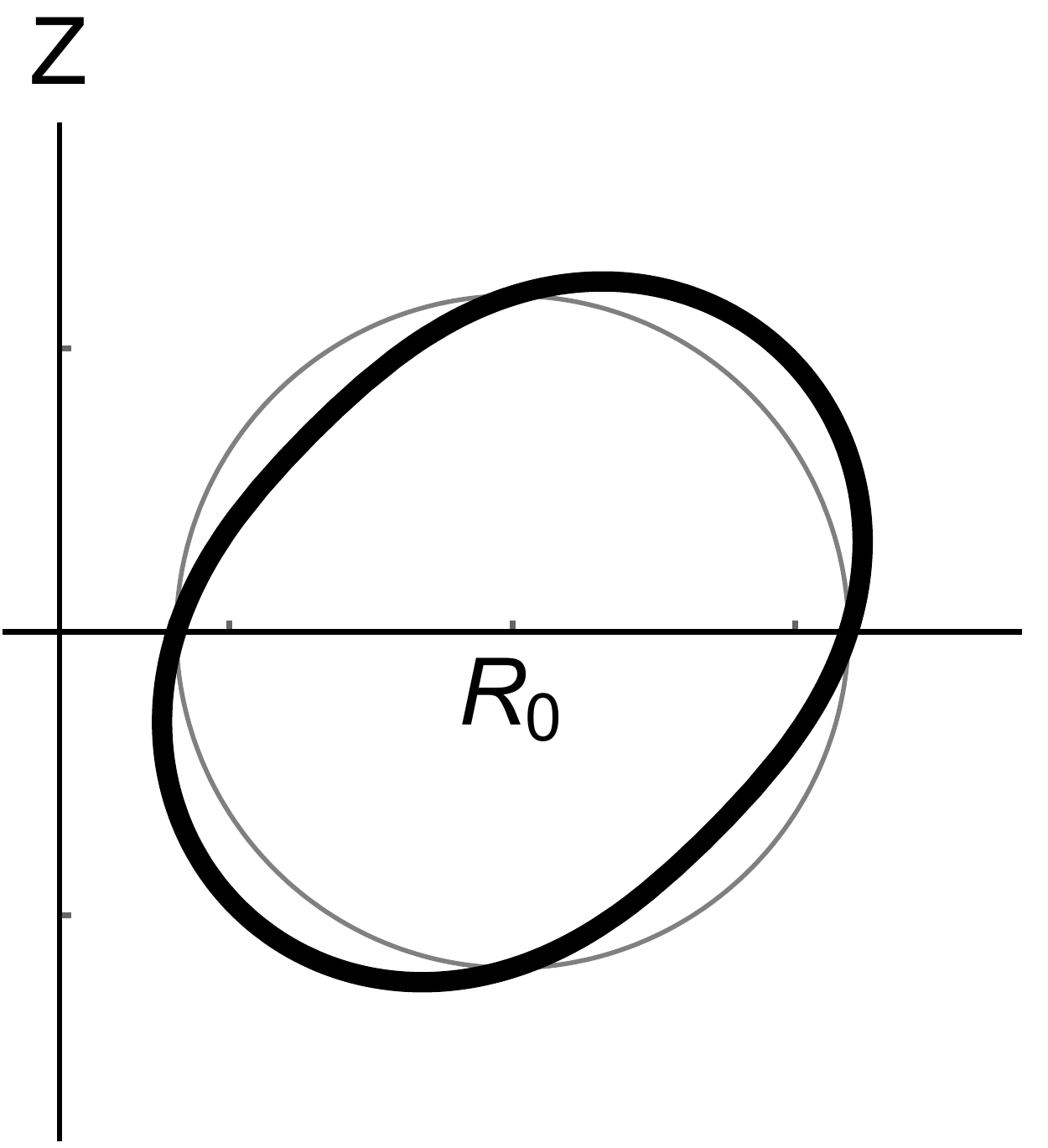}
 \includegraphics[width=0.18\textwidth]{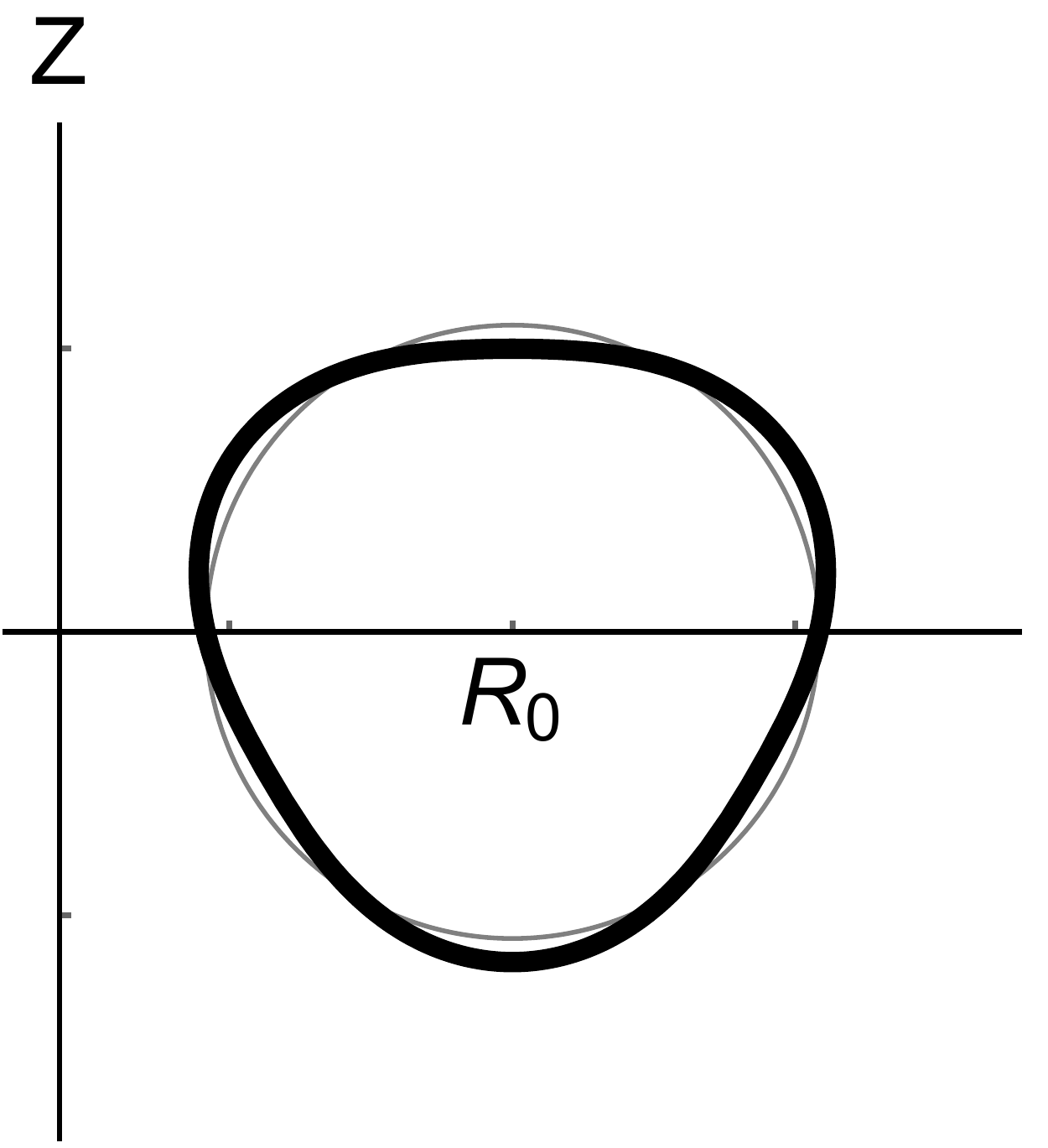}
 \includegraphics[width=0.18\textwidth]{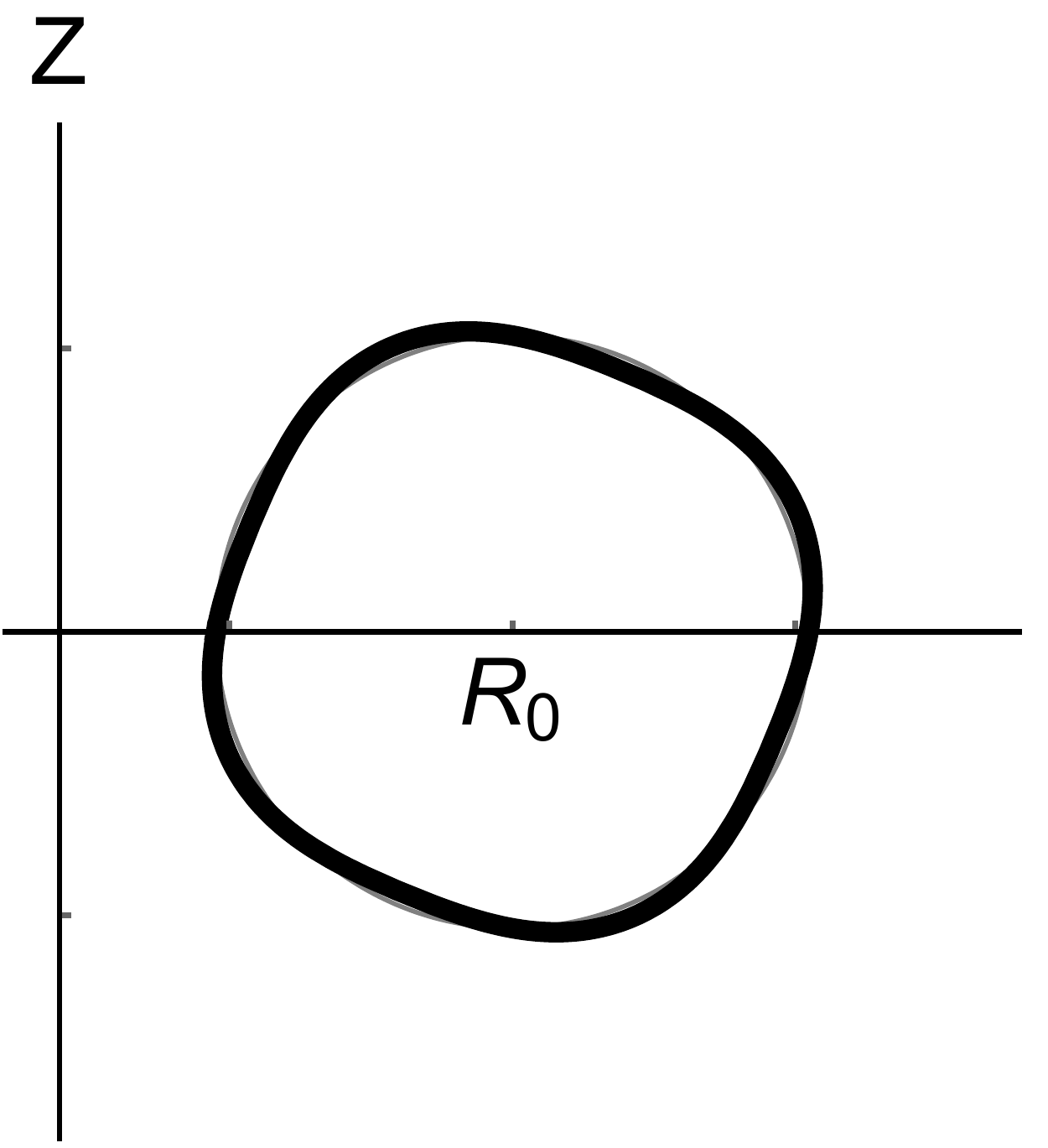}
 \includegraphics[width=0.18\textwidth]{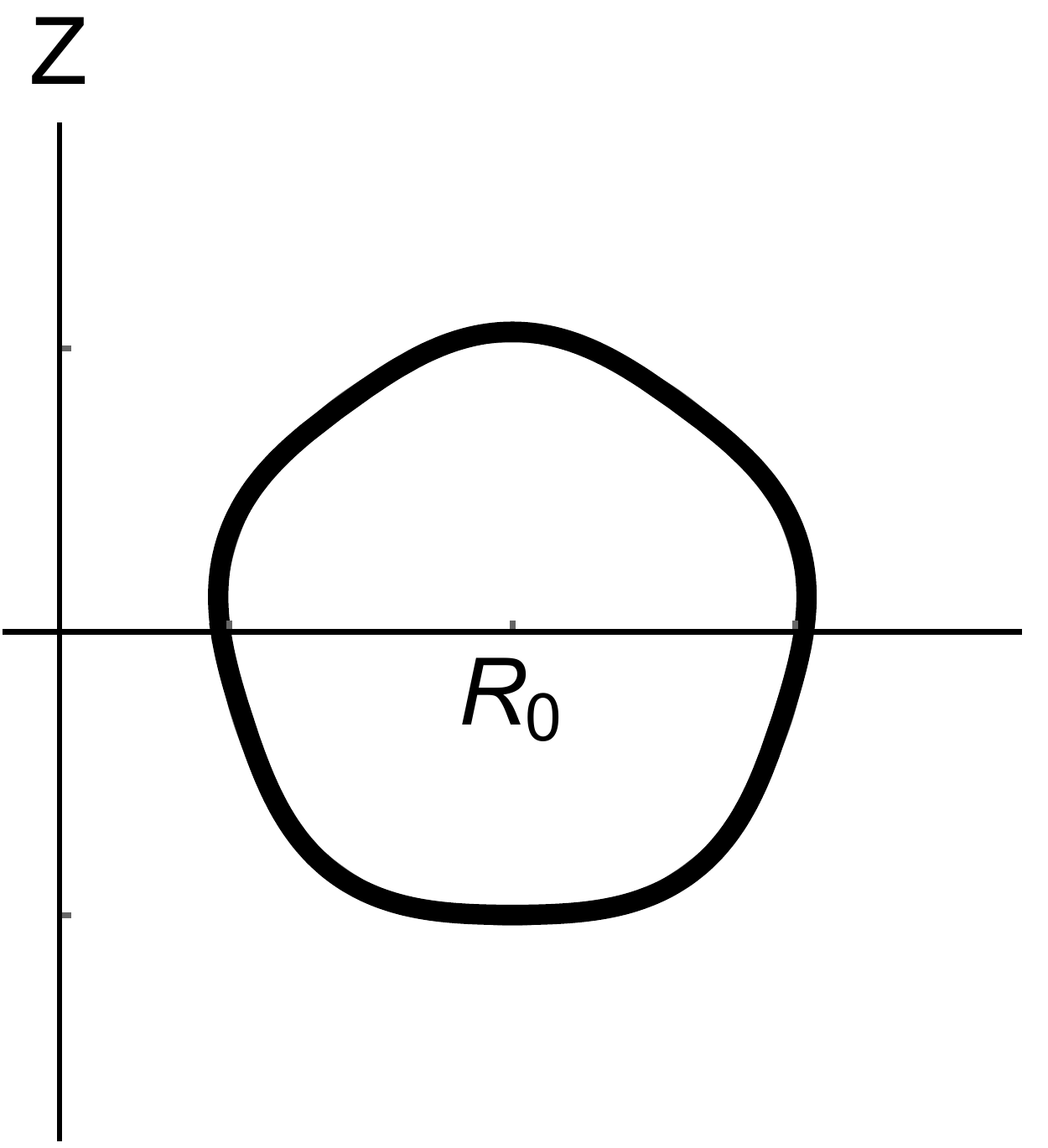}
 \includegraphics[width=0.18\textwidth]{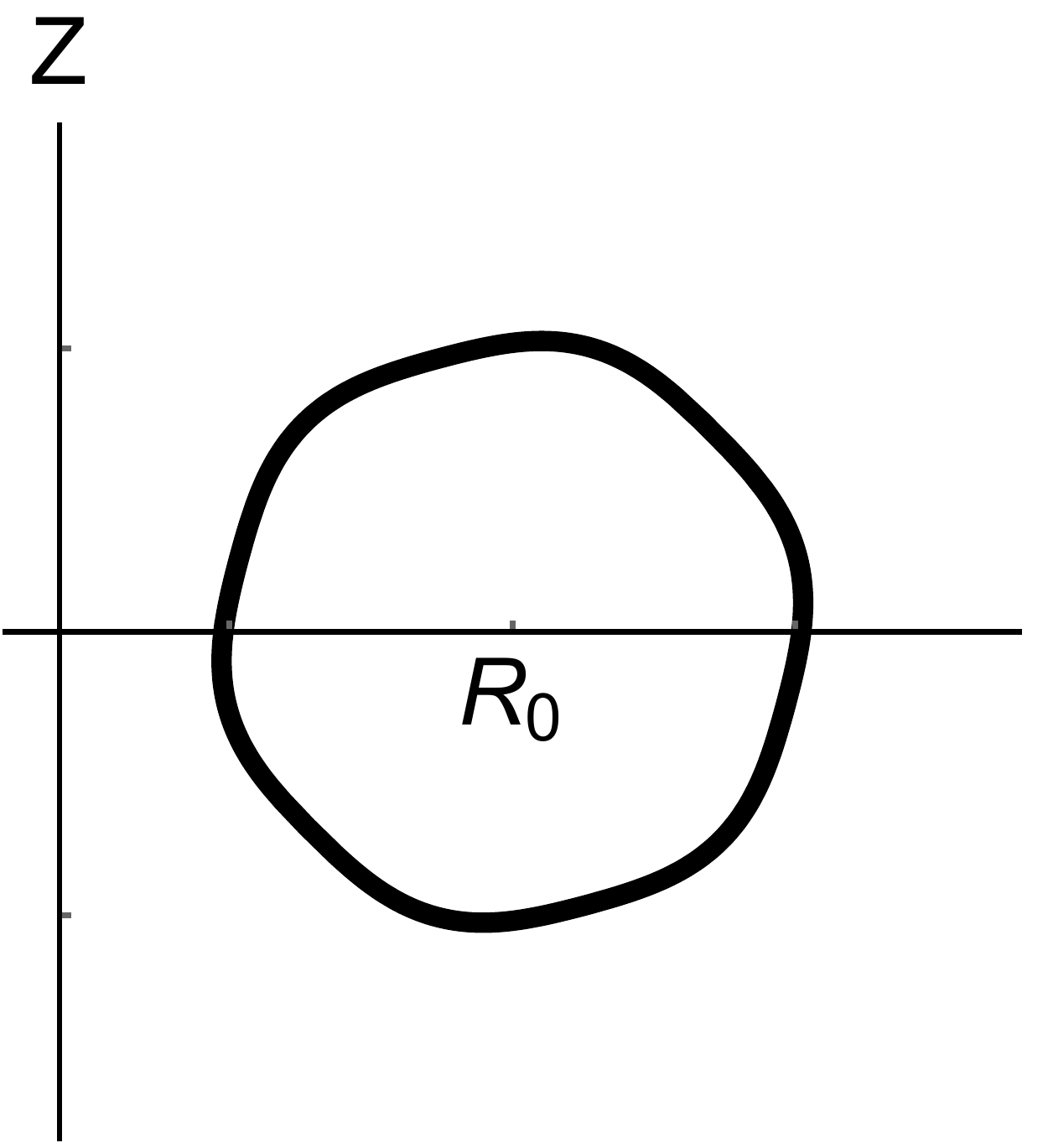}
 \includegraphics[width=0.0111\textwidth]{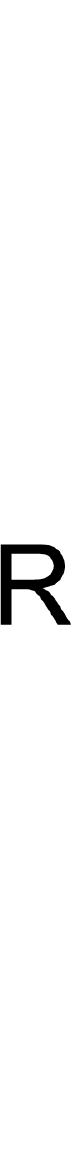}
 
 
  \includegraphics[width=0.18\textwidth]{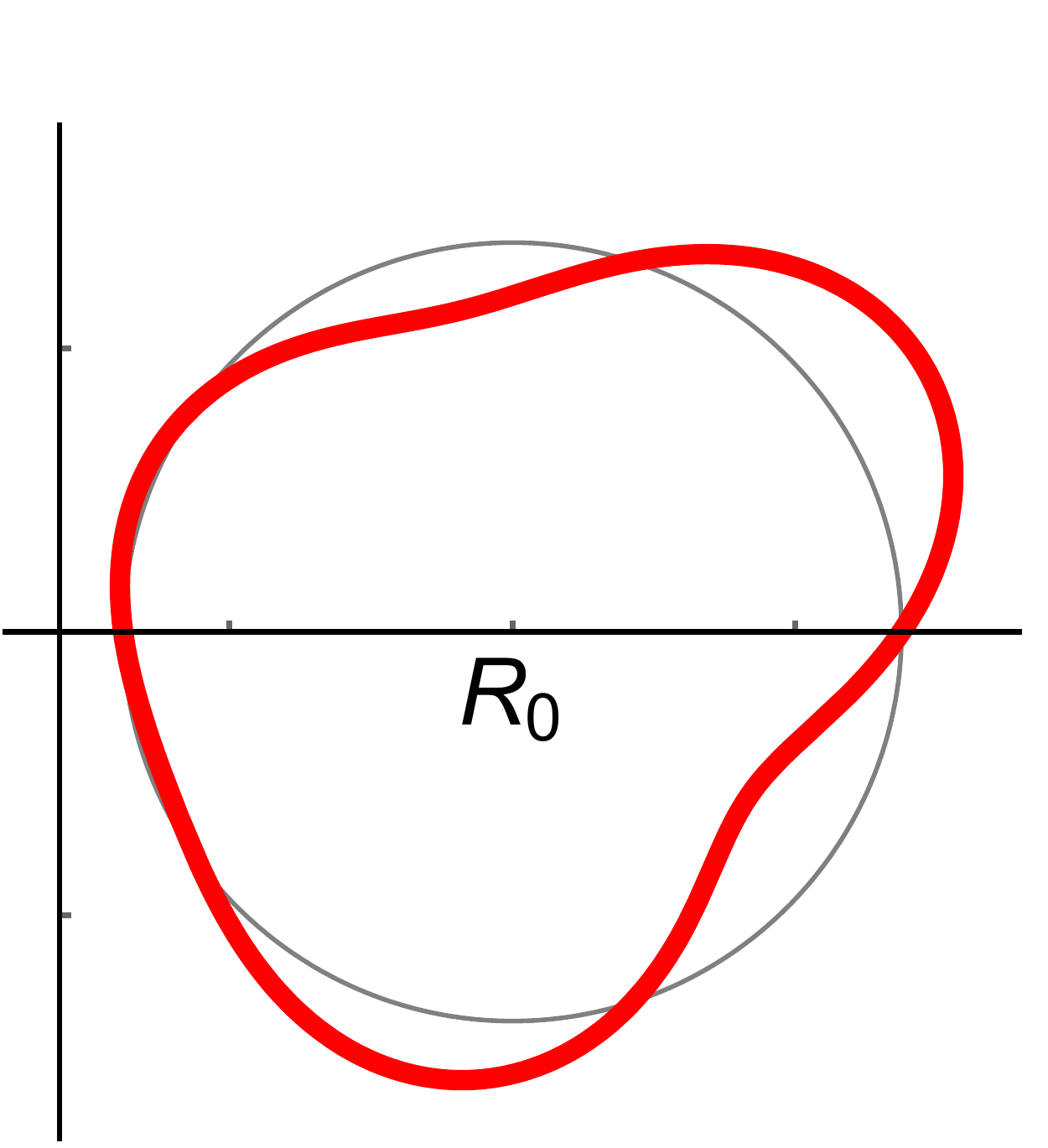}
 \includegraphics[width=0.18\textwidth]{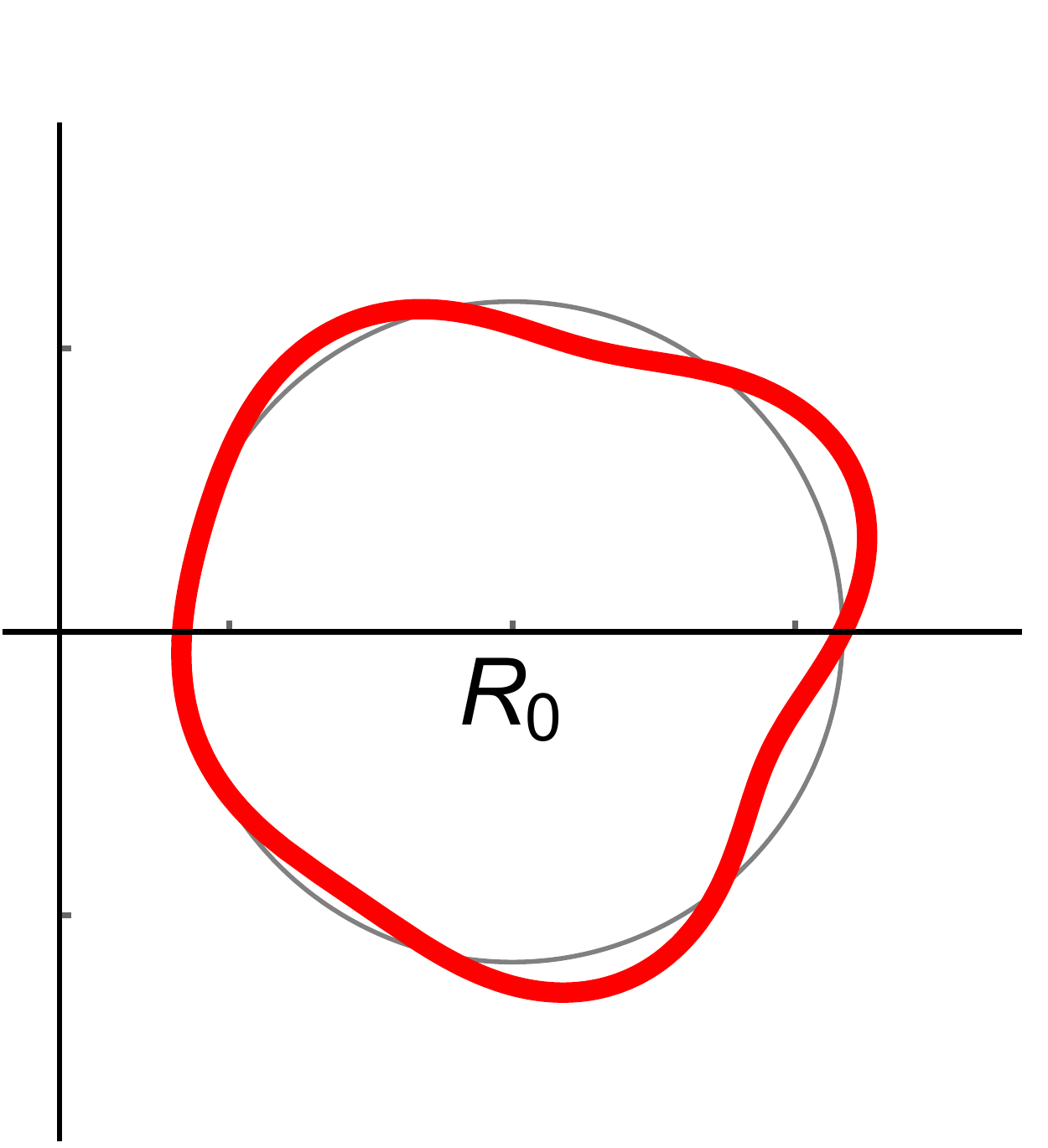}
 \includegraphics[width=0.18\textwidth]{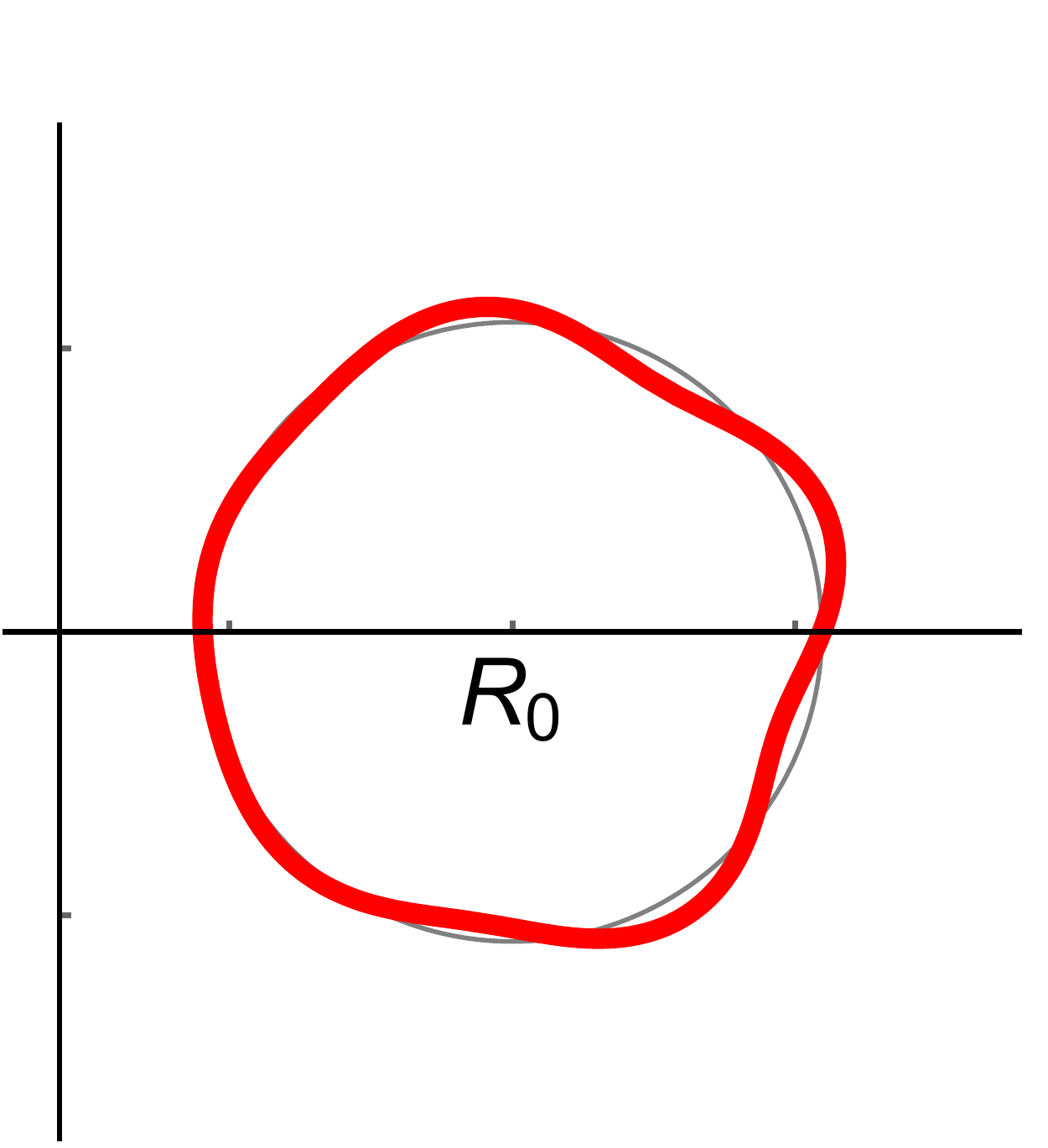}
 \includegraphics[width=0.18\textwidth]{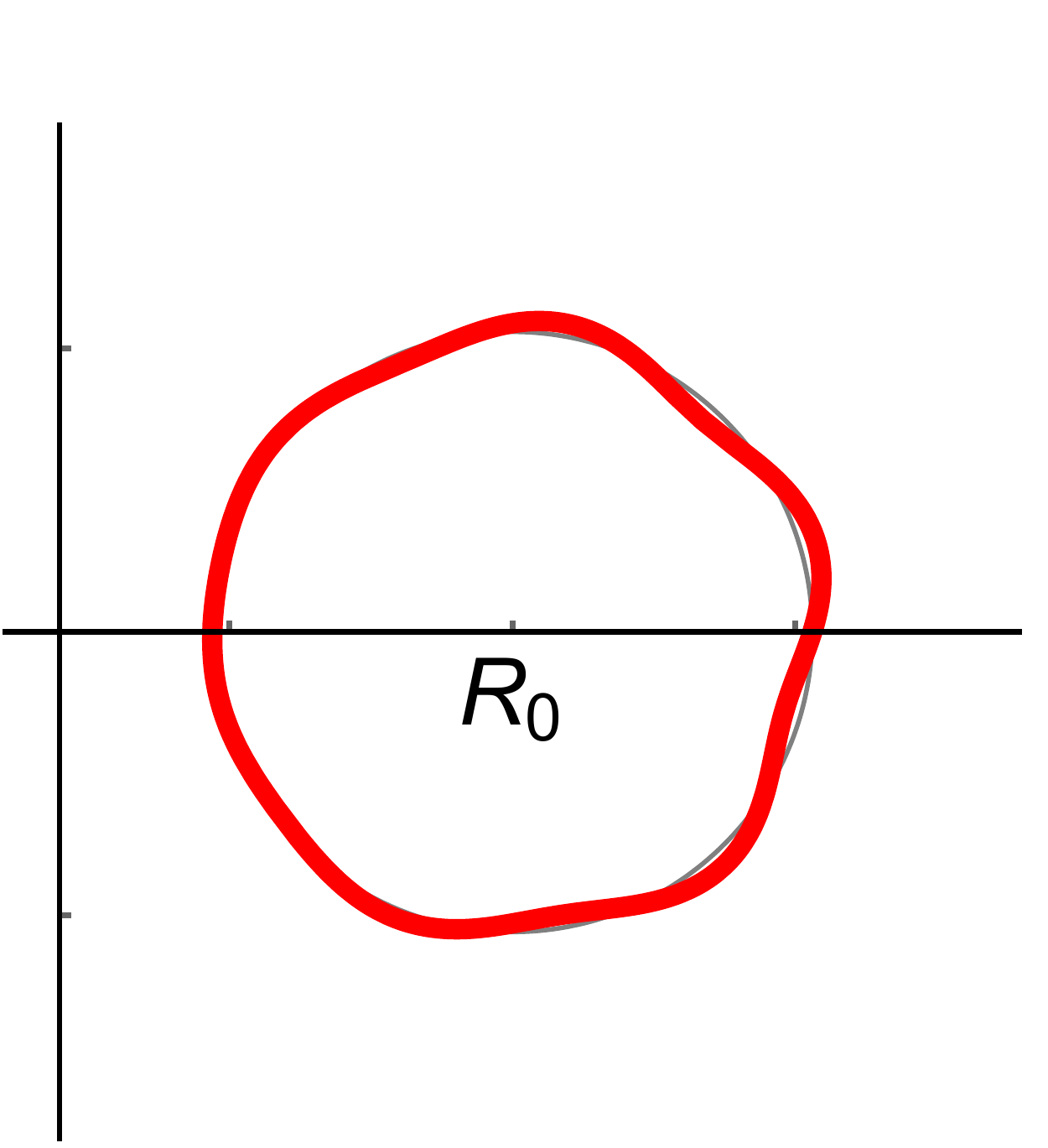}
 \includegraphics[width=0.18\textwidth]{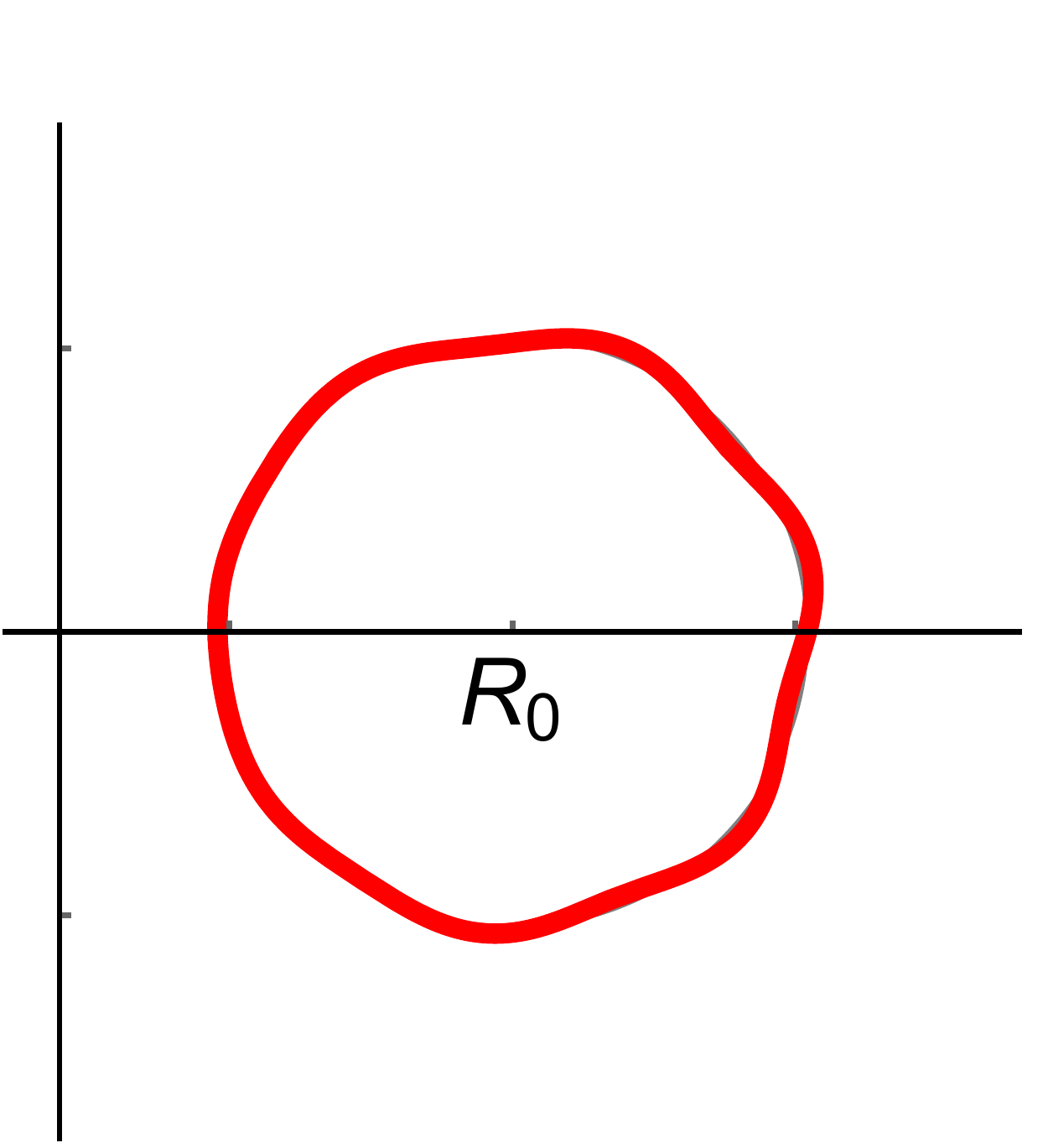}
 \includegraphics[width=0.0111\textwidth]{figs/xAxisLabelR.pdf}

 \caption{The $m=2$ through $m=6$ flux surface geometries in the mirror symmetric (solid) and non-mirror symmetric (red) scans, with circular flux surfaces shown for comparison (gray).}
 \label{fig:simGeo}
\end{figure}

In this section we will present numerical results to test the analytic conclusions of sections \ref{subsec:practicalNonMirrorSymShaping}, \ref{subsec:genNonMirrorSymShaping}, and \ref{subsec:mirrorSymShaping}. We use GS2 \cite{DorlandETGturb2000}, a local $\delta f$ gyrokinetic code, to calculate the nonlinear turbulent fluxes generated by a given geometry. We investigate the influence of the shape of the flux surface of interest by scanning $m$, the poloidal shaping mode number. We will compare the results of these numerical scans to the analytic scalings with $m \gg 1$ for mirror symmetric (see section \ref{subsec:mirrorSymShaping}) and non-mirror symmetric (see sections \ref{subsec:practicalNonMirrorSymShaping} and \ref{subsec:genNonMirrorSymShaping}) geometries.

All simulations are electrostatic and collisionless with deuterium ions and kinetic electrons. Unless specified, all parameters are fixed at Cyclone base case values \cite{DimitsCycloneBaseCase2000}: a minor radius of $r_{\psi 0} / a= 0.54$, a major radius of $R_{0} / a = 3$, a safety factor of $q = 1.4$, a magnetic shear of $\hat{s} = 0.8$, a temperature gradient of $a / L_{T s} = 2.3$, and a density gradient of $a / L_{n s} = 0.733$. Since the non-mirror symmetric geometries have strong flux surface shaping these simulations needed to be run using $a / L_{T s} = 3.0$ to ensure that the turbulence was driven unstable. To estimate the impact of this on our results, a single mirror symmetric case was run at $a / L_{T s} = 3.0$, in addition to the run with $a / L_{T s} = 2.3$. Changing the temperature gradient was found to alter the ratio of the momentum to energy flux by less than a 5\%. All simulations used at least $48$ poloidal grid points, $127$ radial wavenumber grid points, $22$ poloidal wavenumber grid points, $12$ energy grid points, and $10$ untrapped pitch angle grid points (i.e. $\lambda \equiv w_{\perp}^{2} / \left( w^{2} B \right)$).

The geometry for both scans is shown in figure \ref{fig:simGeo} and is specified by equations \refEq{eq:gradShafranovSolExpanded_rMulti} and \refEq{eq:gradShafranovSolDerivFinalMulti} from the Miller local equilibrium. The mirror symmetric scan has only one mode, $m$, while the non-mirror symmetric scan adds a second mode, $n = m + 1$. In section \ref{subsec:practicalNonMirrorSymShaping} we ordered $\Delta_{m} - 1 \sim m^{-2}$, so we will set the strength of the shaping such that $m^{2} \left( \Delta_{m} - 1 \right) = 1.5$ and $m^{2} \left( \Delta_{n} - 1 \right) = 1.5$ (if needed) is constant in the scan. Assuming a constant current profile allows us to calculate the change in the flux surface shape with radius from $\Delta_{m}$ and $\Delta_{n}$ (see \ref{app:MHDequil}). For the mirror symmetric simulations we chose the tilt angle to be $\theta_{t m} = \pi / \left( 2 m \right)$, the angle halfway between the neighboring up-down symmetric configurations (at $\theta_{t m} = 0$ and $\theta_{t m} = \pi/m$). For the non-mirror symmetric cases we must also specify $\theta_{t n} = \theta_{t m} - \pi / \left( 2 m n \right)$, which is the tilt angle halfway between two neighboring mirror symmetric configurations (at $\theta_{t n} = \theta_{t m}$ and $\theta_{t n} = \theta_{t m} - \pi / \left( m n \right)$). We also ran simulations with $\theta_{t n} = \theta_{t m} + \pi / \left( 2 m n \right)$, but those geometries did not drive significant momentum flux at any $m$.

In general, from the argument in section \ref{subsec:genNonMirrorSymShaping}, we would predict the momentum flux in an up-down asymmetric geometry to scale as $\Pi_{s} / Q_{s} \sim m^{-1} R_{0} / v_{th i}$. Indeed, we expect this to be the case for the non-mirror symmetric scan, as we confirmed in section \ref{subsec:practicalNonMirrorSymShaping}. However, section \ref{subsec:mirrorSymShaping} shows the mirror symmetric scan is a special case where the momentum flux almost entirely cancels, giving the scaling $\Pi_{s} / Q_{s} \sim \Exp{- \beta m^{\gamma}}$. Therefore we expect the momentum flux from the non-mirror symmetric runs to decay much more slowly as $m$ is increased, compared to the the mirror symmetric simulations.

As with the momentum flux, we expect that the energy flux in non-mirror symmetric configurations should converge to that of circular flux surfaces like $m^{-1}$. In mirror symmetric configurations we expect the energy flux to have the same $m^{-1}$ scaling, as opposed to the exponential scaling expected for the momentum flux. This is because the up-down symmetric terms in the geometric coefficients (e.g. the first term in equations \refEq{eq:alphaDriftO1}, \refEq{eq:gradPsiDotGradAlphaO1}, and \refEq{eq:gradAlphaSqO1}) cause energy transport, whereas they do not cause momentum transport.

\begin{figure}
 \centering
 \includegraphics[width=0.8\textwidth]{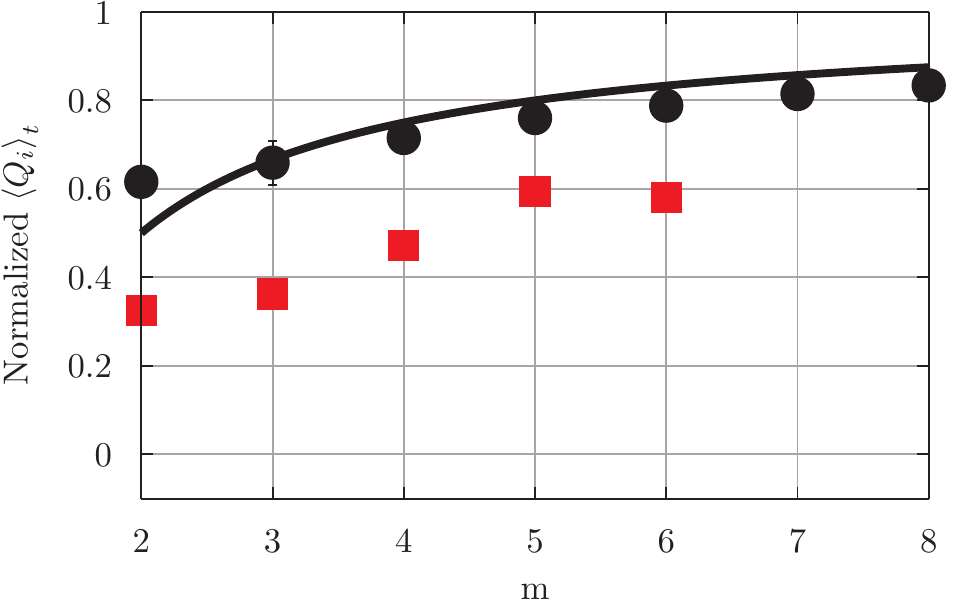}
 \caption{The radial ion energy flux from mirror symmetric (black, circles) and non-mirror symmetric (red, squares) flux surfaces, normalized to the energy flux of a circular flux surface. Also shown is the $m^{-1}$ scaling (black, solid) expected for both geometry scans.}
 \label{fig:energyFlux}
\end{figure}

\begin{figure}
 \centering
 \includegraphics[width=0.8\textwidth]{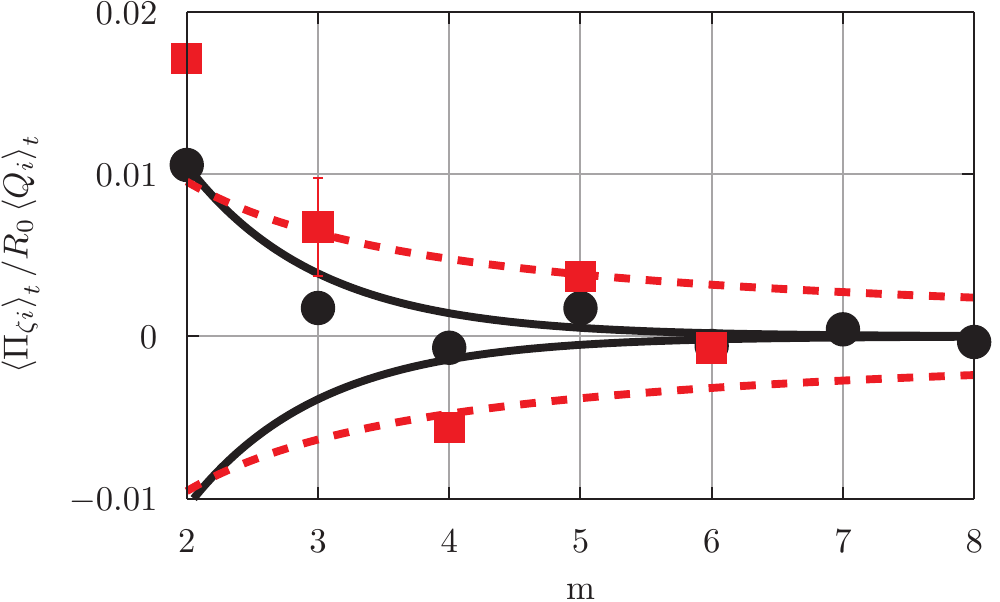}
 \caption{The ratio of ion toroidal momentum flux to ion energy flux from mirror symmetric (black, circles) and non-mirror symmetric (red, squares) flux surfaces, with a single set of error bars representative of the error in all data points. Also shown are the scalings of $\Exp{-m}$ for the mirror symmetric scan (black, solid) and $m^{-1}$ for the non-mirror symmetric scan (red, dotted).}
 \label{fig:momHeatFluxRatioAbs}
\end{figure}

Figure \ref{fig:energyFlux} shows the time-averaged ion energy flux calculated by GS2 for the two scans, which are both consistent our theoretical expectations. In figure \ref{fig:momHeatFluxRatioAbs}, we see the time-averaged ratio of the ion momentum and energy fluxes from the GS2 simulations. This ratio gives an estimate of how strong of a gradient in rotation the flux can sustain \cite{BallMomUpDownAsym2014}. We see that the mirror symmetric configurations nicely agree with the analytic theory. We note that section \ref{subsec:mirrorSymShaping} only demonstrates that the momentum flux from mirror symmetric configurations cannot scale polynomially. It does not predict the scaling must be $\Exp{-m}$, as opposed to $\Exp{-m/2}$ or $\Exp{-m^{2}}$ for example. However, as shown in figure \ref{fig:momHeatFluxRatioAbs}, $\Exp{-m}$ fits the data fairly well.

Additionally, figure \ref{fig:momHeatFluxRatioAbs} shows that the non-mirror symmetric configurations produce more rotation and decay more slowly with increasing $m$ compared to the mirror symmetric scan (as we expected). However, at low $m$ the scan does not match the predicted polynomial scaling well. It seems reasonable to attribute the departure from the ideal analytic theory to a failure to fully satisfy our assumption that $m \gg 1$. Extending the scan to larger $m$ is difficult as these simulations are much more expensive because they require higher poloidal resolution and produce smaller momentum fluxes, which then take longer to discern from the noisy turbulence.

\section{Conclusions}
\label{sec:conclusions}

This paper presents two independent arguments concerning the intrinsic momentum flux generated by turbulence in up-down asymmetric magnetic equilibria. In both arguments we use a generalization of the Miller local equilibrium to specify different up-down asymmetric geometries and look for the effect on the symmetry properties of the gyrokinetic model.

In sections \ref{subsec:practicalNonMirrorSymShaping} and \ref{subsec:genNonMirrorSymShaping} we introduce up-down asymmetric and non-mirror symmetric shaping with strength $\Delta_{m} - 1$ into the flux surface. We then look at the limit of flux surface shaping with large poloidal mode number $m$ and show that the gyrokinetic coefficients lose their symmetry to $O \left( m^{3} \left( \Delta_{m} - 1 \right)^{2} \right)$ due to the effect of the local magnetic shear. Next, we expand the gyrokinetic and quasineutrality equations to show that the symmetry breaking in the geometric coefficients causes momentum flux that generally scales as $\Pi_{s} / Q_{s} \sim m^{3} \left( \Delta_{m} - 1 \right)^{2} R_{0} / v_{th i}$.

In section \ref{subsec:mirrorSymShaping}, we use the gyrokinetic symmetry presented in reference \cite{BallMirrorSymArg2016} to argue that tokamaks with mirror symmetric flux surfaces are a special case. We find that all the symmetry-breaking terms identified in the previous argument exactly cancel and the momentum flux turns out to be smaller than would have been expected. This demonstrates that the momentum flux from tokamaks with mirror symmetric poloidal cross-sections is exponentially small in $m$, even when they are up-down asymmetric.

In order to interpret the results of these analytic arguments we will distinguish between ``geometric'' effects and ``shaping'' effects. Geometric effects are those that give a poloidal dependence to the geometric coefficients, apart from the linear dependence built into $\alpha$ due to magnetic shear. Shaping effects are the subset of the geometric effects that are specified in the flux surface shape or its radial derivative (i.e. the $m \neq 0$ terms in equations \refEq{eq:fluxSurfaceSpec} and \refEq{eq:fluxSurfaceChangeSpec}). Using this terminology we see that the tokamak has an inherent $m=1$ geometric effect due to toroidicity. However, we can see it is distinct from the $m=1$ shaping effect (i.e. the Shafranov shift) by looking at the drift coefficients to lowest order in aspect ratio. The $m=1$ toroidal geometric effect only appears in the two magnetic drift coefficients (see equations \refEq{eq:gradparO0} and \refEq{eq:alphaDriftO0}), while the $m=1$ shaping effect affects all six (see equations \refEq{eq:gradparO1} through \refEq{eq:gradAlphaSqO1}). This toroidal geometric effect is not present in other magnetic geometries like the screw pinch.

From reference \cite{ParraUpDownSym2011} we know that if the magnetic geometry does not include at least two geometric effects with different tilt angles the momentum flux must be small in $\rho_{\ast} \ll 1$. For example, to generate rotation in a tokamak we could use an up-down asymmetric shaping effect and the toroidal geometric effect (see top row of figure \ref{fig:simGeo}) or non-mirror symmetric shaping, which is two shaping effects with different tilt angles (see bottom row of figure \ref{fig:simGeo}). On the other hand, in a screw pinch, the only option to generate momentum flux is non-mirror symmetric shaping.

From sections \ref{subsec:practicalNonMirrorSymShaping} and \ref{subsec:genNonMirrorSymShaping} we know that the momentum flux from two geometric effects with similar mode numbers is polynomially small (for purely concave flux surfaces) in either mode number. This motivates using low order shaping effects (e.g. elongation, triangularity) to create rotation. Additionally, using the argument of reference \cite{BallMirrorSymArg2016} and section \ref{subsec:mirrorSymShaping}, we found that the momentum flux from two geometric effects is exponentially small in the difference between the mode numbers of the two effects. This motivates using low order shaping effects that have similar mode numbers and also distinguishes mirror and non-mirror symmetric configurations. In mirror symmetric tokamaks, the coupling between the toroidal geometric effect and shaping effects is the only mechanism that generates rotation. Non-mirror symmetric tokamaks have this same mechanism, but also allow two shaping effects to directly couple and generate rotation.

The results of this paper confirm that using low order shaping effects (e.g. elongation, triangularity) are best for creating rotation \cite{BallMomUpDownAsym2014, BallMastersThesis2013, BallShapingPenetration2015} and establish a distinction between mirror and non-mirror symmetric configurations. This suggests that non-mirror symmetric configurations may be able to generate higher levels of rotation.

\ack

The authors would like to thank M Barnes for suggestions and discussions that helped improve the quality of this paper. This work was funded in part by the RCUK Energy Programme (grant number EP/I501045). Computing time for this work was provided by the Helios supercomputer at IFERC-CSC under the projects SPIN, MULTEIM, and GKMSC. The authors also acknowledge the use of ARCHER through the Plasma HEC Consortium EPSRC grant number EP/L000237/1 under the projects e281-gs2 and e281-rotation.

\appendix

\section{Up-down asymmetric global MHD equilibrium}
\label{app:MHDequil}

Here we will use the ideal MHD model \cite{FreidbergIdealMHD1987} in the absence of rotation to identify a simple, but physical tokamak equilibrium. In  the context of this global equilibrium we will calculate $d \Delta_{m} / d r_{\psi}$ and $d \theta_{t m} / d r_{\psi}$, the local parameters that we kept unspecified in subsection \ref{subsec:MillerEquil}.

Ideal MHD equilibrium in tokamaks is governed by the Grad-Shafranov equation \cite{GradGradShafranovEq1958}, given by equation \refEq{eq:gradShafranov}. To simplify the mathematics we will take the large aspect ratio limit, $\epsilon \equiv a / R_{0} \ll 1$, of the typical ohmically heated tokamak ordering \cite{FreidbergIdealMHD1987},
\begin{align}
   \frac{B_{p}}{B_{0}} \sim \epsilon, \hspace{3em} \frac{2 \mu_{0} p}{B_{0}^{2}} \sim \epsilon^{2} .
\end{align}
Here the flux surface label $a$ (known as the tokamak minor radius) is the minimum distance of the flux surface from the magnetic axis. In this limit equation \refEq{eq:gradShafranov} becomes
\begin{align}
   \frac{1}{r} \frac{\partial }{\partial r} \left( r \frac{\partial \psi}{\partial r} \right) + \frac{1}{r^{2}} \frac{\partial^{2} \psi}{\partial \theta^{2}} = \mu_{0} j_{\zeta} R_{0} ,\label{eq:gradShafranovLowestOrder}
\end{align}
where $r$ is the typical cylindrical minor radial coordinate and $\theta$ is the typical cylindrical poloidal angle, $R_{0}$ is the tokamak major radius, $\psi$ is the poloidal flux, $j_{\zeta}$ is the toroidal current density, and $\zeta$ is the toroidal coordinate. We will let $j_{\zeta}$ be a constant, which simplifies the problem immensely and is a reasonable approximation of many experiments. Since the toroidal current is constant we can remove it by scaling the poloidal flux, using $\psi_{N} \equiv 4 \psi / \left( \mu_{0} j_{\zeta} R_{0} \right) $, to find
\begin{align}
   \frac{1}{r} \frac{\partial }{\partial r} \left( r \frac{\partial \psi_{N}}{\partial r} \right) + \frac{1}{r^{2}} \frac{\partial^{2} \psi_{N}}{\partial \theta^{2}} = 4 . \label{eq:gradShafranovLowestOrderNormalized}
\end{align}

The solution to this equation is given by cylindrical harmonics. Restricting ourselves to solutions with the magnetic axis at the origin, i.e. $\psi_{N} \left( r = 0, \theta \right) = 0$ and $\left. \Nabla \psi_{N} \right|_{r=0} = 0$, we arrive at
\begin{align}
   \psi_{N} \left( r, \theta \right) = r^{2} + \sum_{m=2}^{\infty} r^{m} \Big( C_{m} \Cos{m \theta} + S_{m} \Sin{m \theta} \Big) . \label{eq:gradShafranovSol}
\end{align}
Here the $m=2$ terms correspond to elongation, $m=3$ to triangularity, $m=4$ to squareness, etc. There are no $m=1$ terms because to lowest order in aspect ratio there is no Shafranov shift. The coefficients $C_{m}$ and $S_{m}$ are set by boundary conditions arising from the placement and currents of external shaping coils. Equation \refEq{eq:gradShafranovSol} can be rewritten as
\begin{align}
   \psi_{N} \left( r, \theta \right) = r^{2} + \sum_{m=2}^{\infty} E_{m} r^{m} \Cos{m \left( \theta + \theta_{t m} \right)} , \label{eq:gradShafranovSolSimple}
\end{align}
where $E_{m} \equiv \sqrt{C_{m}^{2} + S_{m}^{2}}$ and $\theta_{t m} \equiv - m^{-1} \ArcTan{S_{m} / C_{m}}$ are constants that signify the magnitude of the shaping and the tilt angle respectively. We immediately note that $\theta_{t m}$ is a constant, so $d \theta_{t m} / d a = d \theta_{t m} / d r_{\psi} = 0$. Equation \refEq{eq:gradShafranovSolSimple} is the global MHD equilibrium for a large aspect ratio, constant current profile tokamak. It is general to arbitrary flux surface shaping, which is specified by the Fourier coefficients $E_{m}$ and the tilt angles $\theta_{t m}$.

Next we will take the global equilibrium of equation \refEq{eq:gradShafranovSolSimple} and derive the corresponding local equilibrium. First, analogously to the flux surface label $a$, we will define $b$ to be the maximum distance of the flux surface from the magnetic axis. When only one shaping effect is present, we see from equation \refEq{eq:gradShafranovSolSimple} that
\begin{align}
   \psi_{N} \left( r, \theta \right) =& r^{2} + E_{m} r^{m} \Cos{m \left( \theta + \theta_{t m} \right)} \label{eq:gradShafranovSolPsi_r} \\
   =& a^{2} + E_{m} a^{m} \label{eq:gradShafranovSolPsi_a} \\
   =& b^{2} - E_{m} b^{m} . \label{eq:gradShafranovSolPsi_b}
\end{align}
From equations \refEq{eq:gradShafranovSolPsi_a} and \refEq{eq:gradShafranovSolPsi_b} we can solve for
\begin{align}
   E_{m} = \frac{\Delta_{m}^{2} - 1}{\Delta_{m}^{m} + 1} a^{2-m} , \label{eq:Em}
\end{align}
where $\Delta_{m} \equiv b / a$ is a generalization of the elongation (typically denoted by $\kappa$) for any single cylindrical harmonic. Then, substituting this into equations \refEq{eq:gradShafranovSolPsi_r} and  \refEq{eq:gradShafranovSolPsi_a}, we find
\begin{align}
   \frac{r^{2}}{a^{2}} + \frac{\Delta_{m}^{2} - 1}{\Delta_{m}^{m} + 1} \left( \frac{r}{a} \right)^{m} \Cos{m \left( \theta + \theta_{t m} \right)} = \frac{\Delta_{m}^{m} + \Delta_{m}^{2}}{\Delta_{m}^{m} + 1} . \label{eq:gradShafranovSolr}
\end{align}
Expanding in the limit of weak shaping, $\Delta_{m} - 1 \ll 1$, and assuming that the flux surface is circular to lowest order gives
\begin{align}
   r \left( a, \theta \right) = a \left[ 1 + \frac{\Delta_{m} - 1}{2} \Big( 1 - \Cos{m \left( \theta + \theta_{t m} \right)} \Big) \right] . \label{eq:gradShafranovSolExpanded_r}
\end{align}
This specifies the shape of the flux surface at minor radius $a$ with only one shaping effect $m$. Expanding equation \refEq{eq:fluxSurfaceSpec} in $\Delta_{m} - 1 \ll 1$ and comparing with equation \refEq{eq:gradShafranovSolExpanded_r} we see that
\begin{align}
   a \left( r_{\psi} \right) = r_{\psi} \left( 1 - \frac{\Delta_{m} - 1}{2} \right) , \label{eq:minorRadiusConvert}
\end{align}
so
\begin{align}
   r \left( r_{\psi}, \theta \right) = r_{\psi} \left( 1 - \frac{\Delta_{m} - 1}{2} \Cos{m \left( \theta + \theta_{t m} \right)} \right) . \label{eq:gradShafranovSolExpanded_rPsi}
\end{align}

However, the Miller local equilibrium model also requires the radial derivative of the flux surface shape as an input in order to calculate the poloidal field. Directly differentiating equation \refEq{eq:gradShafranovSolExpanded_rPsi}, remembering that $\Delta_{m}$ can change from flux surface to flux surface, we find
\begin{align}
   \frac{\partial r}{\partial r_{\psi}} = 1 - \left( \frac{\Delta_{m} - 1}{2} + \frac{r_{\psi}}{2} \frac{d \Delta_{m}}{d r_{\psi}} \right) \Cos{m \left( \theta + \theta_{t m} \right)} .\label{eq:gradShafranovSolDeriv}
\end{align}
We can derive that
\begin{align}
   \frac{d \Delta_{m}}{d r_{\psi}} = \frac{\Delta_{m} - 1}{r_{\psi}} \left( m - 2 \right) \label{eq:shapingRadialDeriv}
\end{align}
to lowest order in $\Delta_{m} - 1 \ll 1$ by differentiating equation \refEq{eq:Em} implicitly, remembering that $E_{m}$ is a constant and that $a$ and $r_{\psi}$ are related by equation \refEq{eq:minorRadiusConvert}. This validates the ordering $d \Delta_{m} / d r_{\psi} \sim m \left( \Delta_{m} - 1 \right) / r_{\psi 0}$ and gives
\begin{align}
   \frac{\partial r}{\partial r_{\psi}} = 1 - \frac{\Delta_{m} - 1}{2} \left( m - 1 \right) \Cos{m \left( \theta + \theta_{t m} \right)} . \label{eq:gradShafranovSolDerivFinal}
\end{align}

In this work we will study the effects of multiple shaping effects simultaneously. We will parameterize the geometry of these configurations by simply superimposing the different effects, in keeping with equations \refEq{eq:gradShafranovSolExpanded_rPsi} and \refEq{eq:gradShafranovSolDerivFinal}, as
\begin{align}
   r_{0} \left( \theta \right) =& r_{\psi 0} \left( 1 - \sum_{m} \frac{\Delta_{m} - 1}{2} \Cos{m \left( \theta + \theta_{t m} \right)} \right) \label{eq:gradShafranovSolExpanded_rMulti} \\
   \left. \frac{\partial r}{\partial r_{\psi}} \right|_{\psi_{0}} =& 1 - \sum_{m} \frac{\Delta_{m} - 1}{2} \left( m - 1 \right) \Cos{m \left( \theta + \theta_{t m} \right)} . \label{eq:gradShafranovSolDerivFinalMulti}
\end{align}
For a constant current profile, large aspect ratio, weak plasma shaping, and any value of $m$, equations \refEq{eq:fluxSurfaceSpec} and \refEq{eq:fluxSurfaceChangeSpec} reduce to these two equations.

\section{General calculation of geometric coefficients within Miller local equilibrium}
\label{app:genGeoCoeff}

In this appendix we will calculate the eight geometric coefficients (i.e. $\hat{b} \cdot \Nabla \theta$, $B$, $v_{d s \psi}$, $v_{d s \alpha}$, $a_{s ||}$, $\left| \Nabla \psi \right|^{2}$, $\Nabla \psi \cdot \Nabla \alpha$, and $\left| \Nabla \alpha \right|^{2}$) that appear in the gyrokinetic equations. Here we will use the normal cylindrical poloidal angle $\theta$, but the expressions are general to an arbitrary poloidal angle. In order to calculate these coefficients for the local equilibrium specification (given in subsection \ref{subsec:MillerEquil}) we must work within the local Miller geometry model \cite{MillerGeometry1998}. This means that we begin knowing the shape of the flux surface of interest (i.e. $R \left( \theta \right)$ and $Z \left( \theta \right)$), how it changes with minor radius (i.e. $\left. \partial R / \partial r_{\psi} \right|_{\theta}$ and $\left. \partial Z / \partial r_{\psi} \right|_{\theta}$), and four flux functions (e.g. the toroidal flux function, the safety factor, the magnetic shear, and the pressure gradient) evaluated on the flux surface of interest. With only this information we can calculate the toroidal and poloidal magnetic fields using
\begin{align}
   \vec{B}_{\zeta} =& \frac{I \left( \psi \right)}{R} \hat{e}_{\zeta} \label{eq:millerTorField} \\
   \vec{B}_{p} =& \Nabla \zeta \times \Nabla r_{\psi} \frac{d \psi}{d r_{\psi}} \label{eq:millerPolField} ,
\end{align}
where $d \psi / d r_{\psi}$ can be calculated to be
\begin{align}
   \frac{d \psi}{d r_{\psi}} = \frac{I \left( \psi \right)}{2 \pi q} \left. \oint_{0}^{2 \pi} \right|_{\psi} d \theta \left( R^{2} \Nabla r_{\psi} \cdot \left( \Nabla \theta \times \Nabla \zeta \right) \right)^{-1} \label{eq:dpsidrpsi}
\end{align}
from the definition of the safety factor. These gradients can be found from equation \refEq{eq:gradIdentity}. Using only this information we can calculate $\hat{b} \cdot \Nabla \theta$, $B$, $v_{d s \psi}$ (see equation \refEq{eq:driftVelPsi}), $a_{s ||}$ (see equation \refEq{eq:parAccelDef}), and $\left| \Nabla \psi \right|^{2}$.

However to calculate $v_{d s \alpha}$ (see equation \refEq{eq:driftVelAlpha}), $\Nabla \psi \cdot \Nabla \alpha$, and $\left| \Nabla \alpha \right|^{2}$ requires considerably more work as we must know $\Nabla \alpha$, $\left. \partial B_{\zeta} / \partial \psi \right|_{\theta}$, and $\left. \partial B_{p} / \partial \psi \right|_{\theta}$. Starting with $\left. \partial B_{p} / \partial \psi \right|_{\theta}$, we see from equation \refEq{eq:millerPolField} we see that it will depend on second order radial derivatives, which are not inputs to the Miller local equilibrium. The Miller model deals with this by calculating them through the Grad-Shafranov equation (see equation \refEq{eq:gradShafranov}). We can rearrange equation \refEq{eq:gradShafranov} to get
\begin{align}
   \frac{R^{2}}{J} \frac{\partial}{\partial \psi} \left( J B_{p}^{2} \right) + \frac{R^{2}}{J} \frac{\partial}{\partial \theta} \left( \frac{J}{R^{2}} \Nabla \psi \cdot \Nabla \theta \right) = - \mu_{0} R^{2} \frac{d p}{d \psi} - I \frac{d I}{d \psi} ,
\end{align}
where
\begin{align}
   J \equiv \left| \Nabla \psi \cdot \left( \Nabla \theta \times \Nabla \zeta \right) \right|^{-1} = \left( \vec{B} \cdot \Nabla \theta \right)^{-1} = \frac{1}{B_{p}} \dlpdtheta \label{eq:varthetaJacobian}
\end{align}
is the Jacobian and the arc length $l_{p}$ is defined such that equation \refEq{eq:dLpdtheta} holds. Simplifying further and using equation \refEq{eq:gradIdentity} we finally find that
\begin{align}
   \frac{\partial B_{p}}{\partial \psi} =& - \frac{\mu_{0}}{B_{p}} \frac{d p}{d \psi} - \frac{I}{R^{2} B_{p}} \frac{d I}{d \psi} - B_{p} \left( \dlpdtheta \right)^{-1} \frac{\partial}{\partial \psi} \left( \dlpdtheta \right) \label{eq:dBpdpsi} \\
   &+ \left( \dlpdtheta \right)^{-1} \frac{\partial}{\partial \theta} \left( B_{p} \left( \dlpdtheta \right)^{-1} \frac{\partial \vec{r}}{\partial \psi} \cdot \frac{\partial \vec{r}}{\partial \theta} \right) . \nonumber
\end{align}
Note that we have not yet determined $d I / d \psi$, but will do so below.

Next we directly differentiate equation \refEq{eq:alphaDef} to find
\begin{align}
   \Nabla \alpha =& \left( - \left. \int_{\theta_{\alpha}}^{\theta} \right|_{\psi} d \theta' \frac{\partial A_{\alpha}}{\partial \psi} + A_{\alpha} \left( \psi, \theta_{\alpha} \right) \frac{d \theta_{\alpha}}{d \psi} \right) \Nabla \psi - A_{\alpha} \left( \psi, \theta \right) \Nabla \theta + \Nabla \zeta , \label{eq:gradAlphaInitial}
\end{align}
where $A_{\alpha}$ is the integrand in the definition of $\alpha$ (see equation \refEq{eq:IalphaDef}). All quantities in equation \refEq{eq:gradAlphaInitial} are known except for the radial derivative of $A_{\alpha}$. We can calculate it by using the product rule on equation \refEq{eq:IalphaDef} to find
\begin{align}
   \frac{\partial A_{\alpha}}{\partial \psi} =& A_{\alpha} \left[ \left( 1 + \frac{I^{2}}{R^{2} B_{p}^{2}} \right) \frac{1}{I} \frac{d I}{d \psi} + \frac{\mu_{0}}{B_{p}^{2}} \frac{d p}{d \psi} - \frac{2}{R} \frac{\partial R}{\partial \psi} \right. \label{eq:dIntegranddpsi} \\
   &- \left. \frac{1}{B_{p}} \left( \dlpdtheta \right)^{-1} \frac{\partial}{\partial \theta} \left( B_{p} \left( \dlpdtheta \right)^{-1} \frac{\partial \vec{r}}{\partial \psi} \cdot \frac{\partial \vec{r}}{\partial \theta} \right) + 2 \left( \dlpdtheta \right)^{-1} \frac{\partial}{\partial \psi} \left( \dlpdtheta \right) \right] , \nonumber
\end{align}
where we have made use of equation \refEq{eq:dBpdpsi}. This form is acceptable for the purposes of this paper, but we will rearrange it into a form that is more physically illuminating. To do so we will first write
\begin{align}
   \frac{\partial \vec{r}}{\partial \psi} = \frac{1}{R B_{p}} \left( \dlpdtheta \right)^{-1} \frac{\partial \vec{r}}{\partial \theta} \times \hat{e}_{\zeta} + \left( \dlpdtheta \right)^{-2} \left( \frac{\partial \vec{r}}{\partial \psi} \cdot \frac{\partial \vec{r}}{\partial \theta} \right) \frac{\partial \vec{r}}{\partial \theta} 
\end{align}
using only equation \refEq{eq:dLpdtheta}, equation \refEq{eq:varthetaJacobian}, and vector identities such as $\partial \vec{r} / \partial \psi \cdot \left( \partial \vec{r} / \partial \theta \times \partial \vec{r} / \partial \zeta \right) = \left( \Nabla \psi \cdot \left( \Nabla \theta \times \Nabla \zeta \right) \right)^{-1}$. This allows us to see that
\begin{align}
   - \frac{2}{R} \frac{\partial R}{\partial \psi} =& - \frac{2}{R} \frac{\partial \vec{r}}{\partial \psi} \cdot \Nabla R \\
   =& - \frac{2}{R^{2} B_{p}} \left( \dlpdtheta \right)^{-1} \frac{\partial Z}{\partial \theta} + R^{2} \frac{\partial}{\partial \theta} \left( \frac{1}{R^{2}} \right) \left( \dlpdtheta \right)^{-2} \frac{\partial \vec{r}}{\partial \psi} \cdot \frac{\partial \vec{r}}{\partial \theta} . \label{eq:dRdpsi}
\end{align}
Combining this result with the second-to-last term in equation \refEq{eq:dIntegranddpsi} allows us to use the product rule several times to find
\begin{align}
   - \frac{2}{R} \frac{\partial R}{\partial \psi} &- \frac{1}{B_{p}} \left( \dlpdtheta \right)^{-1} \frac{\partial}{\partial \theta} \left( B_{p} \left( \dlpdtheta \right)^{-1} \frac{\partial \vec{r}}{\partial \psi} \cdot \frac{\partial \vec{r}}{\partial \theta} \right) \label{eq:dRdpsiRearrange} \\
   &= - \frac{2}{R^{2} B_{p}} \left( \dlpdtheta \right)^{-1} \frac{\partial Z}{\partial \theta} + R^{2} B_{p} \left( \dlpdtheta \right)^{-1} \frac{\partial}{\partial \theta} \left( \frac{1}{R^{2} B_{p}} \left( \dlpdtheta \right)^{-1} \frac{\partial \vec{r}}{\partial \psi} \cdot \frac{\partial \vec{r}}{\partial \theta} \right) \nonumber \\
   &- 2 \left( \dlpdtheta \right)^{-1} \frac{\partial}{\partial \theta} \left( \left( \dlpdtheta \right)^{-1} \frac{\partial \vec{r}}{\partial \theta} \right) \cdot \frac{\partial \vec{r}}{\partial \psi} - 2 \left( \dlpdtheta \right)^{-2} \frac{\partial}{\partial \theta} \left( \frac{\partial \vec{r}}{\partial \psi} \right) \cdot \frac{\partial \vec{r}}{\partial \theta} . \nonumber
\end{align}
Using equation \refEq{eq:dLpdtheta} we see that the last term of equation \refEq{eq:dRdpsiRearrange} exactly cancels the final term appearing in equation \refEq{eq:dIntegranddpsi}. This shows that we can rewrite equation \refEq{eq:dIntegranddpsi} as
\begin{align}
   \frac{\partial A_{\alpha}}{\partial \psi} =& A_{\alpha} \left[ \frac{1}{I} \frac{d I}{d \psi} + \frac{I}{R^{2} B_{p}^{2}} \frac{d I}{d \psi} + \frac{\mu_{0}}{B_{p}^{2}} \frac{d p}{d \psi} - \frac{2}{R^{2} B_{p}} \left( \dlpdtheta \right)^{-1} \frac{\partial Z}{\partial \theta} \right. \label{eq:dIntegranddpsiRearrange} \\
   &- \left. 2 \left( \dlpdtheta \right)^{-1} \frac{\partial}{\partial \theta} \left( \left( \dlpdtheta \right)^{-1} \frac{\partial \vec{r}}{\partial \theta} \right) \cdot \frac{\partial \vec{r}}{\partial \psi} \right] + \frac{\partial}{\partial \theta} \left( \frac{I}{R^{2} B_{p}} \left( \dlpdtheta \right)^{-1} \frac{\partial \vec{r}}{\partial \psi} \cdot \frac{\partial \vec{r}}{\partial \theta} \right) \nonumber .
\end{align}
Lastly, substituting this into equation \refEq{eq:gradAlphaInitial} and defining the poloidal curvature according to equation \refEq{eq:poloidalCurv} produces
\begin{align}
   \Nabla \alpha =& \left( - \left. \int_{\theta_{\alpha}}^{\theta} \right|_{\psi} d \theta' A_{\alpha} \left( \psi, \theta' \right) \left[ \frac{1}{I} \frac{d I}{d \psi} + \frac{I}{R^{2} B_{p}^{2}} \frac{d I}{d \psi} + \frac{\mu_{0}}{B_{p}^{2}} \frac{d p}{d \psi} - \frac{2}{R^{2} B_{p}} \left( \dlpdthetaPrime \right)^{-1} \frac{\partial Z}{\partial \theta'} \right. \right. \nonumber \\
   &+ \left. \left. \frac{2 \kappa_{p}}{R B_{p}} \right] + \left[ \frac{A_{\alpha} \left( \psi, \theta' \right)}{R^{2} B_{p}^{2}} \Nabla \psi \cdot \Nabla \theta' \right]_{\theta' = \theta_{\alpha}}^{\theta' = \theta} + A_{\alpha} \left( \psi, \theta_{\alpha} \right) \frac{d \theta_{\alpha}}{d \psi} \right) \Nabla \psi  \label{eq:gradAlphaFinal} \\
   &- A_{\alpha} \left( \psi, \theta \right) \Nabla \theta + \Nabla \zeta . \nonumber
\end{align}
The first term inside the integral represents the change in the field line pitch that results from changing the toroidal field flux function on neighboring flux surfaces. The second term in the integral accounts for the modification to the flux surface equilibrium that results from a radial gradient in the toroidal flux function just as the third term expresses the effect the pressure gradient has on the equilibrium. The fourth term corresponds to how the toroidal magnetic field weakens as the major radial location changes and the last term in the integral accounts for the flux expansion (and weakening of the poloidal magnetic field) that occurs at regions of large poloidal curvature \cite{BallShapingPenetration2015}. The term immediately following the integral accounts for the particulars of how $\theta$ is defined, but we note this term vanishes if contours of constant $\theta$ are perpendicular to the flux surface of interest. The last term in the $\Nabla \psi$ coefficient is a consequence of changing which field line is labeled $\alpha = 0$ from flux surface to flux surface. The final two terms of equation \refEq{eq:gradAlphaFinal} reflect the nonuniform spacing of the field lines in the poloidal direction and the uniform spacing in the toroidal direction respectively.

Equation \refEq{eq:gradAlphaFinal} allows us to calculate $\Nabla \psi \cdot \Nabla \alpha$, and $\left| \Nabla \alpha \right|^{2}$, but we must remember that we still lack an expression for $d I / d \psi$. This can be calculated by taking the radial gradient of the safety factor in order to get the magnetic shear,
\begin{align}
   \frac{d q}{d \psi} =& \frac{1}{2 \pi} \left. \oint_{0}^{2 \pi} \right|_{\psi} d \theta \frac{\partial A_{\alpha}}{\partial \psi} .
\end{align}
This turns out to be very closely related to $\Nabla \alpha$, so we can use equation \refEq{eq:dIntegranddpsiRearrange} to find
\begin{align}
   \frac{d I}{d \psi} =& I \left( q + \frac{1}{2 \pi} \left. \oint_{0}^{2 \pi} \right|_{\psi} d \theta A_{\alpha} \left( \psi, \theta \right) \left[ \frac{I^{2}}{R^{2} B_{p}^{2}} \right] \right)^{-1} \\
   &\times \left( \frac{d q}{d \psi} - \frac{1}{2 \pi} \left. \oint_{0}^{2 \pi} \right|_{\psi} d \theta A_{\alpha} \left( \psi, \theta \right) \left[ \frac{\mu_{0}}{B_{p}^{2}} \frac{d p}{d \psi} - \frac{2}{R^{2} B_{p}} \left( \dlpdtheta \right)^{-1} \frac{\partial Z}{\partial \theta} + \frac{2 \kappa_{p}}{R B_{p}} \right] \right) . \nonumber
\end{align}
Lastly we can directly differentiate equation \refEq{eq:millerTorField} to find
\begin{align}
   \frac{\partial B_{\zeta}}{\partial \psi} =& \frac{1}{R} \frac{d I}{d \psi} - \frac{I}{R^{2}} \left( \frac{d \psi}{d r_{\psi}} \right)^{-1} \frac{\partial R}{\partial r_{\psi}} , \label{eq:dBtordpsi}
\end{align}
where we remember that $\partial R / \partial r_{\psi}$ is an input to the Miller model. This fully determines $v_{d s \alpha}$, defined by equation \refEq{eq:driftVelAlpha}.

The expressions in this section allow us to directly calculate all of the the gyrokinetic geometric coefficients within the framework of the Miller local equilibrium model.

\section{Maximum achievable flux surface shaping}
\label{app:maxShaping}

If we try to create flux surfaces with extreme shaping, we will eventually introduce x-points into the plasma, opening the flux surfaces. Since open field lines cannot confine fusion plasmas, this provides a fundamental limit on the strength of plasma shaping (which will prove useful in section \ref{subsec:genNonMirrorSymShaping}). To quantify this we will take equation \refEq{eq:gradShafranovSolSimple} from our analysis of the constant current profile and require that $\Nabla \psi_{N} = 0$. This gives us the condition that $E_{m} = 2 b_{x}^{2-m} / m$, where $b_{x}$ is the radial location of all $m$ of the x-points. Substituting this into equations \refEq{eq:gradShafranovSolPsi_a} and \refEq{eq:gradShafranovSolPsi_b}, we arrive at 
\begin{align}
   \Delta_{x}^{-2} + \frac{2}{m} \Delta_{x}^{-m} = 1 - \frac{2}{m} , \label{eq:sepShapingCond}
\end{align}
where $\Delta_{x} \equiv b_{x} / a_{x}$ is the strongest flux surface shaping possible and $a_{x}$ is the minimum distance of the separatrix from the magnetic axis. This can be solved exactly using numerical methods or approximated analytically as
\begin{align}
   \Delta_{x} - 1 = \frac{1.2785}{m} + O \left( m^{-2} \right) , \label{eq:sepShapingSol}
\end{align}
in the limit that $m \gg 1$. The numerical constant in equation \refEq{eq:sepShapingSol} is the solution $x$ to
\begin{align}
   x - \Exp{-x} = 1 .
\end{align}
Hence, we can conclude that, given a constant current profile, $\Delta_{m} - 1 \sim m^{-1}$ is the strongest possible scaling. Any scaling stronger than this will necessarily introduce x-points into the plasma.

\section{Non-mirror symmetric geometric coefficients}
\label{app:nonMirrorGeoCoeffs}

In this section we give the full gyrokinetic geometric coefficients to lowest and next order in $m \ll 1$ for the geometry investigated in section \ref{subsec:practicalNonMirrorSymShaping}. These coefficients are accurate to lowest order in aspect ratio, given the ordering of equations \refEq{eq:weakShapingOrdering} and \refEq{eq:changeShapeOrdering}. In deriving these coefficients the following quantities are useful as waypoints:
\begin{align}
   \frac{\partial R}{\partial \theta} =& - r_{\psi 0} \left[ \Sin{\theta} - \frac{1}{2} \Cos{\theta} \Big( m \left( \Delta_{m} - 1 \right) \Sin{z_{m s}} + n \left( \Delta_{n} - 1 \right) \Sin{z_{n s}} \Big) \right. \\
   &- \left. \frac{1}{2} \Sin{\theta} \Big( \left( \Delta_{m} - 1 \right) \Cos{z_{m s}} + \left( \Delta_{n} - 1 \right) \Cos{z_{n s}} \Big) \right] + O \left( m^{-3} r_{\psi 0} \right) \nonumber \\
   \frac{\partial Z}{\partial \theta} =& r_{\psi 0} \left[ \Cos{\theta} + \frac{1}{2} \Sin{\theta} \Big( m \left( \Delta_{m} - 1 \right) \Sin{z_{m s}} + n \left( \Delta_{n} - 1 \right) \Sin{z_{n s}} \Big) \right. \\
   &- \left. \frac{1}{2} \Cos{\theta} \Big( \left( \Delta_{m} - 1 \right) \Cos{z_{m s}} + \left( \Delta_{n} - 1 \right) \Cos{z_{n s}} \Big) \right] + O \left( m^{-3} r_{\psi 0} \right) \nonumber
\end{align}
\begin{align}
   \Nabla r_{\psi} =& \left[ \Cos{\theta} + \frac{r_{\psi 0}}{2} \Cos{\theta} \left( \frac{d \Delta_{m}}{d r_{\psi}} \Cos{z_{m s}} + \frac{d \Delta_{n}}{d r_{\psi}} \Cos{z_{n s}} \right) \right. \nonumber \\
   &+ \left. \frac{1}{2} \Sin{\theta} \Big( m \left( \Delta_{m} - 1 \right) \Sin{z_{m s}} + n \left( \Delta_{n} - 1 \right) \Sin{z_{n s}} \Big) \right] \hat{e}_{R} \\
   &+ \left[ \Sin{\theta} + \frac{r_{\psi 0}}{2} \Sin{\theta} \left( \frac{d \Delta_{m}}{d r_{\psi}} \Cos{z_{m s}} + \frac{d \Delta_{n}}{d r_{\psi}} \Cos{z_{n s}} \right) \right. \nonumber \\
   &- \left. \frac{1}{2} \Cos{\theta} \Big( m \left( \Delta_{m} - 1 \right) \Sin{z_{m s}} + n \left( \Delta_{n} - 1 \right) \Sin{z_{n s}} \Big) \right] \hat{e}_{Z} + O \left( m^{-2} \right) \nonumber \\
      \Nabla \theta =& \frac{1}{r_{\psi 0}} \Big( - \Sin{\theta} \hat{e}_{R} + \Cos{\theta} \hat{e}_{Z} \Big) + O \left( \frac{m^{-2}}{r_{\psi 0}} \right)
\end{align}
\begin{align}
   \frac{\partial R}{\partial r_{\psi}} =& \Cos{\theta} - \frac{r_{\psi 0}}{2} \Cos{\theta} \left( \frac{d \Delta_{m}}{d r_{\psi}} \Cos{z_{m s}} + \frac{d \Delta_{n}}{d r_{\psi}} \Cos{z_{n s}} \right) + O \left( m^{-2} \right) \\
   \frac{\partial Z}{\partial r_{\psi}} =& \Sin{\theta} - \frac{r_{\psi 0}}{2} \Sin{\theta} \left( \frac{d \Delta_{m}}{d r_{\psi}} \Cos{z_{m s}} + \frac{d \Delta_{n}}{d r_{\psi}} \Cos{z_{n s}} \right) + O \left( m^{-2} \right)
\end{align}
\begin{align}
   \left( \frac{\partial A_{\alpha}}{\partial \psi} \right)_{\text{orthog}} =& B_{0} \left( \frac{d \psi}{d r_{\psi}} \right)^{-2} \left[ \hat{s}' + \frac{r_{\psi 0}}{2} \left( m^{2} \left( \Delta_{m} - 1 \right) \frac{d \Delta_{m}}{d r_{\psi}} + n^{2} \left( \Delta_{n} - 1 \right) \frac{d \Delta_{n}}{d r_{\psi}} \right) \right. \nonumber \\
   &- \left( m^{2} \left( \Delta_{m} - 1 \right) + \left( 3 \hat{s}' - 2 \right) \frac{r_{\psi 0}}{2} \frac{d \Delta_{m}}{d r_{\psi}} \right) \Cos{z_{m s}} \nonumber \\
   &- \left( n^{2} \left( \Delta_{n} - 1 \right) + \left( 3 \hat{s}' - 2 \right) \frac{r_{\psi 0}}{2} \frac{d \Delta_{n}}{d r_{\psi}} \right) \Cos{z_{n s}} \\
   &+ \frac{r_{\psi 0}}{2} \left( m^{2} \left( \Delta_{m} - 1 \right) \frac{d \Delta_{m}}{d r_{\psi}} \Cos{2 z_{m s}} + n^{2} \left( \Delta_{n} - 1 \right) \frac{d \Delta_{n}}{d r_{\psi}} \Cos{2 z_{n s}} \right) \nonumber \\
   &+ \frac{r_{\psi 0}}{2} \left( m^{2} \left( \Delta_{m} - 1 \right) \frac{d \Delta_{n}}{d r_{\psi}} + n^{2} \left( \Delta_{n} - 1 \right) \frac{d \Delta_{m}}{d r_{\psi}} \right) \nonumber \\
   &\times \left. \Big( \Cos{z_{m s} + z_{n s}} + \Cos{z_{m s} - z_{n s}} \Big) \right] + O \left( \frac{m^{-2}}{r_{\psi 0}^{2} B_{0}} \right) \nonumber
\end{align}
\begin{align}
   \left. \int_{\theta_{0}}^{\theta} \right|_{\psi} d \theta' \left( \frac{\partial A_{\alpha}}{\partial \psi} \right)_{\text{orthog}} =& B_{0} \left( \frac{d \psi}{d r_{\psi}} \right)^{-2} \left[ \hat{s}' \theta + \frac{r_{\psi 0}}{2} \left( m^{2} \left( \Delta_{m} - 1 \right) \frac{d \Delta_{m}}{d r_{\psi}} + n^{2} \left( \Delta_{n} - 1 \right) \frac{d \Delta_{n}}{d r_{\psi}} \right) \theta \right. \nonumber \\
   &- \frac{1}{m} \left( m^{2} \left( \Delta_{m} - 1 \right) + \left( 3 \hat{s}' - 2 \right) \frac{r_{\psi 0}}{2} \frac{d \Delta_{m}}{d r_{\psi}} \right) \Sin{z_{m s}} \nonumber \\
   &- \frac{1}{n} \left( n^{2} \left( \Delta_{n} - 1 \right) + \left( 3 \hat{s}' - 2 \right) \frac{r_{\psi 0}}{2} \frac{d \Delta_{n}}{d r_{\psi}} \right) \Sin{z_{n s}} \\
   &+ \frac{r_{\psi 0}}{4} \left( m \left( \Delta_{m} - 1 \right) \frac{d \Delta_{m}}{d r_{\psi}} \Sin{2 z_{m s}} + n \left( \Delta_{n} - 1 \right) \frac{d \Delta_{n}}{d r_{\psi}} \Sin{2 z_{n s}} \right) \nonumber \\
   &+ \frac{r_{\psi 0}}{2} \left( m^{2} \left( \Delta_{m} - 1 \right) \frac{d \Delta_{n}}{d r_{\psi}} + n^{2} \left( \Delta_{n} - 1 \right) \frac{d \Delta_{m}}{d r_{\psi}} \right) \nonumber \\
   &\times \left. \left( \frac{1}{m + n} \Sin{z_{m s} + z_{n s}} + \frac{1}{m - n} \Sin{z_{m s} - z_{n s}} \right) \right] + O \left( \frac{m^{-2}}{r_{\psi 0}^{2} B_{0}} \right) \nonumber
\end{align}
\begin{align}
   \frac{\partial \alpha}{\partial \psi} =& - B_{0} \left( \frac{d \psi}{d r_{\psi}} \right)^{-2} \Bigg[ \hat{s}' \theta + \frac{r_{\psi 0}}{2} \left( m^{2} \left( \Delta_{m} - 1 \right) \frac{d \Delta_{m}}{d r_{\psi}} + n^{2} \left( \Delta_{n} - 1 \right) \frac{d \Delta_{n}}{d r_{\psi}} \right) \theta \nonumber \\
   &- \frac{1}{2} \Big( m \left( \Delta_{m} - 1 \right) \Sin{z_{m s}} + n \left( \Delta_{n} - 1 \right) \Sin{z_{n s}} \Big) \\
   &+ \frac{r_{\psi 0}}{2 \left( n - m \right)} \left( m^{2} \left( \Delta_{m} - 1 \right) \frac{d \Delta_{n}}{d r_{\psi}} + n^{2} \left( \Delta_{n} - 1 \right) \frac{d \Delta_{m}}{d r_{\psi}} \right) \nonumber \\
   &\times \Sin{\left( n - m \right) \theta - m \left( \theta_{t m} - \theta_{t n} \right)} \Bigg] + O \left( \frac{m^{-2}}{r_{\psi 0}^{2} B_{0}} \right) \nonumber
\end{align}
\begin{align}
   \Nabla \alpha =& - B_{0} \left( \frac{d \psi}{d r_{\psi}} \right)^{-1} \Bigg\{ \nonumber \\
   & \Bigg[ - \Sin{\theta} + \hat{s}' \theta \Cos{\theta} + \frac{r_{\psi 0}}{2} \left( m^{2} \left( \Delta_{m} - 1 \right) \frac{d \Delta_{m}}{d r_{\psi}} + n^{2} \left( \Delta_{n} - 1 \right) \frac{d \Delta_{n}}{d r_{\psi}} \right) \theta \Cos{\theta} \nonumber \\
   &- \frac{1}{2} \Big( \Cos{\theta} - \hat{s}' \theta \Sin{\theta} \Big) \Big( m \left( \Delta_{m} - 1 \right) \Sin{z_{m s}} + n \left( \Delta_{n} - 1 \right) \Sin{z_{n s}} \Big) \nonumber \\
   &+ \frac{r_{\psi 0}}{2} \Big( \Sin{\theta} + \hat{s}' \theta \Cos{\theta} \Big) \left( \frac{d \Delta_{m}}{d r_{\psi}} \Cos{z_{m s}} + \frac{d \Delta_{n}}{d r_{\psi}} \Cos{z_{n s}} \right) \nonumber \\
   &+ \frac{r_{\psi 0}}{2 \left( n - m \right)} \left( m^{2} \left( \Delta_{m} - 1 \right) \frac{d \Delta_{n}}{d r_{\psi}} + n^{2} \left( \Delta_{n} - 1 \right) \frac{d \Delta_{m}}{d r_{\psi}} \right) \Cos{\theta} \nonumber \\
   &\times \Sin{\left( n - m \right) \theta - m \left( \theta_{t m} - \theta_{t n}} \right) \Bigg] \hat{e}_{R} \\
   &+ \Bigg[ \Cos{\theta} + \hat{s}' \theta \Sin{\theta} + \frac{r_{\psi 0}}{2} \left( m^{2} \left( \Delta_{m} - 1 \right) \frac{d \Delta_{m}}{d r_{\psi}} + n^{2} \left( \Delta_{n} - 1 \right) \frac{d \Delta_{n}}{d r_{\psi}} \right) \theta \Sin{\theta} \nonumber \\
   &- \frac{1}{2} \Big( \Sin{\theta} + \hat{s}' \theta \Cos{\theta} \Big) \Big( m \left( \Delta_{m} - 1 \right) \Sin{z_{m s}} + n \left( \Delta_{n} - 1 \right) \Sin{z_{n s}} \Big) \nonumber \\
   &- \frac{r_{\psi 0}}{2} \Big( \Cos{\theta} - \hat{s}' \theta \Sin{\theta} \Big) \left( \frac{d \Delta_{m}}{d r_{\psi}} \Cos{z_{m s}} + \frac{d \Delta_{n}}{d r_{\psi}} \Cos{z_{n s}} \right) \nonumber \\
   &+ \frac{r_{\psi 0}}{2 \left( n - m \right)} \left( m^{2} \left( \Delta_{m} - 1 \right) \frac{d \Delta_{n}}{d r_{\psi}} + n^{2} \left( \Delta_{n} - 1 \right) \frac{d \Delta_{m}}{d r_{\psi}} \right) \Sin{\theta} \nonumber \\
   &\times \Sin{\left( n - m \right) \theta - m \left( \theta_{t m} - \theta_{t n}} \right) \Bigg] \hat{e}_{Z} \Bigg\} + O \left( \frac{m^{-2}}{r_{\psi 0}} \right) . \nonumber
\end{align}
Here $\theta_{0}$ is defined such that the resulting integral does not have a term that is constant in poloidal angle.

The $O \left( 1 \right)$ geometric coefficients are simply those of a circular tokamak and are given by
\begin{align}
  \left( \hat{b} \cdot \Nabla \theta \right)_{0} =& \overline{\left( \hat{b} \cdot \Nabla \theta \right)}_{0} = \frac{1}{r_{\psi 0} R_{0} B_{0}} \frac{d \psi}{d r_{\psi}} \label{eq:gradparO0} \\
  v_{d s \psi 0} =& \overline{v}_{d s \psi 0} = - \frac{1}{R_{0} \Omega_{s}} \frac{d \psi}{d r_{\psi}} \Sin{\theta} \label{eq:psiDriftO0} \\
  v_{d s \alpha 0} =& \overline{v}_{d s \alpha 0} = \frac{B_{0}}{R_{0} \Omega_{s}} \left( \frac{d \psi}{d r_{\psi}} \right)^{-1} \left( \Cos{\theta} + \hat{s}' \theta \Sin{\theta} \right) \label{eq:alphaDriftO0} \\
  \left| \Nabla \psi \right|^{2}_{0} =& \overline{\left| \Nabla \psi \right|}^{2}_{0} = \left( \frac{d \psi}{d r_{\psi}} \right)^{2} \\
  \left( \Nabla \psi \cdot \Nabla \alpha \right)_{0} =& \overline{\left( \Nabla \psi \cdot \Nabla \alpha \right)}_{0} = - B_{0} \hat{s}' \theta \\
  \left| \Nabla \alpha \right|^{2}_{0} =& \overline{\left| \Nabla \alpha \right|^{2}}_{0} = B_{0}^{2} \left( \frac{d \psi}{d r_{\psi}} \right)^{-2} \left( 1 + \hat{s}'^{2} \theta^{2} \right) \label{eq:gradAlphaSqO0} \\
  \left( J_{0} \left( k_{\perp} \rho_{s} \right) \right)_{0} =& \overline{\left( J_{0} \left( k_{\perp} \rho_{s} \right) \right)}_{0} = J_{0} \left( k_{\perp 0} \rho_{s} \right) , \label{eq:FLRO0}
\end{align}
where $\hat{s}'$ is defined by equation \refEq{eq:shiftedShatDef} and
\begin{align}
  k_{\perp 0} \rho_{s} \equiv& \sqrt{\frac{2 m_{s} \mu}{Z_{s}^{2} e^{2} B_{0}}} \sqrt{k_{\psi}^{2} \left| \Nabla \psi \right|^{2}_{0} + 2 k_{\psi} k_{\alpha} \left( \Nabla \psi \cdot \Nabla \alpha \right)_{0} + k_{\alpha}^{2} \left| \Nabla \alpha \right|^{2}_{0}} . \label{eq:FLR0def}
\end{align}
Note that all of the coefficients are independent of the short spatial scale coordinate, $z$.

To $O \left( m^{-1} \right)$ the geometric coefficients are
\begin{align}
  \left( \hat{b} \cdot \Nabla \theta \right)_{1} =& \frac{1}{2 R_{0} B_{0}} \frac{d \psi}{d r_{\psi}} \left( \frac{d \Delta_{m}}{d r_{\psi}} \Cos{z_{m s}} + \frac{d \Delta_{n}}{d r_{\psi}} \Cos{z_{n s}} \right) \label{eq:gradparO1} \\
  v_{d s \psi 1} =& \frac{1}{2 R_{0} \Omega_{s}} \frac{d \psi}{d r_{\psi}} \Bigg[ \Cos{\theta} \Big( m \left( \Delta_{m} - 1 \right) \Sin{z_{m s}} + n \left( \Delta_{n} - 1 \right) \Sin{z_{n s}} \Big) \nonumber \\
  &- r_{\psi 0} \Sin{\theta} \left( \frac{d \Delta_{m}}{d r_{\psi}} \Cos{z_{m s}} + \frac{d \Delta_{n}}{d r_{\psi}} \Cos{z_{n s}} \right) \Bigg] \label{eq:psiDriftO1} \\
  v_{d s \alpha 1} =& \frac{B_{0}}{2 R_{0} \Omega_{s}} \left( \frac{d \psi}{d r_{\psi}} \right)^{-1} \Bigg[ r_{\psi 0} \left( m^{2} \left( \Delta_{m} - 1 \right) \frac{d \Delta_{m}}{d r_{\psi}} + n^{2} \left( \Delta_{n} - 1 \right) \frac{d \Delta_{n}}{d r_{\psi}} \right) \theta \Sin{\theta} \nonumber \\
  &- \Big( \Sin{\theta} + \hat{s}' \theta \Cos{\theta} \Big) \Big( m \left( \Delta_{m} - 1 \right) \Sin{z_{m s}} + n \left( \Delta_{n} - 1 \right) \Sin{z_{n s}} \Big) \nonumber \\
  &- r_{\psi 0} \Big( \Cos{\theta} - \hat{s}' \theta \Sin{\theta} \Big) \left( \frac{d \Delta_{m}}{d r_{\psi}} \Cos{z_{m s}} + \frac{d \Delta_{n}}{d r_{\psi}} \Cos{z_{n s}} \right) \label{eq:alphaDriftO1} \\
  &+ \frac{r_{\psi 0}}{\left( n - m \right)} \left( m^{2} \left( \Delta_{m} - 1 \right) \frac{d \Delta_{n}}{d r_{\psi}} + n^{2} \left( \Delta_{n} - 1 \right) \frac{d \Delta_{m}}{d r_{\psi}} \right) \Sin{\theta} \nonumber \\
  &\times \Big( \Sin{\left( n - m \right) \theta} \Cos{m \left( \theta_{t m} - \theta_{t n} \right)} - \Cos{\left( n - m \right) \theta} \Sin{m \left( \theta_{t m} - \theta_{t n} \right)} \Big) \Bigg] \nonumber \\
  \left| \Nabla \psi \right|^{2}_{1} =& r_{\psi 0} \left( \frac{d \psi}{d r_{\psi}} \right)^{2} \left( \frac{d \Delta_{m}}{d r_{\psi}} \Cos{z_{m s}} + \frac{d \Delta_{n}}{d r_{\psi}} \Cos{z_{n s}} \right) \label{eq:gradPsiSqO1} \\
    \left( \Nabla \psi \cdot \Nabla \alpha \right)_{1} =& - B_{0} \Bigg[ \frac{r_{\psi 0}}{2} \left( m^{2} \left( \Delta_{m} - 1 \right) \frac{d \Delta_{m}}{d r_{\psi}} + n^{2} \left( \Delta_{n} - 1 \right) \frac{d \Delta_{n}}{d r_{\psi}} \right) \theta \nonumber \\
    &- \Big( m \left( \Delta_{m} - 1 \right) \Sin{z_{m s}} + n \left( \Delta_{n} - 1 \right) \Sin{z_{n s}} \Big) \nonumber \\
    &+ r_{\psi 0} \hat{s}' \theta \left( \frac{d \Delta_{m}}{d r_{\psi}} \Cos{z_{m s}} + \frac{d \Delta_{n}}{d r_{\psi}} \Cos{z_{n s}} \right) \label{eq:gradPsiDotGradAlphaO1} \\
    &+ \frac{r_{\psi 0}}{2 \left( n - m \right)} \left( m^{2} \left( \Delta_{m} - 1 \right) \frac{d \Delta_{n}}{d r_{\psi}} + n^{2} \left( \Delta_{n} - 1 \right) \frac{d \Delta_{m}}{d r_{\psi}} \right) \nonumber \\
    &\times \Big( \Sin{\left( n - m \right) \theta} \Cos{m \left( \theta_{t m} - \theta_{t n} \right)} - \Cos{\left( n - m \right) \theta} \Sin{m \left( \theta_{t m} - \theta_{t n} \right)} \Big) \Bigg] \nonumber \\
  \left| \Nabla \alpha \right|^{2}_{1} =& B_{0}^{2} \left( \frac{d \psi}{d r_{\psi}} \right)^{-2} \Bigg[ r_{\psi 0} \left( m^{2} \left( \Delta_{m} - 1 \right) \frac{d \Delta_{m}}{d r_{\psi}} + n^{2} \left( \Delta_{n} - 1 \right) \frac{d \Delta_{n}}{d r_{\psi}} \right) \hat{s}' \theta^{2} \nonumber \\
  &- 2 \hat{s}' \theta \Big( m \left( \Delta_{m} - 1 \right) \Sin{z_{m s}} + n \left( \Delta_{n} - 1 \right) \Sin{z_{n s}} \Big) \nonumber \\
  &- r_{\psi 0} \left( 1 - \hat{s}'^{2} \theta^{2} \right) \left( \frac{d \Delta_{m}}{d r_{\psi}} \Cos{z_{m s}} + \frac{d \Delta_{n}}{d r_{\psi}} \Cos{z_{n s}} \right) \label{eq:gradAlphaSqO1} \\
  &+ \frac{r_{\psi 0}}{n - m} \left( m^{2} \left( \Delta_{m} - 1 \right) \frac{d \Delta_{n}}{d r_{\psi}} + n^{2} \left( \Delta_{n} - 1 \right) \frac{d \Delta_{m}}{d r_{\psi}} \right) \hat{s}' \theta \nonumber \\
  &\times \Big( \Sin{\left( n - m \right) \theta} \Cos{m \left( \theta_{t m} - \theta_{t n} \right)} - \Cos{\left( n - m \right) \theta} \Sin{m \left( \theta_{t m} - \theta_{t n} \right)} \Big) \Bigg] \nonumber \\
  \left( J_{0} \left( k_{\perp} \rho_{s} \right) \right)_{1} =& - k_{\perp 1} \rho_{s} J_{1} \left( k_{\perp 0} \rho_{s} \right) , \label{eq:FLRO1avg}
\end{align}
where $z_{m s}$ and $z_{n s}$ are defined by equations \refEq{eq:zmsDef} and \refEq{eq:znsDef} and
\begin{align}
  k_{\perp 1} \rho_{s} \equiv& \frac{k_{\perp 0} \rho_{s}}{2} \frac{k_{\psi}^{2} \left| \Nabla \psi \right|_{1}^{2} + 2 k_{\psi} k_{\alpha} \left( \Nabla \psi \cdot \Nabla \alpha \right)_{1} + k_{\alpha}^{2} \left| \Nabla \alpha \right|_{1}^{2}}{k_{\psi}^{2} \left| \Nabla \psi \right|_{0}^{2} + 2 k_{\psi} k_{\alpha} \left( \Nabla \psi \cdot \Nabla \alpha \right)_{0} + k_{\alpha}^{2} \left| \Nabla \alpha \right|_{0}^{2}} . \label{eq:FLRO1def}
\end{align}
From the last terms in each of equations \refEq{eq:alphaDriftO1}, \refEq{eq:gradPsiDotGradAlphaO1}, and \refEq{eq:gradAlphaSqO1} we see that (even after averaging over $z$) $v_{d s \alpha 1}$, $\left( \Nabla \psi \cdot \Nabla \alpha \right)_{1}$, and $\left| \Nabla \alpha \right|^{2}_{1}$ are all up-down asymmetric.

\section*{References}
\bibliographystyle{unsrt}
\bibliography{references.bib}

\end{document}